\documentclass[11pt,letterpaper]{JHEP3}
\usepackage{graphicx}
\usepackage{amsmath}
\usepackage{amssymb}

\title{Vector Bosons in the Randall-Sundrum 2 and Lykken-Randall 
  models and unparticles}

\author{Alexander Friedland, Maurizio Giannotti, 
Michael L. Graesser \\ 
Theoretical Division, T-2, MS B285, Los Alamos National Laboratory, Los Alamos, NM 87540 \\
E-mails: \email{friedland@lanl.gov, maurizio@lanl.gov, graesser@lanl.gov}}

\abstract{ Unparticle behavior is shown to be realized in the
  Randall-Sundrum 2 (RS 2) and the Lykken-Randall (LR) brane scenarios
  when brane-localized Standard Model currents are coupled to a
  massive vector field living in the five-dimensional warped
  background of the RS 2 model. By the AdS/CFT dictionary these
  backgrounds exhibit certain properties of the unparticle CFT at
  large $N_c$ and strong 't Hooft coupling.  Within the RS 2 model we
  also examine and contrast in detail the scalar and vector
  position-space correlators at intermediate and large distances.
  Unitarity of brane-to-brane scattering amplitudes is seen to imply a
  necessary and sufficient condition on the positivity of the bulk
  mass, which leads to the well-known unitarity bound on vector
  operators in a CFT. }

\preprint{LA-UR-08-08093}


\begin{document} 

\section{Introduction}

Since its introduction the unparticle physics scenario of Georgi
\cite{GeorgiI,GeorgiII} has attracted a considerable amount of
attention. The premise of this scenario is the existence of
interactions between Standard Model (SM) and a hidden
conformal field theory (CFT) sector.
A key distinction compared to
earlier models coupling the SM (and its supersymmetric
generalizations) to approximate CFTs ({\it e.g.},
\cite{conformal-sequestering}) is that Georgi's hidden sector CFT is
conformal {\em below} the TeV scale. At low energies accessible to the
experiments, there are effective couplings
between SM currents and CFT operators.  As an example, a vector
current in the SM, $j^{\mu}_{SM}$, is coupled to a vector operator
${\cal O}_\mu$ in the CFT via
\begin{equation}
  \frac{c_0}{\Lambda^{d_V-1}} j^{\mu}_{SM} {\cal O}_{\mu},
  \label{unparticle-coupling}
\end{equation}
where $c_0$ is a dimensionless coupling and $d_V$ is the conformal
dimension of ${\cal O}_\mu$, not necessarily an integer. 

The resulting phenomenology can be quite interesting and qualitatively
different from the commonly considered scenarios of new physics, in
which new particles have definite masses \cite{GeorgiI,GeorgiII}.
States in the CFT can be excited either through energetic collisions
between, or in the decays of, SM particles.  For example,
SM-unparticle interactions could lead to processes with unparticles
${\cal U}$ in the final state, {\it e.g.}, $q+\bar{q}\rightarrow G +
{\cal U}$, $t \rightarrow c + {\cal U}$ \cite{GeorgiI}, as well as
provide additional channels for processes between the SM particles,
{\it e.g.}, in the $e^+e^- \rightarrow \mu^+\mu^-$ scattering
\cite{GeorgiII}.  For a representative list of references of various
signatures of the unparticles in
collider physics, astrophysics, neutrino oscillations, {\it etc}, see, {\it
  e.g.}, \cite{Cheung:2008xu}.

In addition to phenomenological signatures, as stressed by Georgi
himself \cite{GeorgiColloquium}, there are many interesting
theoretical issues surrounding unparticles that deserve investigation.
In fact, over the last two years, many thought-provoking discussions
of the subject have emerged. For example, it was shown how the
unparticle spectrum could be discretized and how the effect could be
modeled with warped extra dimensions \cite{stephanov}. This
discretization and its connection to the ``hidden valley'' framework
\cite{hiddenvalley} was further discussed in \cite{Strassler:2008bv}.
The connection between unparticles and QCD-like theories, including
an approximate power-law scaling of the QCD spectral function, was
discussed in \cite{neubert}. 

The unparticle scenario inspired an intriguing proposal for solving
the ``little hierarchy problem'' by promoting the Higgs Boson to a
``UnHiggs'' having a large anomalous dimension and a gapped continuous
mass spectrum \cite{Stancato:2008mp,
  Falkowski:2008yr,Falkowski:2009uy}. An ``unparticle action'' that
can be used to describe unparticle physics in a range of conformal
dimensions \cite{terning2} was proposed in \cite{coloredterning}, with
several consistency checks using 't Hooft anomaly matching performed
in \cite{Galloway:2008sq}.

Several crucial observations about unparticles were made by
Grinstein, Intriligator and Rothstein (GIR) \cite{GIR}, as described
in details below. Finally, the work of Georgi and Kats
\cite{Georgi:2008pq,Georgi:2009xq} explored several important conceptual
issues in unparticle physics, such as the process of dimensional
transmutation and unparticle self-interactions, using an exactly solvable
realization in two dimensions.

The goal of this paper is to seek a model that has unparticle behavior in
four spacetime dimensions. The motivation is two-fold. First,
it is an important issue of principle: having a concrete model
of this type would provide a laboratory for addressing conceptual
questions in unparticle physics.
Second, such a model can be used as
a framework for phenomenological studies, and may help to avoid
certain pitfalls.

Of course, it must be kept in mind that certain properties can -- and,
in fact, as we discuss later, do -- vary between different
realizations of unparticles. At the same time, certain others are
universal, being consequences of the basic principles, such as
conformal invariance or dimensional analysis. These universal
properties must be reproduced by any candidate realization of the
unparticle physics scenario.

What are these universal properties? First of all, the ``unparticle
propagator'' should have the conformal scaling behavior and also,
importantly, a certain phase.  Refs.~\cite{GeorgiI,GeorgiII} obtained
these results by
imposing scale invariance on the spectral function,
\begin{eqnarray}
  \label{eq:spectralGeorgi}
  \langle {\cal O}(p) {\cal O}(-p) \rangle \propto
  \int_0^\infty dM^2  \frac{(M^2)^{d-2}}{p^2-M^2+i\epsilon}=
  \frac{\pi(-p^2-i\epsilon)^{d-2}}{\sin d\pi} 
= \frac{\pi(p^2)^{d-2} e^{-i(d-2)\pi}}{\sin d\pi}.
\end{eqnarray}
Next, as noted by GIR and also in \cite{Nakayama:2007qu}, the
dimension $d$ of the unparticle propagator should satisfy the CFT
unitarity bounds \cite{Mack}.  Furthermore, GIR noted that values
$d\ge2$ at which the integral in (\ref{eq:spectralGeorgi}) diverges
are allowed. For those values, the unparticle scenario must
additionally contain contact interactions between the SM fields. These
contact interactions are \emph{necessary} to cure the divergence in
the spectral integral and, moreover, are very important
phenomenologically, as they \emph{dominate} over the unparticles in
SM-SM scattering processes.  Finally, the tensor structure of the
unparticle propagator is fixed by the conformal group
\cite{OsbornPetkou,ADSCFT_Vector}. In particular, in position space,
the CFT vector two-point function is
\begin{eqnarray}\label{eq:GIRcorr_position}
    \langle {\cal O}_\mu(x) {\cal O}_\nu(0)\rangle 
    = \frac{1}{2\pi^2} \frac{\eta_{\mu\nu}-2 x_\mu x_\nu/x^2}{x^{2d_V}},
\end{eqnarray}
which in momentum space becomes (for $p^2>0$) \cite{GIR}
\begin{eqnarray}
  \label{eq:GIRcorr}
  \langle {\cal O}_{\mu}(p) {\cal O}_{\nu}(-p) \rangle& =& \frac{(d_V-1)\Gamma(2-d_V)}{4^{d_V-1}\Gamma(d_V+1)}
  (-p^2)^{d_V-2}
  \left[
    -  \eta_{\mu\nu} +\frac{2(d_V-2)}{d_V-1}\frac{p_\mu p_\nu}{p^2}
  \right].
\end{eqnarray}

We propose here that the models based on warped extra spacetime
dimensions, specifically the famous Randall-Sundrum 2 (RS 2,
\cite{RSII}) and Lykken-Randall (LR, \cite{lykken-randall}) brane
constructions, with the SM fields on the brane and new fields in the
bulk in fact realize unparticle physics.  We will show, using a simple
example of the bulk vector field, that both of these models reproduce
all the requisite properties listed above. 

Right at the outset, we would like to make the following two comments.
Firstly, ours is not the first assertion that
holographic\footnote{{\it I.e.}, those based on the AdS/CFT correspondence.}
constructions could realize unparticle physics \cite{stephanov,
  Strassler:2008bv, terning2,Falkowski:2008fz}. The issue is
whether such constructions yield theories that are merely similar to
unparticle physics (``unparticle-like''), or are genuine
realizations of it. At the moment, there does not seem to
be a consensus in the literature on this point. To the best of our
knowledge, ours is the first systematic analysis that
establishes \emph{all} of the unparticle properties in these setups.

Secondly, these holographic models are different from the framework
for the unparticle scenario originally envisioned in
\cite{GeorgiI,GeorgiII}.  The latter involves a purely
four-dimensional Banks-Zaks (${\cal BZ}$) \cite{BanksZaks} sector
coupled to the SM by messenger fields at a high mass scale, $M_{\cal
  U}$. If below $M_{\cal U}$ the ${\cal BZ}$ couplings flow into an
infrared fixed point -- at the ``transmutation" scale $\Lambda_{\cal
  U} \ll M_{\cal U}$ -- the hidden CFT sector is obtained\footnote{The
  scale $\Lambda$ appearing in Eq.~(\ref{unparticle-coupling}) is then
  a phenomenological scale, depending on both $M_{\cal U}$ and
  $\Lambda_{\cal U}$.}.  It is important to stress
that, conceptually, there is nothing inherently superior or
inferior about one framework versus another.  In fact, they model
the unparticle sector in different regimes. The ${\cal BZ}$ realization teaches us about the unparticle
sector in the weak (perturbative) regime, and can be used quite
effectively, as demonstrated by GIR. Instead, the RS 2/LR realizations
allow us to extend their results to strong coupling (large $N_c$).  
We will return to this important point at the end of the paper.


From the practical standpoint, the RS 2/LR constructions make it
possible to study what would be a quantum behavior in the CFT sector
with classical equations in the bulk. This makes many of the key
unparticle effects, such as the contact terms, the production of
unparticles, and the CFT unitarity bounds, particularly transparent
and intuitive. It also allows us to easily go beyond simply confirming
these properties. With little additional effort, we find several
interesting effects: (i) We see how the contact terms are resolved at
short distances. (ii) We show that, unlike in the scalar case, a
vector in AdS cannot have a negative mass squared. (iii) Finally, we
explore an interesting interplay between long-distance (pure CFT) and
low-momentum-transfer (CFT subdominant) behaviors.

A brief outline of this paper is as follows. In
Sect.~\ref{sect:leterature}, we review some of the relevant work on
the AdS/CFT correspondence and vector fields in warped backgrounds.
Sect.~\ref{sect:preliminary considerations} contains a preliminary
discussion of the spectral function, as well as of bulk fields in flat
extra dimensions. This discussion is intended as a precursor to our
analysis of the RS 2 and LR models. Sect.~\ref{RS2-LR} derives the
bulk field equations (Section \ref{sec:EOM}) and the boundary
conditions (Section \ref{sec:brane-bc}) for the RS 2 and LR models.

The main analysis for the RS 2 model is presented in
Sect.~\ref{sec:RS2}. The propagator is derived in Sects.~\ref{sec:RS2
  transverse polarization}, \ref{sec:RS2 longitudinal polarization},
\ref{sec:greensfunction}. The unparticle properties are established in
Sect.~\ref{sect:RS2
  analysis} and the position space propagator is studied in
Sect.~\ref{sect:position}. Sect.~\ref{sec:tensionlessbrane} discusses
features of the brane-to-brane propagator for SM observers localized
to a LR brane. We conclude in Sect.~\ref{conclusions}.

This paper is a continuation and an extension of
\cite{Friedland:2009iy} where some of the main results for the RS 2 model
were stated in a condensed form. The reader may wish to consult that
paper for a short summary and overall discussion. Most of the relevant
derivations are omitted there and presented here for the first time.
Additionally, the results for the LR model here are new.

\section{Literature overview}
\label{sect:leterature}

That the RS 2 model has a connection to a CFT is very well known,
having been established ten years ago by Maldacena (unpublished),
Witten \cite{WittenSantaBarbara} and later by \cite{Gubser:1999vj},
\cite{GiddingsKatzRandall}, \cite{ArkaniHamed:2000ds},
\cite{PerezVictoria:2001pa}, \cite{Contino:2004vy} and others. The
holographic interpretation of the LR model has also been discussed,
for example in \cite{ArkaniHamed:2000ds}. It should not then be {\it a
  priori} surprising that models based on warped extra dimensions are
related to unparticle physics.

The connections of RS 2 and LR to conformal field theories of course
relies on the celebrated AdS/CFT correspondence
\cite{Maldacena,Gubser:1998bc,Witten}. In fact, as shown in
\cite{Witten}, any field theory on AdS$_{d+1}$ is linked to a
conformal field theory on the boundary. At the root of this amazing
fact is the rescaling freedom one has when extending the metric to the
AdS boundary (as clearly summarized in \cite{ADSCFT_Vector}). It
should be mentioned, however, that in the RS 2 and LR models the brane
is not at the boundary of the AdS space. This, obviously, means that
in the UV there is no CFT description.  Moreover, in the low-energy
regime, the situation is subtle. The brane in this case is ``close''
to the boundary -- hence some AdS/CFT properties should be present --
but the connection is not completely trivial. As seen explicitly later
in this paper, the theory one obtains on the brane is not a pure CFT.
Rather, the leading interaction has a contact nature, which,
however, is exactly the property of unparticle physics 
\cite{GIR}.

To analyze the RS 2/unparticle connection, we will consider a scenario
with SM fields on the brane and a vector field in the bulk. For our
purpose we then need to know the properties of the massive vector
field in the RS 2 and LR models, particularly the complete
brane-to-brane propagator (with both transverse and longitudinal
parts). Somewhat surprisingly to us, a complete study of this problem
is lacking in the literature.
Refs.~\cite{Davoudiasl:1999tf,ArkaniHamed:2000ds}, for example, only
consider vector fields with zero bulk mass. Ref.~\cite{Pomarol:1999ad}
does examine the massive case, but only the transverse modes of the
vector field are considered.

The reason why relatively little attention has been focused on vector
fields in the original RS 2 setup perhaps has to do with
phenomenological motivations. A considerable effort has been 
focused on models with a vector zero-mode on the brane, which could be
identified with a gauge boson.  As shown in \cite{Bajc:1999mh}, unlike
a scalar, a vector in the RS 2 background does not have a zero-mode
bound to the brane purely by gravity\footnote{A field theoretic
  mechanism of confining vector fields is discussed in
  \cite{Dvali:1996xe}.}.  The two possible extensions to overcome this
considered in the literature involve adding a term on the brane that
cancels the mass \cite{GhorokuNakamura,BatellGherghetta} and adding
extra compact dimensions
\cite{CohenKaplan,GherghettaShaposhnikov,RubCharge,ourprl}.  While
some of the steps in these analyses are common with our
problem\footnote{Ref. \cite{GhorokuNakamura} studies all four
  polarizations at intermediate steps in the calculation.  The
  analysis is not taken as far as here, however. In particular, the
  CFT tensor structure is not explicitly restored and unitarity is not
  discussed.  Ref.  \cite{BatellGherghetta} investigates massive bulk
  vector bosons by utilizing the Stuckelberg mechanism; the
  longitudinal component is presented as a scalar degree of freedom
  but not studied. In light of our results, particularly the unitarity
  bounds, some of the analysis in these models should  perhaps be
  reexamined. }, the full propagator for the original RS 2 setup --
and the unparticle properties that are obtained from it -- do not
readily follow from these studies.

Another direction of phenomenological interest was to investigate a
similarity between AdS and QCD. 
The paper \cite{ADSQCD1} on this topic implicitly contains the
longitudinal polarization of the axial correlator as the Higgs field
in the bulk. Only the transverse propagator for the vector correlator
is given, however. Ref.  \cite{ADSQCD2} also studies the vector and
axial current correlators using AdS/QCD. Only the bulk vector boson
mass for the axial vector correlator is non-vanishing, but obtaining
an analytic expression for this correlator was not possible because
the bulk mass has a non-trivial profile in the bulk.

Although the comprehensive analysis of the massive vector propagator
in the RS2/LR models, as we have in mind here, has not been done before,
some important ingredients can be found in the literature in other
contexts. In particular, the AdS/CFT correspondence for a massive
vector field is beautifully treated in \cite{ADSCFT_Vector}, along
with the fermion case (see also \cite{Contino:2004vy}) and
vector-spinor interactions.  The analysis in
that paper considers \emph{both} the longitudinal and transverse
polarization and the correct CFT tensor structure is obtained. The
calculations are performed in a
Euclidean setup, with the brane at the boundary of AdS. The
philosophy of the analysis is somewhat different from ours, so the
contact terms are subtracted and unitarity not discussed.
The observation that the Minkowski version of the (scalar)
brane-to-brane propagator contains an imaginary part is discussed in
\cite{Dubovsky:2000am}. An important connection is made to the process
of escape of the bulk field into extra dimensions.
The imaginary part of the Minkowski propagator, or more precisely the phase of
its nonanalytic part  (see later), is also noted in
\cite{PerezVictoria:2001pa}. The contact terms also appear
there (without discussion of their short-distance behavior).
Finally, Ref.~\cite{terning2}, in the context of bulk fermions,
discusses the appearance of the contact terms and, in particular, the
improved convergence of the spectral integral upon their subtraction.

Other issues, particularly the unitarity considerations that require
the positivity of the bulk mass, the resolution of contact terms at
short distances and the position space behavior of the correlator,
have not been discussed, to the best of our knowledge. This questions
are essential for demonstrating the models we consider realize
the unparticle scenario and/or for understanding its properties.

\section{Preliminary Considerations}
\label{sect:preliminary considerations}

\subsection{Regulating the spectral representation}
\label{sect:spectral}

Reference \cite{GeorgiII} argues that by scale invariance the unparticle propagator in four space-time dimensions must have the spectral representation of the form
\begin{eqnarray}
  \label{eq:spectralintegral}
  \langle {\cal O}(p) {\cal O}(-p) \rangle \propto
  \int_0^\infty dM^2  \frac{(M^2)^{d-2}}{p^2-M^2+i\epsilon}.
\end{eqnarray}
For the moment, we consider the scalar case, as the vector case will be shown to contain additional subtleties.

The integral in Eq.~(\ref{eq:spectralintegral}) converges in the interval $1<d<2$, where it is evaluated \cite{GeorgiII} to be
\begin{eqnarray}
  \label{eq:integralevaluated}
  \frac{\pi}{\sin d\pi}(-p^2-i\epsilon)^{d-2} = \frac{\pi}{\sin d\pi}(p^2)^{d-2} e^{-i(d-2)\pi}.
\end{eqnarray}
This clearly shows the right conformal behavior\footnote{The Fourier transform of Eq.~(\protect{\ref{eq:integralevaluated}}) to position space, by dimensional analysis, behaves like
$1/(x^2)^{d}$, indicating that $d$ is indeed the conformal dimension
({\it cf.} Eqs.~(\ref{eq:GIRcorr_position}) and
(\ref{eq:GIRcorr})).}.

First we explore the nature of the divergences at $d=1$ and $2$. 
As $d\rightarrow1$, the integral diverges in the infrared (IR). This means that in this limit the propagator is dominated by the lightest modes in the spectrum. Indeed, as $d \rightarrow 1+$ the 
factor  $(-p^2-i\epsilon)^{d-2}$ approach the spectral representation of a single massless particle
\cite{GeorgiI}. To see this explicitly, one can renormalize the  coupling of the states by an overall factor $(\sin d\pi)/\pi$. Then, as $d\rightarrow 1+$, the $M \ne 0$ states decouple and one recovers the single-particle spectral representation of a massless particle because $\delta (x) \sim \hbox{lim}_{\epsilon \rightarrow 0} ~\epsilon x^{\epsilon-1}$
\cite{GeorgiI}. 
The value $d=1$ is known to be the unitarity bound on the conformal dimension of a scalar.
In the limit $d\rightarrow2$ the divergence is instead in the ultraviolet (UV). The factor $(-p^2-i\epsilon)^{d-2}$  in this limit becomes a constant, which is a $\delta$-function contact term in position space, as it should be for an interaction dominated by ultra-heavy states.

For $d>2$ the problem is that in Eq.~(\ref{eq:spectralintegral}) the upper limit of integration is \emph{extended to infinity}, even though as we mentioned in the Introduction the underlying model may not be a conformal theory above some scale $\Lambda$. An implicit assumption made in using 
Eq.~(\ref{eq:spectralintegral}) is that the interactions involving exchange of momentum $p$ ($p\equiv\sqrt{p^2}$) is dictated by modes with masses not much greater than $\sim p$. This assumption works for $1<d<2$, but breaks down for $d\ge2$, when the contributions of the heavy states ($M\gg p$) dominate the integral.

Since primary scalar operators in a CFT can have operator dimensions greater than 2, there 
should be  a sense in which Eq.~(\ref{eq:integralevaluated}) can be continued beyond the original interval of convergence. In fact, the simplest procedure is to cut-off the integral over the spectral function,
with $\Lambda \gg p$,  which leads to a correlator that is sensitive to the physics at the cut-off  \cite{terning2}. We shall see in Section \ref{sec:RS2}  that the RS 2 model naturally implements such features (though the regulation is more complicated and not a rigid cutoff); ultimately it is through softening the UV behavior of the wavefunctions of the KK states at the origin. 

A way to understand the consequences of regulating the spectral integral is to begin, instead, with the position space correlator (see also CMT \cite{terning2} for an equivalent conclusion using a different regularization method). 
Suppose the CFT correlation function in position space has the form $a/(x^2)^d +b\delta^{(4)}(x)$. Here, $a$ and $b$ are numerical coefficients and $b$ in particular could be divergent as the upper limit of the integration in Eq.~(\ref{eq:spectralintegral}) is taken to infinity. Upon Fourier transforming this when $d$ is not an integer, one gets $c(-p^2+ i \epsilon)^{d-2} + const$. The way to drop this constant is to differentiate
the propagator with respect to $p^2$ and integrate it back. Let us
apply this procedure to the integral in Eq.~(\ref{eq:spectralintegral}), after first regulating the upper limit with a cutoff.
Upon
differentiation we get
\begin{eqnarray}\label{eq:integral_prime}
  \frac{\partial}{\partial p^2}  \langle {\cal O}(p) {\cal O}(-p) \rangle 
 = - \int_0^{\Lambda^2} d M^2
\frac{(M^2)^{d-2}}{(p^2-M^2+i\epsilon)^2}
\stackrel{\Lambda^2 \rightarrow \infty}{\longrightarrow} \frac{-\pi(d-2)}{\sin\pi d} (-p^2+i\epsilon)^{d-3}.
\end{eqnarray}
The integral now \emph{converges for}
$1<d<3$ when the cutoff is sent to infinity. This means the UV divergence of Eq.~(\ref{eq:spectralintegral})
for $2<d<3$ is indeed confined to the $\delta$-function contact term. Next we integrate back to 
get 
\begin{eqnarray} 
 \langle {\cal O}(p) {\cal O}(-p) \rangle = 
 \frac{ \pi}{\sin d\pi}(p^2)^{d-2} e^{-i(d-2)\pi} + a_0
\end{eqnarray} 
with $a_0$ depending on both the cutoff and the subtraction point $(p^2=-\mu^2)$.  

The next steps are obvious. Differentiating the integral twice and then
integrating back twice gets rid of contact terms of the type
$\delta^{(4)}(x)$ and $\partial^2\delta^{(4)}(x)$ (in the Fourier space,
constant and $p^2$ terms) leaving the non-analytic contribution. The integral obtained after
the two differentiations,
\begin{eqnarray}\label{eq:integral_doubleprime}
  2 \int_0^\infty d M^2
\frac{(M^2)^{d-2}}{(p^2-M^2+i\epsilon)^3}
= \frac{\pi(d-2)(d-3)}{\sin\pi d} (-p^2+i\epsilon)^{d-4},
\end{eqnarray}
converges for $1<d<4$.

In general, for noninteger $d$ we then have 
\begin{eqnarray}
  \label{eq:spectralintegralcontact}
  \int_0^{\Lambda^2} dM^2  \frac{(M^2)^{d-2}}{p^2-M^2+i\epsilon} & = & \frac{ \pi}{\sin d\pi}(p^2)^{d-2} e^{-i(d-2)\pi} \left(1+ \cdots \right) 
  \nonumber \\
& &  + a_0 + a_1 p^2 + ... + a_{[d-2]} (p^2)^{[d-2]} + \cdots,
 \label{correlator_cutoff}
\end{eqnarray}
where $[d]$ denotes the greatest integer less than $d$. The coefficients $a_n$ diverge as $\Lambda^{2([d]-2-n)}$ with the cut-off of the integral, and 
we have only kept terms in the series that diverge in the limit that the cut-off is sent to infinity. 
(When the spectral integral is regulated with a cutoff, subdominant non-analytic terms of order $(p^2)^{d-2 +n} \Lambda^{-2n}$ are typically present. They are however not important for any of the discussions in this paper.) 

The integral therefore yields a nonanalytic part (the first term and all its subleading terms), plus a series of contact terms. 
As we can see, for $d>2$ the latter generically dominate the interaction, whereas for $1<d<2$ they do not.  That is, for $d>2$ the regulated integral is not dominated by the modes with $M \sim p$, but instead by  the modes living near the UV cutoff. 


Note that the apparent singularities at integer dimension 
are resolved: they are pushed into the contact terms, which are renormalized anyway 
by the counterterms \cite{GIR}. However, a non-analytic term always 
survives and has a finite coefficient. This can be seen by expanding (\ref{correlator_cutoff}) about any integer dimension to get a logarithm as the finite correction.  Explicitly, we see that in Eqs.~(\ref{eq:integral_prime}) and (\ref{eq:integral_doubleprime}). For $d=2$, Eq.~(\ref{eq:integral_prime}) becomes $p^{-2}$, so that upon integrating it back over $p^2$ we get $\ln p^2$. For $d=3$ the argument is exactly the same using Eq.~(\ref{eq:integral_doubleprime}). Thus, the nonanalytic (CFT) part of the propagator does not disappear at integer dimensions, but becomes a logarithm \cite{GIR}. 
In fact, this connection will be precisely realized when we analyze the RS 2 setup. Mathematically it occurs there 
because  of the properties of the expansions of the Bessel functions $Y_\nu(x)$, which have a branch point at $x=0$ with a log cut for integer $n$ and a power-law cut otherwise. 

One last observation is that while the CFT term has both real and imaginary parts, as discussed in \cite{GeorgiII}, the contact terms are purely real. This has transparent physical meaning: the imaginary part indicates creation of on-shell particles in the intermediate state, as will be discussed in detail later. Explicitly, the integral in Eq.~(\ref{eq:spectralintegral}) receives an imaginary part from the infinitesimal semicircle around the pole $M^2=p^2+i\epsilon$. In contrast, the contact terms originate from the exchange of massive ($M\gg p$) states, which cannot be produced on-shell.

\subsection{5d flat space}

\subsubsection{Scalar field}
\label{sec:flat space example}

To begin our analysis of extra dimensional models and their connection to the spectral representation of ``unparticles", let us consider the simplest case: a scalar field living in flat five-dimensional space. The tree-level momentum space Green's function is 
\begin{eqnarray}
  \label{eq:flat_scalar_5d}
  \frac{1}{p^2 - p_5^2 - m^2_5 +i \epsilon},
\end{eqnarray}
where $p_{\mu} \equiv (p_0, p_1,p_2,p_3)$, $p^2 \equiv p^{\mu} p_{\mu}$ is the four-dimensional momentum invariant, $p_5$ is the momentum along the extra dimension,  and $m_5$ is the bulk mass of the scalar. 

Now suppose there is a  4-dimensional Minkowski defect -- a brane - located 
at $x_5 = 0$. To find  the correlation function between two points on the brane  we need to Fourier transform back
to position space along the $x_5$ direction and evaluate the result at $x_5=0$. This gives 
\begin{eqnarray}
  \label{eq:flat_scalar_integrate}
  \Delta^{flat}(p^2) =
  \int_{-\infty}^{\infty}\frac{dp_5}{2\pi} 
  \frac{1}{p^2-m^2_5 -p_5^2 +i \epsilon} = -\frac{i}{2}\frac{1}{\sqrt{p^2-m^2_5}}.
\end{eqnarray}
Curiously, observe that for $m_5=0$ this integral has exactly the form of Eq.~(\ref{eq:spectralintegral}) with $p_5$ playing the role of $M$. We learn that coupling sources on the brane to an otherwise free massless scalar in a 5-dimensional flat space provides at tree level a spectral function  with $d=3/2$. For a finite volume the spectral representation becomes the sum over the Kaluza-Klein (KK) modes along the fifth dimension. For $m_5\ne0$ the theory has a mass gap. In this case, for $p^2\gg m_5^2$ the theory is ``approximately unparticles''. 

This connection between the spectral representation of ``unparticles" and models with large extra dimensions has been noted before. Ref.~\cite{Cheung:2007PRD} in particular compares the phase space integral over the KK modes to the spectral integral for unparticles and, for scalars, derives the tree-level relationship $d=n/2+1$ for a model with $n$ extra dimensions, which is also, not surprisingly, the engineering dimension of a scalar  in $D=4+n$ dimensions. Ref. \cite{Krasnikov:2007fs}  also notes the connection between unparticles and fermions coupled to scalar fields having a continuously distributed 
mass. Such a scenario can arise from fields living in extra dimensions coupled to four-dimensional fermions localized at a brane in a higher-dimensional space  \cite{Krasnikov:1994fw}. 

For us, $n=1$ and hence $d$ in the interval $1<d<2$. As already discussed, there are no UV divergences in this case and no resulting contact terms. In fact, we can see that in Eq.~(\ref{eq:flat_scalar_integrate}) the contributions from $p_5>\sqrt{p^2-m_5^2}$ and $p_5<\sqrt{p^2-m_5^2}$ cancel each other out in the integral. Only the infinitesimal semicircle around the pole contributes, giving for $p^2>m_5^2$ a purely imaginary answer and for $p^2<m_5^2$ a purely real answer. The imaginary part of the Green's function points to the KK states escaping the brane \cite{Dubovsky:2000am}. For $p<m_5$, no states asymptotically far from the brane ($z\rightarrow\infty$) can be
excited, hence the Green's function is purely real. In the complex $p_5$ plane, the propagator has a cut, corresponding to the continuum of states with $p^2>m_5^2$.

\subsubsection{Vector field and unitarity}
\label{flat-space unitarity}

Now, let us consider the case of a massive vector field. The momentum space Green's function of the Proca equation
$\eta^{MN}\partial_M F_{NR}+m^2_5 A_R =0$ in flat space is
\begin{eqnarray}
  \label{eq:flat_proca_5d}
  \frac{-\eta_{MN} + P_M P_N/m^2_5}{p^2-p^2_5 - m^2_5 +i \epsilon},
\end{eqnarray}
where $P \equiv (p_\mu, p_5)$.  To find the
brane-to-brane Green's function, we again Fourier transform along the $x_5$ direction, evaluate at the location of the brane $(x_5=0)$, and consider the components along the brane,
\begin{eqnarray}
  \label{eq:flat_proca_integrate}
  \Delta_{\mu\nu}^{flat}(p^2)  &=&
  \int_{-\infty}^{\infty}\frac{dp_5}{2\pi}
  \frac{-\eta_{\mu\nu} + p_\mu p_\nu/m^2_5}{p^2-m^2_5 -p_5^2 +i \epsilon} \nonumber \\
  & =& - \left(-\eta_{\mu\nu} + p_\mu p_\nu/m^2_5\right) \frac{i}{2}\frac{1}{\sqrt{p^2-m^2_5}}.
  \end{eqnarray}
The tensor in the
numerator can be decomposed 
as follows:
\begin{eqnarray}
  \label{eq:tensor_decompose}
  -\eta_{\mu\nu} + \frac{p_\mu p_\nu}{m^2_5}=
  -\eta_{\mu\nu} + \frac{p_\mu p_\nu}{p^2}
  +\frac{p_\mu p_\nu}{p^2}\frac{p^2-m^2_5}{m^2_5}~.
  \label{tensor decomposition}
\end{eqnarray}
which leads to 
\begin{eqnarray}
  \label{eq:flat_proca_brane}
  \Delta_{\mu\nu}^{flat}(p^2) &=&-
  \left(-\eta_{\mu\nu} +\frac{p_\mu p_\nu}{p^2}\right)
\frac{1}{2}\frac{i}{\sqrt{p^2-m^2_5}}
-\frac{p_\mu p_\nu}{p^2}\frac{i}{2m^2_5}\sqrt{p^2-m^2_5}  \\
& \equiv & \left(-\eta_{\mu\nu} +\frac{p_\mu p_\nu}{p^2}\right) \Delta_{flat}^{(T)}(p) 
 -\frac{p_\mu p_\nu}{p^2} \Delta_{flat}^{(L)}(p)
 \label{eq:flat_proca_brane2}
\end{eqnarray}
Seen from a brane observer, the first two terms 
describe a continuum of massive gauge bosons each with 3 degrees of freedom, while the last term (the longitudinal mode in five dimensions) appears as a continuum of scalars. In the bulk, the 
on-shell longitudinal polarization vector is $\epsilon^{(L)}_{A}=(p_{\mu} p_5/p,p)/m_5$ which has a vanishing component along the brane when $p_5=0$, explaining why the last term vanishes when 
$p^2 =m^2_5$. 
In both cases, the cut associated with the square root describes the continuous
spectrum of Kaluza-Klein (KK) modes coupled to the brane. The factor of $i$ describes the production and escape of on-shell KK modes for $p>m_5$.  

An important observation here is that for $p^2>m_5^2$, when the
longitudinal part of the correlator is purely imaginary, the sign is
controlled by the factor $m_5^{-2}$.  For in order to have a
consistent picture of particle creation on the brane and escape into
the extra dimensions ({\it cf}. \cite{Dubovsky:2000am}) and not to
violate unitarity, we must have
\begin{eqnarray}
  \label{eq:flat_unitarity}
    m^2_5 \ge 0.
\end{eqnarray}

To see that more formally, recall that the imaginary part of the forwarding scattering amplitude is 
constrained by unitarity to be non-negative. 
With 
\begin{equation} 
{\cal S} = { 1} + i {\cal T} 
\end{equation} 
perturbative unitarity implies 
\begin{equation} 
\hbox{Im} ~ {\cal T}  \geq 0
\end{equation} 
in the forward scattering channel. 

Now consider \cite{GIR} the forward scattering amplitude of, say, $e \overline{e} \rightarrow e \overline{e}$. 
This 
 is given by a sum of an  $s-$channel and a
$t-$channel contribution. The latter amplitude is purely real since both the propagator (which has $p^2<0$ space-like) and the current amplitudes are purely real. It therefore does not contribute to the imaginary part of the total forward scattering amplitude. 

The contribution from the $s-$channel
is given by
 \begin{equation}
{\cal T} = - \chi^{out} _{\mu} \Delta^{flat} _{\mu \nu}   \chi^{in}_{\nu}
\end{equation}
(the $-$ sign is from the two factors of $i$ appearing at the vertices)
where $p^2$ is time-like. Also, $\chi_{\mu} =(\chi^0,\vec{\chi})$ are the amplitudes of the external currents in the initial and final states, with $\chi^{out}_{\mu}(p) = (\chi^{in }_{\mu}(p))^*$ for forward scattering.

The external currents can be decomposed in their transverse $(p \cdot \chi_T(p)=0)$ 
and longitudinal  $(\chi_L^{\mu}(p) \propto p^{\mu})$ components:
\begin{eqnarray} 
\chi^{\mu}(p) & =& \chi^{\mu}_T(p) + p^{\mu} \chi_L(p) \\ 
\chi_L(p) & = & \frac{ p \cdot \chi(p)}{p^2}  \\
\chi^{\mu}_T(p) & =  & \chi^{\mu}(p) - p^{\mu}  \frac{ p \cdot \chi(p)}{p^2} 
\end{eqnarray} 
In the center-of-mass frame $\chi^{\mu}_L=(\chi^0,\vec{0})$ and $\chi^{\mu} _T = (0,\vec{\chi})$
where $\chi^{\mu}=(\chi^0,\vec{\chi})$. The transverse current is space-like, so its positive definite norm is $\chi^{\dagger}_T(p) \cdot \chi_T(p) \equiv - \eta_{\mu \nu} \chi^{\dagger \mu}_T(p) \chi^\nu_T(p)
\geq 0$.

Then 
\begin{equation} 
0 \leq \hbox{Im} {\cal T} =-\chi^{in \dagger}_T(p) \cdot \chi^{in}_T(p) \hbox{Im} \Delta^{(T)}(p) +|\chi^{in}_L(p) |^2  \hbox{Im} \Delta^{(L)}(p) 
 \end{equation} 
Noting that the transverse and longitudinal polarizations of the external currents
are positive-definite and independent, the unitarity condition $\hbox{Im } {\cal T} \geq 0$ is then
 equivalent to the two conditions $\hbox{Im} \Delta^{(T)}(p) \leq 0$ and $\hbox{Im} \Delta^{(L)}(p) \geq 0$. Inspecting the brane-to-brane vector Green's function (\ref{eq:flat_proca_brane}),  this first condition is seen to be trivially satisfied for all $p^2$. The second condition however requires
$m^2_5  \geq 0$ which is what we wanted to show. 

As we will see, the above arguments directly generalize to curved space. In particular, the longitudinal component will be the source of the unitarity bound in that case as well. Eq.~(\ref{eq:flat_unitarity}) will carry over \emph{unchanged} and will lead to $d_V \ge3$ in that case.

 We close by returning to Eq.~(\ref{eq:flat_proca_brane}) - the brane-to-brane propagator in flat space -  and consider  the  
 $p \gg m_5$ limit. The transverse propagator has a spectral representation corresponding to $d_V=3/2$, so that 
from the phenomenological point of view, an experiment probing the extra dimensional gauge boson in this limit will observe a vector spectral representation with $d_V<3$. In passing we note that in this model $d_V<3$ is not in conflict with the unitarity bounds on primary, vector operators in a conformal theory 
\cite{Mack,Minwalla:1997ka}, simply because when $m^2_5 \neq 0$ the theory is not conformal, and when $m^2_5=0$ the correlator is not gauge invariant.

\subsubsection{Flat space propagator: Green's function approach}
\label{Flat space propagator: Green's function approach}

We now outline an alternative method of obtaining the brane-to-brane propagators of the previous Subsection. 

Recall that the propagator is a Green's function of the equation of motion. For simplicity, let us consider the scalar case. Choosing to put the delta-function perturbation at the origin and 
Fourier transforming along the four brane coordinates, we can write for the Green's function at point 
$x_5$
\begin{equation}
     (p_i^2 + \partial_5^2) \Delta(p_i,x_5)
    - m_5^2 \Delta(p_i,x_5) = \delta (x_5).
\label{eq:scalar_flat_greens}
\end{equation}
Everywhere outside of the origin, the Green's function satisfies the equation of motion, which means it is a superposition of plane waves, $e^{-i (p_i x_i-p_5x_5)}$. The coefficients in the superposition are chosen such as to satisfy the boundary condition set by the delta-function.
We take a symmetric anzatz, $\Delta(p_4,x_5)\rightarrow c_1 e^{i p_5 |x_5|}$, around the brane. The physical picture here is that the particles created by interactions on the brane radiate into extra dimensions.  Substituting this anzatz into Eq.~(\ref{eq:scalar_flat_greens}), we see that off the brane the equation is satisfied so long as $p_5^2=p_4^2-m_5^2$. Integrating across the brane, we see that the derivative $\partial_{x_5}\Delta(p_4,x_5)$ must experience a unit jump. This fixes the constant $c_1$. We have $\partial_{x_5}\Delta(p_4,0+)=c_1 ip_5 e^{i p_5 0}=1/2$, or
\begin{equation}\label{eq:c1_flat}
    \Delta(p_4,0) = \frac{1}{2ip_5}=\frac{-i}{2\sqrt{p_4^2-m_5^2}}.
\end{equation}
This is in complete agreement with Eq.~(\ref{eq:flat_scalar_integrate}), confirming that the two methods are equivalent. The advantage of this second method is that its generalization to the warped RS 2 background is straightforward.

\section{Proca equation in the Randall-Sundrum 2 and Lykken-Randall models}
\label{RS2-LR}

We now turn to the main topic of this paper, the study of a massive
vector field in the warped RS 2 background with SM fields localized on either the UV brane or a probe brane (LR) located in the bulk. As we shall see, unparticle-like behavior is obtained in either scenario by coupling SM currents to the bulk vector boson.

The outline of our analysis is as follows. The equations of motion for the vector boson are derived in Section \ref{sec:EOM}. The boundary conditions for this field are discussed and derived in \ref{sec:brane-bc}. Since the boundary conditions depend on whether the source is localized on the UV brane (RS 2) or on a probe brane (LR), these conditions are discussed separately in \ref{sec:RS 2 BC} and \ref{sec:LR BC}, respectively. 
Then in Sections \ref{sec:RS2 transverse polarization}-\ref{sect:position} and in
Section \ref{sec:tensionlessbrane}
we turn to deriving and analyzing the brane-to-brane
propagators in the RS 2 and LR models, respectively.

\subsection{Equations of motion}
\label{sec:EOM}

The background is a five-dimensional warped AdS space with a single 4-dimensional brane located in the ``UV". This is the well-known RS 2 \cite{RSII}
background. We use the Poincare metric 
\begin{equation} 
ds^2 = a(z)^2 \left( \eta_{\mu \nu} dx ^{\mu} d x^{\nu} - dz^2 \right) 
\label{poincaremetric}
\end{equation} 
where $a=1/(\kappa z)$, 
$\kappa^{-1}$ is the AdS radius of curvature and the signature is $(+----)$. The UV brane is located at the boundary $z =\kappa^{-1}$ where the scale factor is normalized to be one. 

The action is 
\begin{equation}\label{L_warped}
    \int d^4 x d z \sqrt{g}(-F^2/4 +m^2_5A^2/2) +  \int d^4 x \sqrt{g_4} \left({\cal L }_{SM} + 
e_0  j^{\mu} (x) A_{\mu}(x, z= z_{SM}) \right) 
\end{equation} 
When $m_5 \ne0$, the gauge symmetry is explicitly broken and the vector field has \emph{four degrees of freedom} \footnote{Alternatively, it is possible to Higgs the theory by introducing scalar field with a VEV. For our purposes, writing an explicit mass term is sufficient.}. In the AdS/CFT correspondence, the value of $m_5/\kappa$ controls the conformal dimension $d_V$ of the CFT operator \cite{Witten,ADSCFT_Vector}
\begin{equation} 
d_V= 2 + \sqrt{1+m^2_5 /\kappa^2} 
\end{equation} 
We shall see that this prediction remains valid in the RS 2 background, as expected from the evidence presented in  \cite{ArkaniHamed:2000ds} that RS 2 is a good regulator of the CFT.

We will consider two models for the SM fields. In the first, the SM fields are localized on the UV brane at $z=\kappa^{-1}$. In the second, the
the SM fields are localized on a tensionless ``probe" brane located 
in the bulk at $z=z_{SM}> \kappa^{-1}$. This is the Lykken-Randall \cite{lykken-randall} model.  The metric (\ref{poincaremetric}) is therefore valid from the boundary to the horizon at $z \rightarrow \infty$.  

The current $j^{\mu}$ is any gauge-invariant current composed of SM fields. An example is 
\begin{equation} 
j^{\mu}= \overline{Q} _3 \sigma^{\mu} Q_2 + \overline{L} \sigma^{\mu} L + \cdots  =
\overline{t}_L \sigma^{\mu} c_L + \overline{\nu} \sigma^{\mu} \nu + \overline{l}_L \sigma^{\mu} l_L + \cdots
\end{equation}   
This current is not conserved and therefore couples to both the transverse and longitudinal components of the bulk vector boson.
In the action above $e_0$ is the coupling of the SM current to the bulk vector field. If the SM fields are canonically normalized then the current coupling $(e)$ to the bulk vector field does not receive any warp factor suppression and is given by 
\begin{equation} 
e= e_0
\end{equation} 
The parameter $e$ has mass dimension $-1/2$, so it can be written as 
\begin{equation} 
e = \frac{c}{M^{1/2}} 
\end{equation} 
for some mass scale $M$ and dimensionless constant $c$. Physically $M$ represents the scale at which the interaction between the SM current and the bulk vector field is generated. This could for instance occur on the order of  the (inverse) thickness of the brane.

In the following analysis it will be important to include all four polarizations, especially the longitudinal component (which is often neglected in the literature). First a practical reason :  the SM current may not be conserved (which is true for the example above), in which case the longitudinal component does not decouple from the brane. Next, the longitudinal and transverse components make comparable contributions  
to  the tensor structure of the CFT; without the longitudinal component one gets the incorrect tensor structure. But most significantly,  the unitarity bound on the dimension of the vector operator in the CFT follows from considering the longitudinal part of the propagator.  

The equations of motion are 
\begin{equation} 
\partial^{\nu} F_{\nu \mu} + a^{-1} \partial _z (a F_{\mu 5}) + m^2_5 a^2 A_{\mu} = 
-e a j_{\mu} \delta(z-z_{SM}) 
\label{firstEOM}
\end{equation}
and 
\begin{equation} 
\Box A_5 - \partial_z \partial \cdot A + a^2 m^2_5 A_5 =0
\label{bulkEOM2}
\end{equation} 
Here $\Box \equiv \partial^{\mu} \partial _{\mu}$ is the Minkowski-space Laplacian with respect to the global four-dimensional coordinates $x^{\mu}$ and $\partial \cdot A \equiv \eta^{\mu \nu} \partial_{\mu} A_{\nu}$. It will also be useful to Fourier transform functions of $x^{\mu}$ to the momentum space coordinate $p_{\mu}$ that is the conserved momenta associated with the translation symmetry $x^{\mu} \rightarrow x^{\mu} + c^{\mu}$. It is also the momenta observed by a four-dimensional observer. 

As already mentioned, when $m_5 \ne 0$, $A_\mu$ has four polarization
states. Three of these are transverse, defined by $p^\mu
A_\mu^{(T)}=0$. The remaining one has $p \cdot A \neq 0 $ and  is 
related to $A_5$ by projecting the bulk equation of motion (\ref{firstEOM}) onto its longitudinal component and then subtracting (\ref{bulkEOM2}) to obtain (away from the brane) 
\begin{equation}\label{eq:constraint}
    -i p \cdot A = a^{-3} \partial_z (a^3 A_5).
\end{equation}
This equation is the curved space generalization of the transversality
condition $p \cdot  A  = p_5 A_5$ for the solutions of the Proca
equation in flat space.

The analysis is therefore simplified if the components of the Green's function along the brane directions are decomposed into its transverse and longitudinal components as follows, 
 \begin{equation}\label{eq:corr_decompose}
   \Delta_{\mu\nu}(p,z) \equiv 
   \left(-\eta_{\mu\nu} +\frac{p_\mu p_\nu}{p^2}\right)\Delta^{(T)}(p,z)
   -\frac{p_\mu p_\nu}{p^2}\Delta^{(L)}(p,z) 
\end{equation}
with 
\begin{equation} 
\langle T(A_{\mu}(x,z) A_{\nu}(y,z^{\prime})) \rangle \equiv i \Delta_{\mu \nu}(x-y,z) 
\end{equation}
and where the dependence of the propagator on the 
location $z^{\prime}$ of the source in the bulk is left implicit.  The brane-to-brane propagator is obtained after the fact by setting $z=z^{\prime}$.
With this definition of $\Delta_{\mu \nu}$ the analysis of perturbative unitarity is straightforward, simply because $i \Delta_{\mu \nu}$ is the Feynman propagator.  This is also 
the definition we implicitly used in Section \ref{flat-space unitarity}. 
Then with this definition 
\begin{equation} 
A_{\mu}(p,z) = -e \Delta_{\mu \nu}(p,z) j^{\nu}(p) 
\label{A-Delta}
\end{equation} 
so the Green's function is $- \Delta_{\mu \nu}$, which is the standard $(-)$ sign relating Green's functions and Feynman propagators (with the factor of $i$ omitted).  
With this decomposition the equations for $\Delta^{(T)}(p,z)$ and $\Delta^{(L)}(p,z)$ are decoupled. 

From (\ref{A-Delta}) one then has the following relations which are useful for translating boundary conditions on $A_{\mu}$ into boundary conditions on $\Delta^{(T)}$ and $\Delta^{(L)}$, 
\begin{eqnarray} 
A^{(T)}_{\mu}(p,z) & = &  e \Delta^{(T)}(p,z) j^{(T)}_{\mu}(p) \\ 
-i p \cdot A(p,z) & =& \Delta^{(L)}(p,z)  (-ie  p \cdot j(p)) 
\end{eqnarray} 

It is convenient to define the $55$ propagator $\Delta_5$ through 
\begin{eqnarray} 
A_5(p,z) & \equiv &  \Delta _5(p,z) (-ie  p \cdot j(p) )
\end{eqnarray}

There are several ways to proceed. 

From Eq. (\ref{firstEOM}) one obtains an equation for 
the transverse component, 
\begin{equation} 
\Box A^{(T)}_{\mu} - a^{-1} \partial_z ( a \partial_z A^{(T)}_{\mu}) + m^2_5 a^2 
A^{(T)}_{\mu} = -a  e j^T_{\mu} \delta(z- z_{SM}) 
\label{bulkEOM1}
\end{equation}
which in terms of $\Delta^{(T)}$ is simply 
\begin{equation} 
-p^2 \Delta ^{(T)} - a^{-1} \partial_z ( a \partial_z \Delta^{(T)}) 
+ m^2_5 a^2 \Delta^{(T)} = -a  \delta(z- z_{SM}) 
\label{transverseEOM}
\end{equation}
This equation will be solved in Section (\ref{sec:RS2 transverse polarization}) for RS 2 and 
Section (\ref{LR transverse mode}) for LR using the boundary conditions obtained 
in Section (\ref{sec:brane-bc}). 

For the longitudinal mode one has from (\ref{firstEOM}) and (\ref{bulkEOM2}) 
\begin{equation} 
\partial \cdot A = a^{-3} \partial_z ( a^3 A_5) - \frac{1}{a m^2_5}  e \partial \cdot j \delta(z-z_{SM}) 
\label{bulkEOM3}
\end{equation} 
In the bulk this relation becomes
\begin{equation} 
\Delta^{(L)}=  a^{-3} \partial _z ( a^3 \Delta_5) 
\label{L5 relation} 
\end{equation} 
The $A_5$ equation (\ref{bulkEOM2}) is equivalent to 
\begin{equation} 
-p^2 \Delta_5 - \partial_z \Delta^{(L)} + a^2 m^2_5 \Delta_5 =0
\label{DeltaL5EOM}
\end{equation} 
No source appears in this equation because the brane current does not couple to $A_5$. 
In Sections (\ref{sec:RS2 longitudinal polarization}) (RS 2) 
and (\ref{sec:tensionless brane:longmode}) (LR) 
the solution for the longitudinal component will 
be obtained by solving  
these latter two equations in the bulk and applying the boundary conditions discussed in 
Section (\ref{sec:brane-bc}). 

Finally, we mention an equivalent method for solving these equations. One 
can use Eq. (\ref{DeltaL5EOM}) to solve for 
 $\Delta_5$ and substitute it back into Eq. (\ref{firstEOM}), to obtain an equation for 
 $\Delta^{(T)}$ and $\Delta^{(L)}$ only, 
 \begin{equation}
 \label{eq:master_greensfunction}
    (p^2 \eta_{\mu\nu}- p_\mu p_\nu) \Delta^{\mu\rho} + \partial_y\left(a^2\left[\eta_{\mu\nu}-\frac{p_\mu p_\nu}{p^2-m^2_5 a^2}\right]\partial_y \Delta^{\mu\rho} \right)
    - m^2_5 a^2 \Delta_\nu^\rho = - \delta (y) \delta_\nu^\rho,
\end{equation}
(Note: this equation is in the ``RS" coordinate system: $a = e^{- \kappa y}$ with $dy /dz =a$). 
This is the equation presented in our previous work 
\cite{Friedland:2009iy}.

\subsection{Boundary Conditions}
\label{sec:brane-bc}

The boundary conditions for the fields at both the UV boundary and the SM brane (where the source is located) are obtained from the variational principle. That is, surface terms obtained by varying the bulk action are cancelled by contributions arising from the variation of the interactions on the brane. 

To determine the propagator, we need to impose an additional boundary
condition at large $z$, which we choose to be the radiative boundary condition following 
\cite{Balasubramanian:1999ri,GiddingsKatzRandall,Dvali:2000rv, Dubovsky:2000am}.
This condition can be justified from several points of view.
As pointed out in \cite{GiddingsKatzRandall}, the radiative boundary
condition is analogous to the Hartle-Hawking boundary
condition in gravity, with positive frequency waves going towards the
horizon $z=\infty$. Ref.~\cite{Dubovsky:2000am} stressed that this
physically means escape of particles from the brane into the bulk. In
the unparticle picture, this means the SM model particles can
(irreversibly) decay into unparticles. This boundary condition is also the one that leads to a finite action when rotated to Euclidean space  \cite{Witten}.

We divide this discussion  into two parts depending on whether the source is on the UV brane (RS 2) or on a brane at $z =z_{SM} > \kappa^{-1}$ (LR). 

\subsubsection{Source on UV brane} 
\label{sec:RS 2 BC}

The surface term obtained by  varying the action consists of a term from the bulk action and the contribution from the brane current: 
\begin{equation} 
\left(\partial_{\mu} A_5 - \partial_z A_{\mu} + a e  j_{\mu} \right) \delta A^{\mu} |_{z = \kappa^{-1}} = 0 
\label{UVbc}
\end{equation} 
Next we project onto the transverse and longitudinal components and use 
$\delta A_{\mu} \neq 0 $. 

For the transverse mode the boundary condition is simply 
\begin{equation} 
 \partial_z \Delta^{(T)} |_{z = \kappa^{-1}} = \frac{a}{2}
 \label{RS2BCT}
 \end{equation} 
 (The factor of 1/2 is an arbitrary normalization of the current, and $a=1$ on the UV brane.) 
 
Projecting (\ref{UVbc}) onto the longitudinal component gives
\begin{equation} 
 \partial^{\mu} F_{\mu 5} |_{z=\kappa^{-1}} = -  e a \partial \cdot j 
 \end{equation} 
 Substituting the equation of motion (\ref{bulkEOM2})  to eliminate $\partial^{\mu} F_{\mu 5}$ 
 gives 
 \begin{equation} 
 A_5  |_{z=\kappa^{-1}}   = \frac{1}{ a m^2_5} e \partial \cdot j = \frac{1}{a m^2_5} (-i e p \cdot j(p) ) 
 \end{equation} 
 Thus
 \begin{equation} 
 \Delta_5 |_{z=\kappa^{-1}} = \frac{1}{2 a m^2_5}
 \label{RS2BCL}
 \end{equation} 
 
\subsubsection{Source on LR brane} 
\label{sec:LR BC}

Here the boundary conditions on the UV brane follow directly from the preceding discussion, setting the source to zero: 
\begin{eqnarray} 
\label{LRUVT}
 \partial_z \Delta^{(T)} |_{z = \kappa^{-1}} & = & 0 \\
  \Delta_5 |_{z=\kappa^{-1}}  & = & 0 
  \label{LRUV5}
 \end{eqnarray} 
 
At the LR brane we have to allow for ``jumps" or discontinuities in the fields or their derivatives across the brane. The above boundary condition (\ref{UVbc}) is modified at the LR brane to 
\begin{equation} 
\left(\left[\partial_{\mu} A_5 - \partial_z A_{\mu}\right]_{\pm} +  e a_{SM}  j_{\mu} \right) \delta A^{\mu} |_{z = z_{SM}} = 0 
\label{modifiedBC}
\end{equation} 
where $[X]_{\pm} \equiv X | _{z \rightarrow z^+_{SM}} -  X | _{z \rightarrow z^-_{SM}}$
denotes the difference of $X$ across the SM brane. 

On the brane $\delta A_{\mu} \neq 0$ and is $A_{\mu}$ is chosen to be continuous across the brane since it couples to a source. Therefore 
\begin{equation} 
[\Delta_{\mu \nu} ]_{\pm} =0
\label{transverseBC-3}
\end{equation} 
For the transverse modes one obtains from  (\ref{modifiedBC}) and (\ref{transverseBC-3}) simply 
\begin{equation} 
[\partial_z \Delta^{(T)}]_{\pm} = a_{SM}  ~,~ [\Delta^{(T)}]_{\pm} =0 
\label{ATbc}
\end{equation} 
For the longitudinal mode one first projects (\ref{modifiedBC}) onto the longitudinal component $\delta A^{(L)}_{\mu}$ to find $[\partial^{\mu} F_{\mu 5}]_\pm = -e a_{SM} \partial \cdot j$, or
\begin{equation} 
[p^2 \Delta_5 + \partial _z \Delta^{(L)} ]_{\pm}= a_{SM} \end{equation} 
Using Eq. (\ref{bulkEOM2}), this boundary condition is the same as 
\begin{equation} 
[\Delta_5]_{\pm} = \frac{1}{a_{SM} m^2_5} 
\label{a5bc}
\end{equation}

To obtain a condition for $\partial _z \Delta_5$, note that the bulk equation $-i p  \cdot A = a^{-3} \partial _z (a^3 A_5)$ together with the 
continuity of $\Delta^{(L)}$ implies  $[\partial_z (a^3 A_5)]_{\pm}=0$, giving finally  
\begin{eqnarray} \label{da5bc-0}
[\partial _z \Delta_5] _{\pm} &= & -\frac{3}{a_{SM}} [(\partial _z a) \Delta_5]_\pm \\ 
& =& 3  a_{SM} [\Delta_5]_{\pm} = \frac{3 }{m^2_5} 
\label{da5bc}
\end{eqnarray} 
In stepping from (\ref{da5bc-0}) to (\ref{da5bc})  $\partial _z a$ was assumed to be continuous across the brane. This assumption is true for the LR brane, but not for the UV brane; Eq. (\ref{da5bc}) therefore does not apply to it.  
Evidently the presence of the source leads to a discontinuity in both $\Delta_5$ and its derivative.  

We have now obtained enough boundary conditions to uniquely solve for the transverse and longitudinal propagators. To recap, in the LR model the longitudinal and transverse propagators are solved for in the region between the 
UV brane and LR brane, and in the region between the LR brane and the horizon. For each propagator there will be {\em a priori} four integration parameters; two of these are fixed by the boundary condition at the UV brane and the outgoing wave condition at the horizon. The remaining two parameters are fixed by matching the solutions across the boundary at the LR brane using Eqs. (\ref{a5bc}) and (\ref{da5bc}). 

Equivalently, these boundary conditions can be obtained by matching singularities 
in the bulk equations of motion (\ref{bulkEOM1}), (\ref{bulkEOM2}) and (\ref{bulkEOM3}) with the source term on the brane. For the transverse mode this equivalence is obtained rather easily. For the longitudinal mode one substitutes $\partial \cdot A$ from (\ref{bulkEOM3}) into (\ref{bulkEOM2}), expands 
\begin{equation} 
A_5(p,z) = A^{(2)}_5(p,z) \theta(z-z_{SM}) +   A^{(1)}_5(p,z) \theta(z_{SM}-z) 
\end{equation} 
and matches the discontinuities appearing in the equations of motion to the discontinuities 
($\delta(z-z_{SM})$ and $\partial_z \delta(z-z_{SM})$) appearing from the sources.

\section{Randall-Sundrum 2} 
\label{sec:RS2}

This Section contains the derivation and analysis of the brane-to-brane Green's functions for observers localized on the UV brane. 

The transverse propagator is derived in 
Section \ref{sec:RS2 transverse polarization} and the longitudinal propagator in Section \ref{sec:RS2 longitudinal polarization}. The two propagators are then 
summarized in Section \ref{sec:greensfunction}.  

Then they are analyzed in various regimes. 

First, 
the limit of momenta much below the AdS curvature scale is
considered. In Sect.~\ref{sect:contactRS} it is shown that in this
limit the propagators, upon expansion in series
(Eqs.~(\ref{eq:longitudinal_branetobrane_expand_Mink}) and
(\ref{eq:transverse_branetobrane_expand_Mink})), yield precisely the
unparticle form of Eq.~(\ref{eq:spectralintegralcontact}), {\it i.e.}
contact terms plus the conformal piece. 
The
longitudinal and transverse components of the conformal piece are then
shown to combine into the tensor structure required by conformal
invariance for a gauge invariant vector operator. The imaginary parts of the propagators are seen to receive contributions only from the CFT part, and are interpreted in terms of the production and escape of KK modes. 
Then the contact terms are
explicitly seen to dominate the scattering amplitudes (Sect. \ref{sect:contactRS}). Some phenomenological implications of this feature are then discussed (Sect. \ref{sec:RS 2 phenomenological implications}). The contact interactions are also seen to
cancel the corresponding divergences in the conformal piece at integer
conformal dimensions (Sect.~\ref{sect:cancellations}).  By considering the
sign of the imaginary part of the longitudinal component of the
propagator the unitarity constraints on the conformal dimension are
obtained, as discussed in Sect.~\ref{sect:unitarity}.  

We
then check, in Sect.~\ref{sect:flatRS}, that in the high momentum
limit the brane-to-brane propagators reproduce the flat space result,
Eq.~(\ref{eq:flat_proca_brane}). 

Sect.~\ref{sect:position} considers the position space representation
of the correlator, to see how the flat space behavior at short
distances turns into the conformal behavior at longer distances. The
scalar propagator is also considered, to underscore the similarities and 
differences between the two cases. We also elaborate further on the absence of fundamental contact interactions. We argue that the ``contact" interactions seen at low-energies are not contact at all, but are generated at the scale $x \sim \kappa^{-1}$, as can be also seen explicitly in the high-energy limit of the momentum space propagators. 

Finally, Sect.~\ref{sec:generalD} considers two generalizations from  $D=4$ to arbitrary 
space-time dimension $D$ on the brane. The first is to reconsider the implication of perturbative unitarity. We find $d_V \geq D-1$, which is the correct bound on the dimension of gauge-invariant, primary vector operators \cite{Minwalla:1997ka}. Next, we reconsider the vector spectral representation, finding that the condition for its UV convergence coincides with the condition that in scattering the CFT contribution dominates over the contact interactions, namely: $ d_V < D/2$. Given the above unitarity bound, one finds $D-1 \leq d_V < D/2$ for the CFT to dominate in scattering. By inspection this cannot be realized for all  $D \geq 2$. For the scalar we find the allowed window to be $(D-2)/2 \leq d_S <D/2$.

\subsection{Transverse polarization}
\label{sec:RS2 transverse polarization}

The equation for the transverse propagator obtained from (\ref{bulkEOM1}) and (\ref{transverseEOM}) 
is 
\begin{equation}\label{eq:eom_warped_transverse_2}
  -\partial_z^2 \Delta^{(T)}(p,z)
  + a \partial_z \Delta^{(T)}(p,z)+m^2_5 a^2 \Delta^{(T)}(p,z)-
  p^2 \Delta^{(T)}(p,z)= - a \delta(z-\kappa^{-1}).
\end{equation}
The general solution of this equation in the bulk is
\begin{equation}\label{eq:gen_sol_transverse}
    \Delta^{(T)}(p,z) = c_T(p) z \left[J_\nu (p z ) +
    d_T(p) Y_\nu(p z) \right],
\end{equation}
where $p \equiv \sqrt{p^2}$ and 
\begin{equation}\label{eq:define_nu}
    \nu \equiv  \pm \sqrt{1+m^2_5 /\kappa^2}.
\end{equation}
For $m^2_5 /\kappa^2  \geq -1$ both roots for $\nu$ are purely real.  
However, using the properties of the Bessel functions the solutions for  $\nu<0$ can be expressed in terms of solutions having positive $\nu$ argument.  
For  $m^2_5 /\kappa^2  \leq -1$ both roots for $\nu$ are purely imaginary, but the  solutions with negative and purely imaginary $\nu$ can be mapped to those solutions with positive and purely imaginary $\nu$. Therefore, without any loss of generality we either have $\nu$ purely real positive or purely imaginary positive. 
As  we shall see, the positivity of the real  $\nu$ solutions automatically restricts us to CFT vector operators having dimension $d_V \geq 2$.  {\em All} solutions with purely imaginary $\nu$ will be seen to violate unitarity and are therefore excluded (for a discussion of unitarity see Section \ref{sect:unitarity}).  Moreover, in order that the real $\nu$ solutions satisfy unitarity will further require $\nu \geq 1$, or $d_V \geq 3$.

The Green's function satisfying the radiative condition at large $z$ therefore has the form
\begin{equation}\label{eq:greens_transverse_2}
     \Delta^{(T)}(p,z) =c_T(p) z 
        H_\nu^{(1)} ( p z ) 
\end{equation}

The second boundary condition is imposed at the location of the brane, where the source is located. 
From the boundary condition (\ref{RS2BCT})  
 the derivative of the transverse propagator at the location of the UV brane is 
$\partial_z \Delta^{(T)} |_{z=\kappa^{-1}} = 1/2$.  
 We can now fix $c_T(p)$:
\begin{eqnarray}\label{eq:greens_transverse_3}
     \partial_z \Delta^{(T)} (p,z) &=& c_T(p)  \left[
     p z H^{(1)}_{\nu-1}(pz) - (\nu-1) H^{(1)}_{\nu}(pz)    \right],\\
    \label{eq:greens_transverse_cp}
    \rightarrow c_T(p) &=&\frac{1}{2}\left[
     p H_{\nu-1}^{(1)}(p/\kappa)  - (\nu-1)\kappa H_{\nu}^{(1)}(p/\kappa) 
    \right]^{-1}
\end{eqnarray}

Eqs.~(\ref{eq:greens_transverse_2}) and (\ref{eq:greens_transverse_cp}) define the brane-to-bulk propagator. The brane-to-brane transverse propagator $(z=\kappa^{-1})$ is
\begin{equation}\label{eq:transverse_branetobrane}
          \Delta^{(T)}(p,z=\kappa^{-1}) =\frac{1}{2}\left[
     p H_{\nu-1}^{(1)} (p /\kappa )/H_\nu^{(1)} (p /\kappa) -(\nu-1) \kappa
    \right]^{-1}
    .
\end{equation}

\subsection{Longitudinal polarization}
\label{sec:RS2 longitudinal polarization}

As a warm-up, let's first consider flat space. In the bulk the solution having the outgoing wave boundary condition is simply 
\begin{equation} 
\Delta^{flat}_5(p,z) = c^{flat}_5 e^{i p_5 z} 
\end{equation} 
where $p_5 = \sqrt{p^2-m^2_5}$. The boundary condition (\ref{a5bc}) at $z=0$ implies 
$c^{flat}_5=1/(2 m^2_5)$, so 
\begin{equation} 
\Delta^{flat}_5(p,z) = \frac{1}{2 m^2_5} e^{i p_5 z} 
\end{equation} 
Next, we obtain $\Delta^{(L)}_{flat}$ from the flat space version of (\ref{L5 relation}), 
\begin{equation} 
\Delta^{(L)}_{flat}(p,z) = \partial_z \Delta^{flat}_5(p,z) =  \frac{i p_5}{2 m^2_5}  e^{i p_5 z} 
\end{equation} 
so that the longitudinal brane-to-brane propagator is 
\begin{equation} 
\Delta^{(L)}_{flat}(p,0) = \frac{i p_5}{2 m^2_5} = \frac{i}{2 m^2_5} \sqrt{p^2-m^2_5} 
\end{equation} 
which is precisely (\ref{eq:flat_proca_brane}) and (\ref{eq:flat_proca_brane2}). 
 
Now, let us repeat the same steps for the RS 2 background. Away from the brane Eqs.
(\ref{L5 relation}) and (\ref{DeltaL5EOM}) combine to give 
\begin{equation} 
\partial^2_z \Delta_5(p,z) -3 z^{-1} \partial_z \Delta_5(p,z)   +\left[3 z^{-2} - m^2_5 \kappa^{-2} z^{-2}+p^2
\right] \Delta _5(p,z) =0
\end{equation}
The general solution of this equation is 
\begin{equation} 
\Delta_5(p,z) = c_5(p) z^2 \left[ J_{\nu}(pz) + d_5(p) Y_{\nu}(pz) \right] 
\end{equation} 
 with $\nu$ as before (\ref{eq:define_nu}) and again, without loss of generality we have either  $\nu \geq 0$ and purely real, or $\nu = i \tilde{\nu}$ with $\tilde{\nu} \geq 0$. But as with the transverse mode solutions, these solutions having purely imaginary $\nu$ will be seen to violate unitarity (see Section \ref{sect:unitarity}).  

Again, we choose the radiative boundary condition at $z \rightarrow
\infty$, combining the Bessels into the Hankel function
$H_{\nu}^{(1)}$,
\begin{equation}\label{eq:greens_g}
     \Delta_5(p,z) =c_5(p) z^2   H_\nu^{(1)}(pz) 
\end{equation}

The second boundary condition comes from (\ref{RS2BCL}) 
and is 
\begin{equation} 
\Delta_5 |_{z=\kappa^{-1}}= \frac{1}{2} \frac{1}{m^2_5} 
\end{equation} 
This gives 
\begin{equation} 
\Delta_5(p,z) = \frac{\kappa^2}{2 m^2_5} z^2 \frac{H^{(1)}_{\nu}(pz)}{H^{(1)}_{\nu}(p/\kappa)}
\end{equation} 

Finally, returning to Eq.~(\ref{bulkEOM3}), away from the brane we obtain
\begin{eqnarray}
  \label{eq:DeltaL_answer}
  \Delta^{(L)} (p,z) &=&
   a^{-3}\partial_z\left(a^3 \Delta_5(p,z)\right) \nonumber\\
  &=& \frac{\kappa^2 }{2 m^2_5} z  \frac{H^{(1)}_{\nu}(pz)}{H^{(1)}_{\nu}(p/\kappa)} 
  \left[p z H_{\nu-1}^{(1)} (p z)/H_\nu^{(1)} (p z ) -(\nu+1)
    \right]
\end{eqnarray}

The brane-to-brane Green's function follows from this, since 
$\Delta^{(L)}$  is continuous there,
\begin{eqnarray}
  \label{eq:DeltaL_answer_brane}
  \Delta^{(L)} (p,z=\kappa^{-1})
  &=&
   \frac{1}{2m^2_5}
\left[ p\frac{ H_{\nu-1}^{(1)}(p/\kappa)}{H_{\nu}^{(1)}(p /\kappa)}
    - \kappa (\nu+1)  \right].
\end{eqnarray}

\subsection{Green's function: summary}
\label{sec:greensfunction}

The RS 2 brane-to-brane propagator for $p^2>0$ is
\begin{equation}
\Delta_{\mu \nu}(p) = \left( -\eta_{\mu \nu} + \frac{p_{\mu} p_{\nu}}{p^2} \right) \Delta^{(T)}(p) - \frac{p_{\mu} p_{\nu}}{p^2} \Delta^{(L)}(p) \label{eq:corr_answer_combined}
\end{equation}
where
the transverse and longitudinal propagators are
\begin{eqnarray}
\Delta^{(T)}(p)  &=&  \frac{1}{2}\left[ p  \frac{H_{\nu-1}^{(1)}(p/\kappa)}{H_{\nu}^{(1)}(p /\kappa)}  - \kappa (\nu -1)   \right] ^{-1} ~,
\label{eq:transverseP} \\
\Delta^{(L)} (p)  &=& \frac{1}{2 m^2_5} \left[  p  \frac{H_{\nu-1}^{(1)}(p /\kappa  )}{H_{\nu}^{(1)}(p/\kappa )}  -\kappa(\nu +1) \right].
\label{eq:longitudinalP}
\end{eqnarray}
The order  appearing in these solutions is  
\begin{equation}
\nu = \sqrt{1+ m^2_5 /\kappa^2},
\end{equation}
which without loss of  generality, is either purely real and positive for $m^2_5 /\kappa^2 \geq -1$ or purely imaginary and positive for $m^2_5 /\kappa^2  \leq -1$. Only those solutions with $m^2_5 \geq 0$ will be seen to satisfy unitarity; all others will violate it (see Section \ref{sect:unitarity}).

\subsection{Analysis}
\label{sect:RS2 analysis}

Following Georgi, GIR model unparticles using the Banks-Zaks model
which is a perturbative CFT \cite{BanksZaks}. The Banks-Zaks model is
a $SU(N_c)$ gauge theory with $N_F$ flavors of quarks, where the
number of colors and flavors is large. By choosing $N_F/N_c$
appropriately, the one-loop beta-function $\beta(g) = - \eta N_c
g^3/16\pi^2$ is arranged to be small $(\eta \ll1)$, but still
asymptotically-free. As the coefficient of the two-loop beta-function
is positive, the beta function can vanish to this order with an
appropriate choice of the 't Hooft coupling. Importantly, Banks and
Zaks further show that the beta-function can be made to vanish to all
orders of perturbation theory, with a 't Hooft coupling that can be
made arbitrarily small at the fixed point.

In the microscopic theory GIR couple a SM current directly to a
(gauge-invariant) current formed from the Banks-Zaks quarks.  Assuming
the Banks-Zaks theory flows into its fixed point, such interactions
then lead at low-energy to the unparticle coupling
(\ref{unparticle-coupling}).  GIR then found that quantum corrections
involving the Banks-Zaks quarks generate dimension 8 and higher
contact interactions involving just SM fields. These contact
interactions cannot be neglected since they are suppressed by the same
scale suppressing the SM current - $\cal{BZ}$ current interaction. In
fact, as GIR note, in SM-SM plane wave scattering amplitudes these
contact interactions dominate over the purely CFT contribution.

The SM current-current couplings arise from inserting the Banks-Zaks
quarks into a loop. By inspection, the $O(p^2)$ contribution (i.e,
dimension 8 operator) is logarithmically divergent, which means that
it is present in any regularization scheme. Therefore SM contact
interactions are necessarily present, either initially at the UV
boundary or by RGE operator mixing \cite{GIR}. Since the Banks-Zaks
coupling is perturbative, this microscopic analysis is valid and this
loop is the leading effect.

Does this conclusion, obtained at weak 't Hooft
coupling, generalizes to strong coupling? Two reasons suggest that it
does. From effective field theory we do expect SM-SM contact
interactions mediated by the new physics, simply because any
messengers that generate the interactions between the SM and the CFT
will also generate SM-SM interactions. Moreover, the need to regulate
the spectral representation for operators of dimension $d_V>2$ also
suggests that contact interactions are required.  We now turn to this
and other questions in the RS2 model, using the propagators previously
derived.

\subsubsection{Contact Interactions, Tensor Structure, Phase and Particle Escape}
\label{sect:contactRS}

To begin, consider the limit where the momenta are much smaller than
the AdS curvature, $p\ll\kappa$. Note that the Green's function, Eq. (\ref{eq:corr_answer_combined}), does not have the structure expected for a
conformal theory, Eq.~(\ref{eq:GIRcorr}). Thus, our first task is to extract the CFT part from the full RS 2 propagator.

We first evaluate the longitudinal Green's function,
given in Eq.~(\ref{eq:DeltaL_answer_brane}). Expanding
in powers of $(p/\kappa)$ gives for $p^2  \ll m^2_5 $
\begin{eqnarray}\label{eq:longitudinal_branetobrane_expand_Mink}
          \Delta^{(L)} (p,z=\kappa^{-1}) &\simeq& \frac{\kappa}{2m^2_5}\left[
          -(1+\nu) +
          \frac{(p/\kappa)^2}{2(\nu-1)} +
          \frac{(p/\kappa)^4}{8(\nu-1)^2(\nu-2)} \right.\nonumber\\
       &+&
          \frac{(p/\kappa)^6}{16(\nu-1)^3(\nu-2)(\nu-3)} +
       \frac{(5\nu-11)(p/\kappa)^8}{128(\nu-1)^4(\nu-2)^2(\nu-3)(\nu-4)} 
\nonumber\\
& + &     \frac{(-19+7 \nu)(p/\kappa)^{10}}{256 (\nu-5)(\nu-4)(\nu-3)(\nu-2)^2(\nu-1)^5}   
+   \cdots 
 \nonumber \\
        &+&\left.\frac{2 \pi}{\Gamma(\nu)^2}(i - \cot\pi\nu)\left(\frac{p}{2 \kappa} \right)^{2\nu}\left[1+\cdots \right]\right]
      .
\end{eqnarray}
The ellipses denote terms higher order in $(p/\kappa)^2$. 

First, we note that in performing this expansion we assume that $\nu >1$ and is purely real. The case of when $\nu=1$ requires some care and is dealt with in Section \ref{sec:small mass}. And  
in Section \ref{sect:unitarity} it will be shown that {\em all} solutions having $\nu$ purely imaginary or $0 \leq \nu <1$ violate unitarity, so the restriction to $\nu \geq 1$ (i.e., $m^2_5 \geq 0$) is justified (for $D=4$ space-time dimensions on the brane; see Section \ref{sec:generalD} for the generalization to general $D$). 

Next notice that this expansion has the form of Eq.~(\ref{eq:spectralintegralcontact}). Hence the discussion of Sect.~\ref{sect:spectral} applies here: the terms with integer powers of $p^2$ have the form of  contact interactions, while the nonanalytic term $p^{2\nu}$ represents the contribution of a 
 CFT vector operator having dimension $d_V=2+ \nu$. The analytic terms are the contact interactions between the currents found by \cite{GIR}. Physically, the conformal symmetry is broken in the UV by the presence of the brane and 
the contact interactions are the result of that breaking.


The expansion of the transverse propagator for $(p/ \kappa)^2 \ll (\nu-1)^2$ is
\begin{eqnarray}
\Delta^{(T)}(p,z=\kappa^{-1}) &  \simeq & \frac{1}{2\kappa}\left[-\frac{1}{(\nu-1)} -\frac{1}{2 (\nu-1)^3} \left(\frac{p}{\kappa}\right)^2
-\frac{3\nu-5}{8(\nu-2)(\nu-1)^5}\left(\frac{p}{\kappa}\right)^4+ \cdots \right. \nonumber \\
& &  \left. - \frac{ 2\pi}{(\nu-1)^2\Gamma[\nu]^2}
\left( i - \cot \pi \nu \right)\left( \frac{p}{2 \kappa}\right)^{2 \nu}\left[1 +\cdots \right] \right]
\label{eq:transverse_branetobrane_expand_Mink}
\end{eqnarray}
The preceding discussion on the physical content of the expansion in Eq.~(\ref{eq:longitudinal_branetobrane_expand_Mink}) applies here as well: we see the dominant contact terms and subleading CFT piece.  

In the by now standard computation \cite{Gubser:1998bc,Witten,ADSCFT_Vector} (see also \cite{Aharony:1999ti}) these contact terms are subtracted from the
CFT two-point correlator. The principle behind this is conformal symmetry: the dual CFT gauge theory  {\em is} conformally invariant. 
In contrast to this, in the RS 2 (and also the LR) scenario the location of the UV brane (and probe brane) is fixed, breaking the symmetry. The four-dimensional dual theory is not conformally invariant: it has both a cutoff and gravity, both of which explicitly break  the conformal symmetry \cite{Gubser:1999vj,ArkaniHamed:2000ds}.  Moreover, in the dual description of the LR model the conformal field theory in the UV breaks to the SM and another conformal field theory {\em at a fixed scale} 
$\Lambda =  z^{-1}_{SM}$  in the IR \cite{ArkaniHamed:2000ds}.  In RS 2 (and as we shall see, in LR) the contact terms are therefore physical, and generically non-zero. To cancel them requires a fine-tuning between these contributions from the bulk and additional new contributions from interactions on the brane. In short, in the RS 2 and (minimal) LR models the coefficients of the contact interactions are fixed, but in a more general UV completion these coefficients  are sensitive to the physics above the (local) curvature scale \cite{terning2}.

Next we turn to the tensor structure of the CFT contribution to the propagator. Using both expansions of the propagator, we can combine the leading non-analytic terms. After some algebra, and remembering that $m_5^2=\kappa^2(\nu^2-1)$, we get
\begin{equation}
\frac{  \pi}{\kappa (\nu-1)^2 \Gamma[\nu]^2}(-i + \cot \pi \nu)  \left(-\eta_{\mu \nu} + \frac{2 \nu}{\nu+1} \frac{p_{\mu} p_{\nu}}{p^2} \right) \left(\frac{p}{2\kappa}\right)^{2 \nu}
\label{tensor structure}
 \end{equation}
With the identification $d_V=2+\nu$, this equation has the correct tensor structure and scaling to describe the two-point function of a CFT vector operator of dimension $d_V$, in complete agreement with \cite{GIR}.

As discussed in Sect.~\ref{sect:spectral}, the contact terms should be real, while the CFT piece can have a phase. Eq.~(\ref{eq:longitudinal_branetobrane_expand_Mink}) explicitly confirms this. Moreover, given that $i-\cot \pi\nu=-\exp(-i\pi\nu)/\sin\pi\nu=-\exp(-i\pi (d_V-2))/\sin\pi d_V$, we see that the nonanalytic term has exactly the phase discussed by Georgi in \cite{GeorgiII}, as well as the poles at integer $d_V$. The Bessel functions automatically know about these properties. The RS 2 scenario gives a very clear physical meaning to the {\em imaginary} part of this phase: it is related to the rate of decay into extra dimensions ({\it cf.} \cite{Dubovsky:2000am}).

We end with a final comment on a subtlety of the phase appearing in the non-analytical piece. 
At integer dimension 
$d_V$ the phase of the non-analytic terms vanish : $\exp[-i\pi (d_V-2)] \rightarrow 1$. Physically, however, the imaginary part of the correlator is non-vanishing, since 
the produced bulk KK mode still escapes from the brane, independent of whether or not $d_V$ is an integer.  
Indeed, by inspection of Eq. (\ref{tensor structure}) the imaginary part is seen to be regular for integer dimension $ d_V \geq 3$ \footnote{The case $d_V=3$ (or $\nu=1$) requires some care since the Taylor expansions (\ref{eq:longitudinal_branetobrane_expand_Mink}) and (\ref{eq:transverse_branetobrane_expand_Mink}) do not apply.  But an imaginary part of the correlator is also present in this case  - the reader is referred to Eqs. (\ref{IMT}), (\ref{IML}), (\ref{phiprime}) and the more general discussion found in Section \ref{sect:unitarity}.} .  Thus the {\em imaginary} part of the correlator is always present. 


\subsubsection{Phenomenological Implications} 
\label{sec:RS 2 phenomenological implications}

Let us elaborate on this last point a little further. The rate for this production 
can be computed using the optical theorem and the imaginary part of the forward scattering amplitude obtained from the vector boson propagator, 
\begin{equation} 
\sigma(f_1 f_2 \rightarrow ~escape) = \frac{1}{s} \hbox{Im} A(f_1 f_2 \rightarrow f_1 f_2) \simeq 
\frac{e^2}{\kappa} \left(\frac{p}{\kappa}\right)^{2 \nu}
\end{equation}
 (recall that $e$ denotes the SM current - bulk vector field coupling and it has mass dimension $-1/2$.)
For plane wave scattering on the brane this process describes the continual production of an 
outgoing flux of plane waves of the right mass, moving away from the brane.  For scattering of SM {\em wavepackets}, this cross-section gives the rate for the production of a bulk coherent state, which then escapes into the bulk. 
Once escaped, the bulk particles fall into the horizon and never re-interact with the fields on the brane. 

The purely CFT effects also contribute to SM-SM scattering, but as noted above and previously discussed by \cite{GIR} and \cite{terning2}, they are generically subleading. The contribution of the leading contact interaction to the cross-section for SM-SM scattering
 $f_1 f_2 \rightarrow f_3 f_4 $  at energies $s \ll \kappa^2$
 is 
 \begin{equation} 
 \sigma( f_1 f_2 \rightarrow f_3 f_4 ) \simeq e^4 \frac{s}{\kappa^2} 
 \end{equation} 

The leading CFT contribution to this process comes from its interference with the contact interaction and is easily seen to be subdominant,  
 \begin{equation} 
 \frac{\sigma \left( f_1 f_2 \stackrel{\hbox{CFT}}{\longrightarrow}  f_3 f_4 \right)}{\sigma \left( f_1 f_2 \stackrel{\hbox{contact}}{\longrightarrow}  f_3 f_4 \right)} \simeq \left(\frac{p}{\kappa}\right)^{2 \nu}  \ll 1
 \end{equation} 
where the last equality uses the unitarity constraint $\nu \geq 1$ and assumes $p \ll \kappa$. We then find that for vector operators, the contact operators appear to always dominate plane wave scattering amplitudes. 

The situation for vector bosons therefore differs from  
the case of bulk scalars or bulk fermions propagating on this background. There the CFT contributions can dominate the scattering amplitude if the dimension of the CFT operator is not too big \cite{terning2}. Specifically, for scalar or fermionic operators in the CFT Ref. \cite{terning2} finds that
the CFT part dominates 
 if $d_S <2$ or $d_F <5/2$.  

Next we 
notice that the escape process dominates over the interference process: 
\begin{equation} 
\frac{\sigma( f_1 f_2 \stackrel{\hbox{CFT}}{\longrightarrow}  f_3 f_4 )}{\sigma(f_1 f_2 \rightarrow ~escape)}
\simeq e^2 \kappa \frac{s}{\kappa^2} 
\end{equation} 
This result suggests that the best opportunity to discover unparticle-like behavior is not in SM-SM 
scattering processes \cite{GIR}, but either in direct production such as $t \rightarrow c +nothing$ \cite{GeorgiI}, or associated production. 

We note however that for the former process to be dominated by the CFT behavior 
it is necessary that the SM current coupling to the CFT not include neutrinos. For if it does, the contact interactions mediated by the vector unparticles will then contribute to the same process, giving a background that dominates in rate over the direct production of unparticles.

Associated production \cite{GeorgiI}
\begin{equation} 
 q + \bar{q} \rightarrow gluon + unparticle~, ~gluon+gluon  \rightarrow gluon + unparticle 
\end{equation} 
may be a promising channel in which to search for unparticle-like behavior, since the vector unparticle mediated contact interactions do not contribute.  In the detector this event appears as a monojet. 
Since large extra dimensions \cite{ADD} also produce monojets \cite{Giudice:1998ck}, it would be useful to investigate whether 
 the $p_T$ distribution of the monojet is a useful discriminator.

\subsubsection{Cancellation of divergence in CFT correlator at integer dimension}
\label{sect:cancellations}

Several authors have noted that the coefficient of the CFT propagator in momentum space diverges at integer dimension. By inspection, the coefficient is proportional to $cot ~ \pi \nu$ which indeed diverges. As noted by \cite{GIR},
the contact interactions are necessary to resolve this divergence.

To see this explicitly, first note by inspection of  
the explicit expression for the local terms in Eqs. (\ref{eq:longitudinal_branetobrane_expand_Mink}) and (\ref{eq:transverse_branetobrane_expand_Mink}) that the coefficients of the local terms also diverge when $\nu$ is an integer.  These divergences indeed cancel the divergences
that appear at integer dimension in the CFT contribution to the correlation.  What happens term by term as $\nu \rightarrow n$ is that the divergence in real part of the non-analytic term is cancelled by the divergence in the local term of $O(p^{2n})$. (The cancellation for $\nu \rightarrow 1$ requires more care; see Section \ref{sec:small mass}.) We have explicitly checked this for several of the terms in Eq. (\ref{eq:longitudinal_branetobrane_expand_Mink}).  

For example, consider from (\ref{eq:longitudinal_branetobrane_expand_Mink}) 
the $O(p^{10})$ analytic term in the longitudinal propagator, 
as $\nu \rightarrow 5$.  One has 
\begin{eqnarray} 
\lim_{\nu \rightarrow 5} \frac{(-19+7
  \nu)(p/\kappa)^{10}}{256(\nu-5)(\nu-4)(\nu-3)(\nu-2)^2(\nu-1)^5} 
 =  \frac{(p/\kappa)^{10}}{294 912}   \left(\frac{1}{\nu-5} - \frac{143}{48} \right)
\end{eqnarray}
On the other hand, in this limit the leading non-analytic term becomes 
\begin{eqnarray}
&&\lim_{\nu \rightarrow 5}\left[ \frac{2 \pi}{\Gamma(\nu)^2}(i - \cot\pi\nu)\left(\frac{p}{2 \kappa} \right)^{2\nu} \right] \nonumber \\ 
&& =  -\frac{1}{294 912} \frac{(p/\kappa)^{10}}{\nu-5}  + \frac{1}{294 912}\left(i \pi -2  \log[p/2\kappa]  -2 \gamma_E + \frac{25}{6} \right) (p/\kappa)^{10}
\end{eqnarray}
Explicitly one sees that the pole at $\nu=5$ cancels between the analytic and non-analytic terms. Next note that the appearance of the finite part is consistent with what we expect. First, there is an imaginary part which, as we shall see in Section \ref{sect:unitarity}, has the correct sign required by unitarity. Physically, it corresponds to the production of KK particles which escape from the brane. Next,  
 the leading order non-analytic term is $\log p$ which has a branch cut. This result confirms the findings of \cite{GIR} in the weakly coupled Banks-Zaks theory that a $\log p$ appears at integer 
 vector operator dimension. 
 
The reason for this cancellation  is that
from the AdS side, the dimension of the operator is determined by the value of the five-dimensional gauge boson mass and there is nothing special about values of $m^2_5$ that correspond to integer operator dimension.
In fact, the Green's function is expected to be regular in $m^2_5$, which is confirmed by the explicit solution. Specifically, we see that the solution is given by Hankel functions of order $\nu$ and $\nu-1$, which are entire functions of their order.
The series of contact terms provided by the AdS computation are seen, from the CFT side,  to be necessary in order that physical predictions are smooth functions of the operator dimension. 

Another reason to see that contact interactions might be relevant to fixing this problem is the following. 
{\em The position space correlator does not diverge at integer dimension} (the explicit formula can be found in Eq. \ref{eq:GIRcorr_position} or Section \ref{sect:position}). But 
the only difference between the position space correlator and the Fourier transform of the momentum space CFT propagator (i.e., non-analytic terms) are terms that vanish faster than $x^{-2d_V}$. Examples include terms that in momentum space are precisely contact interactions or a series of contact terms that sum up to have a finite range. In other words, the divergence that appears at integer dimension in the momentum space representation can be regulated by terms local in momentum, without affecting the correlator at large distances. 

In summary, we have seen that in the RS 2 model a number of the conclusions of \cite{GIR} found at weak CFT gauge coupling are also true, {\em viz.vi} AdS/CFT,  at large $N_c$, strong 't Hooft coupling: that contact interactions exist for any operator dimension (including non-integer); that they are required to cancel divergences that otherwise appear at integer dimension; and that for exclusive scattering, e.g. $e e \rightarrow \mu \mu$, they dominate the contribution from the non-analytic terms.

\subsubsection{Unitarity}
\label{sect:unitarity}

In a pure CFT the dimensions of operators are constrained by unitarity, as shown by Mack \cite{Mack}, Minwalla \cite{Minwalla:1997ka} and more recently by Grinstein, Intriligator, and Rothstein \cite{GIR}. Since scattering amplitudes do not exist in a pure CFT because there are no asymptotic states, bounds are obtained either by acting the (super)conformal algebra on states \cite{Mack,Minwalla:1997ka} 
or by using the state operator correspondence and manipulating correlation functions \cite{GIR}. Another physical approach to obtain these bounds is to couple the CFT operators to weakly interacting particles (such as Standard Model particles) through an irrelevant operator \cite{GIR}. The CFT operators contribute to the forward scattering of SM particles and  their physical properties can therefore be constrained by requiring that perturbative unitarity be satisfied.
This constraint leads to the same bounds on the dimensions of the CFT operators \cite{GIR}.

Let's see how this works in the RS 2 model. We will find that 
a necessary and sufficient condition for 
 the brane-to-brane forward scattering amplitude due to an intermediate bulk vector boson to preserve unitarity 
is given by
\begin{equation} 
m^2_5 \geq 0
\end{equation} 
Note that this bound is non-trivial, since a negative mass squared is allowed for scalars propagating in AdS space \cite{Breitenlohner:1982bm}.
Using this result, AdS-CFT predicts $d_V \geq 3$, which is the correct bound on the dimension of primary, gauge invariant operators in 4 dimensions. 
 
To begin,  we momentarily restrict ourselves to $\nu$ real (and, without loss of generality non-negative). As described in Section \ref{sec:flat space example} in the flat space example, following \cite{GIR}, the forward scattering amplitude for 
 $f \overline{f} \rightarrow f \overline{f}$  is given by a sum of an  $s-$channel and a
$t-$channel contribution. The $t-$channel amplitude does not contribute 
to the imaginary part of the amplitude since it is purely real because : i) it requires $p^2<0$ space-like and is therefore given by the Euclidean
brane-to-brane Green's functions, Eqs. (\ref{euclideanT}) and (\ref{euclideanL}) which
are
purely real; and ii) the current amplitudes are purely real for forward scattering. 

Next consider the $t-$channel amplitude when $\nu$ is purely imaginary. Here one has to analytically continue the brane-to-brane Green's functions to complex $\nu$, and make use of the properties of Bessel functions when their orders are complex \cite{Dunster}.  After doing that, 
it turns out that the 
 $t-$channel exchange amplitude is also purely real. 
 
It remains to consider the 
 $s-$channel amplitude which is given by
 \begin{equation}
{\cal T} = - \chi ^{out} _{\mu} \Delta_{\mu \nu}   \chi^{in}_{\nu}
\end{equation}
(as in Section \ref{sec:flat space example}, the $-$ sign is from the two factors of $i$ appearing at the vertices and $\Delta_{\mu \nu}$ is the brane-to-brane vector boson Green's function obtained in the previous sections). Also, $\chi^{out }_\mu = \chi^{in *} _\mu$.

Recall that the unitarity condition $\hbox{Im} ~{\cal T} \geq 0$ is equivalent to the two conditions $\hbox{Im} \Delta^{(T)}(p) \leq 0$ and $\hbox{Im} \Delta^{(L)}(p) \geq 0$. To write the brane-to-brane Green's functions in a more convenient form
the following identity is useful, 
\begin{equation}
x \frac{d}{dx} \log H^{(1)}_{\nu}(x) = x \frac{H^{(1)}_{\nu-1}(x)}{H^{(1)}_\nu(x)}  - \nu
\end{equation}
Then 
\begin{equation}
\Delta^{(T)} (p)= \frac{1}{2}\left[ p  \frac{d}{dx} \log H_{\nu}^{(1)}(x)   +\kappa \right] ^{-1}
\end{equation}
\begin{equation}
\Delta^{(L)} (p) = \frac{1}{2 m^2_5} \left[  p  \frac{d}{dx} \log H_{\nu}^{(1)}(x)   -\kappa  \right]
\end{equation}
with $x \equiv \sqrt{p^2}/\kappa \geq 0$. We note that since the Bessel functions are entire functions of their order  \cite{Besselbook}, these formulae are also valid for $\nu$ purely complex (i.e., $m^2_5 < - \kappa^2$). 

The imaginary part of these Green's functions comes from the phase of the Hankel function, which for any $\nu$ is
\begin{equation}
\phi(\nu,x) = \hbox{Im} \log H^{(1)}_{\nu}(x)
\end{equation}

In terms of these variables one finds 
\begin{eqnarray} \label{IMT}
0 \geq  \hbox{Im} \Delta^{(T)}(p) &=& -\frac{p \phi^{\prime}/2}{( p  \frac{d}{dx} \hbox{Re} \log H_{\nu}^{(1)}(x) +\kappa)^2 + (p \phi^{\prime})^2}
 \\
0 \leq \hbox{Im} \Delta^{(L)}(p) &=& \frac{ p \phi^{\prime}}{2 m^2_5}
\label{IML}
\end{eqnarray}
where $ \phi^{\prime} \equiv \partial \phi(\nu,x) / \partial x$ . These results are completely general, since we have allowed for $m^2_5$ to be positive or negative (i.e, $\nu$ purely real or purely imaginary). 

The desired bound is obtained from looking at the ratio of these two imaginary parts. Unitarity requires that the ratio have a fixed sign, which by inspection is 
\begin{equation} 
0 \geq \frac{ \hbox{Im} \Delta^{(T)}(p)}{ \hbox{Im} \Delta^{(L)}(p)} = -\frac{m^2_5}{( p  \frac{d}{dx} \hbox{Re} \log H_{\nu}^{(1)}(x) +\kappa)^2 + (p \phi^{\prime})^2}
\end{equation} 
Since the denominator of the right-side is positive, this condition implies 
\begin{equation} 
m^2_5 \geq 0
\label{necessarysufficient} 
\end{equation}
Note that this result automatically implies solutions having $\nu$ purely imaginary violate unitarity, for they all have $m^2_5 <-\kappa^2 < 0$. 

It remains to check that the positivity of the mass squared it is sufficient. 
Now the first condition requires $\phi^{\prime} >0$ which seems non-trivial. But it turns out that this condition is 
automatically satisfied. Since for $m^2_5 \geq 0$ the order $\nu$ is purely real, an explicit expression for the phase is easily obtained. It is 
\begin{equation} 
\phi(\nu,x) = \arctan \frac{Y_{\nu}(x)}{J_{\nu}(x)}
\end{equation} 
Then using 
the Wronskian $W[J_{\nu},Y_{\nu}]= J_{\nu} Y^{\prime}_{\nu} - J^{\prime}_{\nu} Y_{\nu} =2 /(\pi x)$ of the two Bessel functions, one obtains after some algebra
\begin{equation}
  \label{phiprime}
  \phi^{\prime}  =  \frac{2}{\pi}\frac{1}{x} \frac{1}{J^2_{\nu}(x)+Y^2_{\nu}(x)} \geq 0
\end{equation}
which is always positive definite. This result, combined with the
above observation on the ratio of the transverse to longitudinal modes
establishes that (\ref{necessarysufficient}) is the necessary and
sufficient unitarity bound on the vector boson mass.
 
In passing we reiterate the importance of the longitudinal component
in obtaining this bound.  For had only the transverse Green's function
been considered, one would have found the weaker condition
\begin{equation} 
  m^2_5  \geq -\kappa^2
\end{equation} 
This condition is reminiscent of the necessary stability bound on a
bulk scalar of mass $m_{\phi}: m^2_{\phi} \geq -4 \kappa^2$
\cite{Breitenlohner:1982bm}.

Using the AdS-CFT identification $d_V=2+ \sqrt{1+m^2_5/\kappa^2}$
(which we have seen remains the same in RS 2), the bound
(\ref{necessarysufficient}) is seen to be equivalent to $d_V \geq 3$
($i.e.,~\nu \geq 1$), which is the same as the unitarity constraint on
the dimension of (primary) vector operators in a CFT.  That this bound
comes from the {\em longitudinal} part of the Green's function is
consistent with the fact that the bound on the CFT operator comes from
requiring positivity of the second descendent operator $\langle
 \partial^{\mu} O_{\mu} (x) \partial^{\nu} O_{\nu}(0) \rangle $
 \cite{GIR}.  We also note that the bound (\ref{necessarysufficient})
 implies that at large distances $x$, the position-space correlator
 must fall at least as fast as $x^{-6}$ (see Eqs.(\ref{eq:a_largex})
 and (\ref{eq:b_largex})). This last statement makes no reference to
 AdS-CFT.

Thus the brane-to-brane scattering amplitude satisfies the unitarity
condition for all values of $p$ if and only if $\nu$ is purely real
and $ \geq 1$.  That this condition is the same as for a vector
operator in a CFT might be at first surprising.  Indeed, the
brane-to-brane propagator is dominated by the contact interactions.
However, the local terms do not have a cut (and therefore no imaginary
part). In the four-dimensional interpretation this is understandable
since they are generated by virtual degrees of freedom having mass $M
\gg p$. On the other hand, the CFT contribution does have a cut and an
imaginary part, so only it contributes to the imaginary part of the
scattering amplitude. Physically, the imaginary part arises because in
four dimensions the SM currents excite CFT states at all momentum
scales. The CFT therefore provides the 
imaginary part of the amplitude.
  

It is worth stressing that the UV brane explicitly breaks the
conformal invariance. Yet the condition $m^2_5 \geq 0$ (or $d_V \geq
3)$ is the same as for a (primary) vector operator in a CFT {\em
  without UV breaking scale}.  This feature is another indication that
the UV brane in RS 2 provides a good UV regulator to the four
dimensional CFT \cite{ArkaniHamed:2000ds} (i.e., it does not violate
conformal invariance at large distances or violate perturbative
unitarity).

\subsubsection{High Energy or Flat space limit}
\label{sect:flatRS}

In the limit of large momenta, $p \gg \kappa$, the geometry looks flat and we expect to recover the flat 
space-time propagator, Eq. (\ref{eq:flat_proca_brane}). In particular, the flat-space propagator in this  limit  has no contact interactions, as expected for an ``unparticle-like'' spectral representation of $d_V=3/2$.
That this is also the case for the RS 2 Green's  
function as given in Sect. \ref{sec:greensfunction}
is technically less obvious. 

First note that the RS 2 Green's function has a similar general form to that of flat space, in particular that the transverse and longitudinal components are almost the inverse of each other. We just need to show that the expressions in the square brackets in Eqs.~(\ref{eq:transverseP}) and (\ref{eq:longitudinalP}) reduce to the square root in Eq.~(\ref{eq:flat_proca_brane}).
Consider first the limit of large $p$ and fixed $m_5$ and $\kappa$. Then from the large
argument expansion of the Hankel functions for fixed $\nu$, namely that for large $x = p/\kappa$,  
\begin{equation} 
H^{(1)}_{\nu}(x) \rightarrow \sqrt{\frac{2}{\pi x}} e^{i( x- \nu \pi/2-\pi/4)} 
\end{equation} 
implying
$H_{\nu-1}(x)/H_{\nu}(x) 
\rightarrow i$,  from which we obtain $ 2m ^2_5 \Delta^{(L)}(p/\kappa)
\rightarrow [2 \Delta^{(T)}(p/\kappa)]^{-1} \rightarrow i p$, agreeing
with Eq.~(\ref{eq:flat_proca_brane}) in the massless limit. Moreover, the corrections are easily seen to 
be of $O(\kappa/p)$, so that in this limit contact interactions are not present. 

More
generally, the massive case can also be reproduced. To do that we need to consider 
$p,~m_5 \gg \kappa$. That is, send 
$\kappa$ to zero, while holding $p$ and $m_5$ finite or, in other words, 
$\kappa\rightarrow0$, $\nu\rightarrow\infty$, such that
$\nu\kappa=m_5$. The large $\nu$ expansion of the
Hankel functions can be found in
Ref.~\cite{GradshteynRyzhik}, on p. 912:
\begin{eqnarray}
     H^{(1)}_\nu(x)=\frac{w}{\sqrt 3}\exp{ \left\{
    i\left[\frac{\pi}{6}+\nu\left(w-\frac{w^3}{3}-{\rm arctan} ~w \right) \right]\right\} }
    H_{1/3}^{(1)}\left(\frac{\nu}{3}w^3 \right) +O \left(\frac{1}{|\nu|} \right),
\end{eqnarray}
where $w=\sqrt{x^2/\nu^2-1}$. In this case, $x=p/\kappa$ and $w=\sqrt{p^2/m^2_5-1}$. Using these results and the asymptotic form 
for $H^{(1)}_{1/3}(z)$, for large
$\nu$ one finds  
\begin{equation} 
H_{\nu-1}(p/\kappa)/H_{\nu}(p/\kappa) \rightarrow \exp[i
\arccos(m_5/p)] 
\end{equation} 
which is independent of $\kappa$ to leading order. Therefore in this limit
\begin{eqnarray}
  \label{eq:flatlimit_massive}
2m^2_5   \Delta^{(L)}(p/\kappa) = [2 \Delta^{(T)}(p/\kappa)]^{-1}  &=&
p \frac{H^{(1)}_{\nu-1}(p/\kappa)} {H^{(1)}_{\nu}(p / \kappa)} -
\kappa \nu
\nonumber\\
 &=&   p
  e^{i\arccos(m_5/p)} - m_5 = i\sqrt{p^2-m^2_5}  \label{highenergyprop},
\end{eqnarray}
which are the correct brane-to-brane Green's functions in flat space.

Recall from our earlier discussion that for the flat space theory the spectral integral converges in the UV, thus requiring no contact terms. That no contact terms are present in RS 2 for $p,~m_5 \gg \kappa$ is also evident from the explicit expression for the high energy propagators (\ref{highenergyprop}).  On the other hand, as we just saw in the previous Section, in the low energy limit the theory has contact terms. Hence, the contact terms are generated at a scale 
$p \sim \kappa$.

\subsection{Scene from position space}
\label{sect:position}

In this Section we investigate the properties of the Green's function in position space. Position-space correlators of the vector boson for a few choices of $m^2_5$ are shown in 
Figure \ref{fig:Greens_position}. 
For comparison, the position-space correlator of a  scalar field for several values of its  bulk mass are shown in Figure \ref{fig:Greens_position_scalar}. The scalar and vector correlators for a  bulk mass much less than the curvature scale are shown in detail in Figures \ref{fig:Greens_position_scalar_detail}, 
\ref{fig:Greens_position_vector_resonance} and \ref{fig:Greens_position_vector_resonance_b} .  
For  all of these plots we have performed a numerical Fourier transform of the full momentum-space propagator. 

The most prominent feature of these plots is their simplicity: at small and large distances the correlators can be described by two power laws, with the transition to a significantly less-divergent power-law as $x$ decreases occurring at the scale $x \approx \kappa^{-1}$. This softening indicates that the contact interactions seen at low-energy at not fundamental, but are instead resolved at the curvature scale.  
As we shall see, at large distances the power laws in these Figures  
are described by the pure CFT contribution, whereas the behavior at short distances is given by the expected 5-dimensional behavior. 
The next striking feature is that the transition between these two regimes is sharp, except for values of $\nu$ that correspond to small bulk masses. 
For small values of the bulk mass parameters one can understand the deviations at intermediate distances from power-law behavior as due to a scalar 
\cite{Dubovsky:2000am} or vector boson resonance. 
As we shall see, all these features can be understood analytically.

\subsubsection{General Features of Vector and Scalar Position Space Green's Functions}

To begin, we
rotate to Euclidean space using Eqs. (\ref{eq:transverse_branetobrane}), (\ref{eq:DeltaL_answer_brane}) and
\begin{equation}
ip H_{\nu-1}^{(1)}(ip/\kappa)/H_{\nu}^{(1)}(ip/\kappa) = - p
K_{\nu-1}(p/\kappa)/K_{\nu}(p/\kappa) 
\end{equation}
where $K$ is the modified
Bessel function. Here $p^2 \geq 0$ in Euclidean space, and, for this Section, we use the signature $(++++)$. Then the Euclidean brane-to-brane Green's functions are 
\begin{equation}
\Delta^{(E)}_{ij}(p) = \left( \delta_{ij} - \frac{p_{i} p_{j}}{p^2} \right) \Delta^{(T)}_E(p)
+\frac{p_{i} p_{j}}{p^2} \Delta^{(L)}_E(p), \label{eq:corr_answer_combined_E}
\end{equation}
where
\begin{eqnarray}
\Delta^{(T)}_E(p)  &=&  \frac{1}{2}\left[ p  \frac{K_{\nu-1}(p/\kappa)}{K_{\nu}(p/\kappa)}  +\kappa(\nu -1) \right] ^{-1},
\label{euclideanT}  \\
\Delta^{(L)}_E (p) &=& \frac{1}{2 m^2_5} \left[  p  \frac{K_{\nu-1}(p/\kappa)}{K_{\nu}(p/\kappa)}  +\kappa(\nu +1) \right].
\label{euclideanL}
\end{eqnarray}

We now Fourier transform the transverse and longitudinal components
to position space.
Explicitly,
we have
\begin{eqnarray}
  \label{eq:position_transverse}
  D^{(T)}_{ij } (x) &=& \int_0^\infty \frac{p^3 dp}{4 \pi^3}
  \int_0^\pi d\theta \sin^2\theta e^{ip x \cos\theta}    \left( \delta_{ij}- \frac{p_{i} p_{j}}{p^2} \right)
 \Delta_E^{(T)}(p),  \\
  \label{eq:position_longitudinal}
  D^{(L)}_{ij} (x) &=& \int_0^\infty \frac{p^3 dp}{4\pi^3 }
  \int_0^\pi d\theta \sin^2\theta e^{ip x \cos\theta}   \left( \frac{p_{i} p_{j}}{p^2} \right)
 \Delta_E^{(L)}(p).
\end{eqnarray}
The integral over $\theta$ can be done analytically using \cite{GradshteynRyzhik}
\begin{equation}
J_{\gamma}(px) = \frac{\left(px/2\right)^\gamma}{\Gamma[\gamma
+\frac{1}{2}] \Gamma[\frac{1}{2}]}
\int_0^\pi
d\theta e^{ip x \cos\theta}  \sin^{2 \gamma} \theta
\end{equation}
We write the total position space brane-to-brane propagator as
\begin{equation}
 D_{ij} (x)=  D^{(T)}_{ij} (x) + D^{(L)}_{ij} (x) =a(x) \delta_{ij} + b(x) \frac{x_{i} x_{j}}{x^2}.
 \end{equation}
Taking the trace of this equations and also multiplying it by $p_{i} p_{j}$ yields two equations, from which we can solve for $a(x)$ and $b(x)$ in terms of $\Delta_T^{(E)}(p)$ and $\Delta_L^{(E)}(p)$. After some algebra, we find
\begin{eqnarray}
a(x) &=&  \int_0^\infty \frac{p^3 dp}{4 \pi^2} \left[\left(\frac{J_1(px)}{(px)} - \frac{J_2(px)}{(px)^2} \right)
 \Delta_E^{(T)}(p)
   + \frac{J_2(px)}{(px)^2}  \Delta_E^{(L)}(p)  \right],  \label{a} \\
 b(x) &=&  \int_0^\infty \frac{p^3 dp}{4 \pi^2} \left(\frac{J_1(px)}{px}- 4 \frac{J_2(px)}{(px)^2} \right)
  \left(  \Delta_E^{(L)}(p) - \Delta_E^{(T)}(p) \right).  \label{b}
\end{eqnarray}
The
remaining integral over $p$ can be done numerically.

\begin{figure}
  \centering
  \includegraphics[angle=0,width=0.83\textwidth]{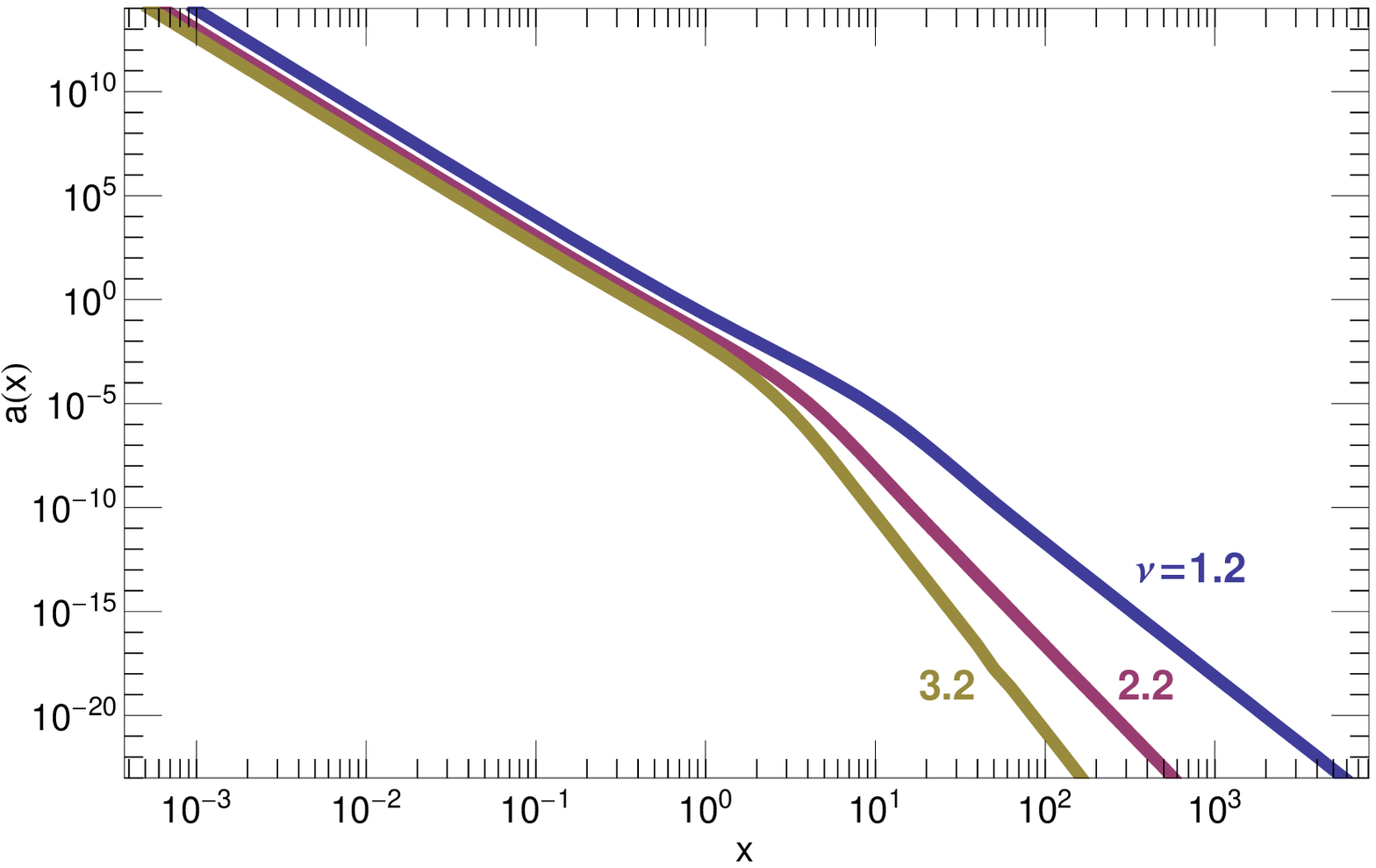}
  \includegraphics[angle=0,width=0.83\textwidth]{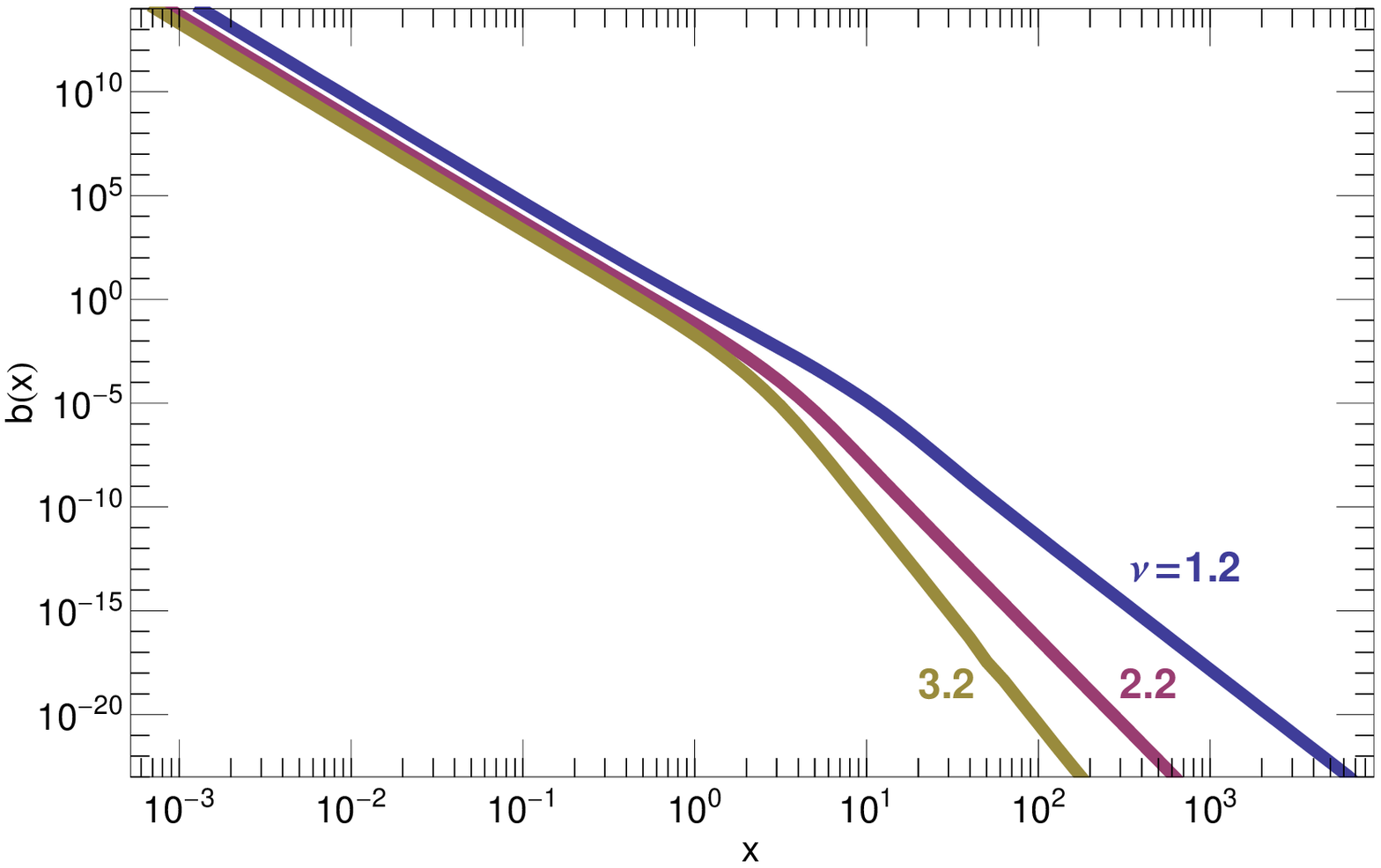}
  \caption{The Euclidean Green's function of the vector field in position space, $ D_{ij} (x)=  a(x) \delta_{ij} + b(x) x_{i} x_{j}/x^2$. Functions $a(x)$ ({\it
      top}) and $-b(x)$ ({\it bottom}) are plotted. The AdS curvature $\kappa$ is set to
    1, {\it i.e.}, the distance $x$ is in units of $\kappa^{-1}$. In
    both cases, three different values of $\nu$ are considered, as
    labeled in the plots. The functions exhibit two
    power law regimes; physically, these correspond to the flat space
    limit ($x < 1$) and to the CFT-dominated  limit ($x \gg 1$), as explained in
    the text.}
  \label{fig:Greens_position}
\end{figure}

The results for three representative choices of $\nu$ (1.2, 2.2, and
3.2) are shown in Figure ~\ref{fig:Greens_position} for both the
$a$ and $b$ components. The AdS curvature $\kappa$ is
set to 1, such that the distance $x$ is in units of
$\kappa^{-1}$. As previously advertised, we see that the position space Green's functions are
composed of two power laws, $D \sim x^\alpha$. The power $\alpha$ for
$x<1$ is independent of the value of $\nu$, and is the same for $a$ and $b$. For $x\gg1$, $\alpha$
depends on $\nu$, but is the same for the
two components\footnote{In this case the relative $O(1)$
  numerical coefficient between $a$ and $b$ is also physically important, as
  will be discussed later.}. As we later show, the $x<1$ regime
corresponds as expected to the 5d flat space limit, while for $x\gg1$ and $\nu$ not close to 1,  the two-point function 
behaves like a pure 4d CFT.

\begin{figure}
  \centering
  \includegraphics[angle=0,width=0.90\textwidth]{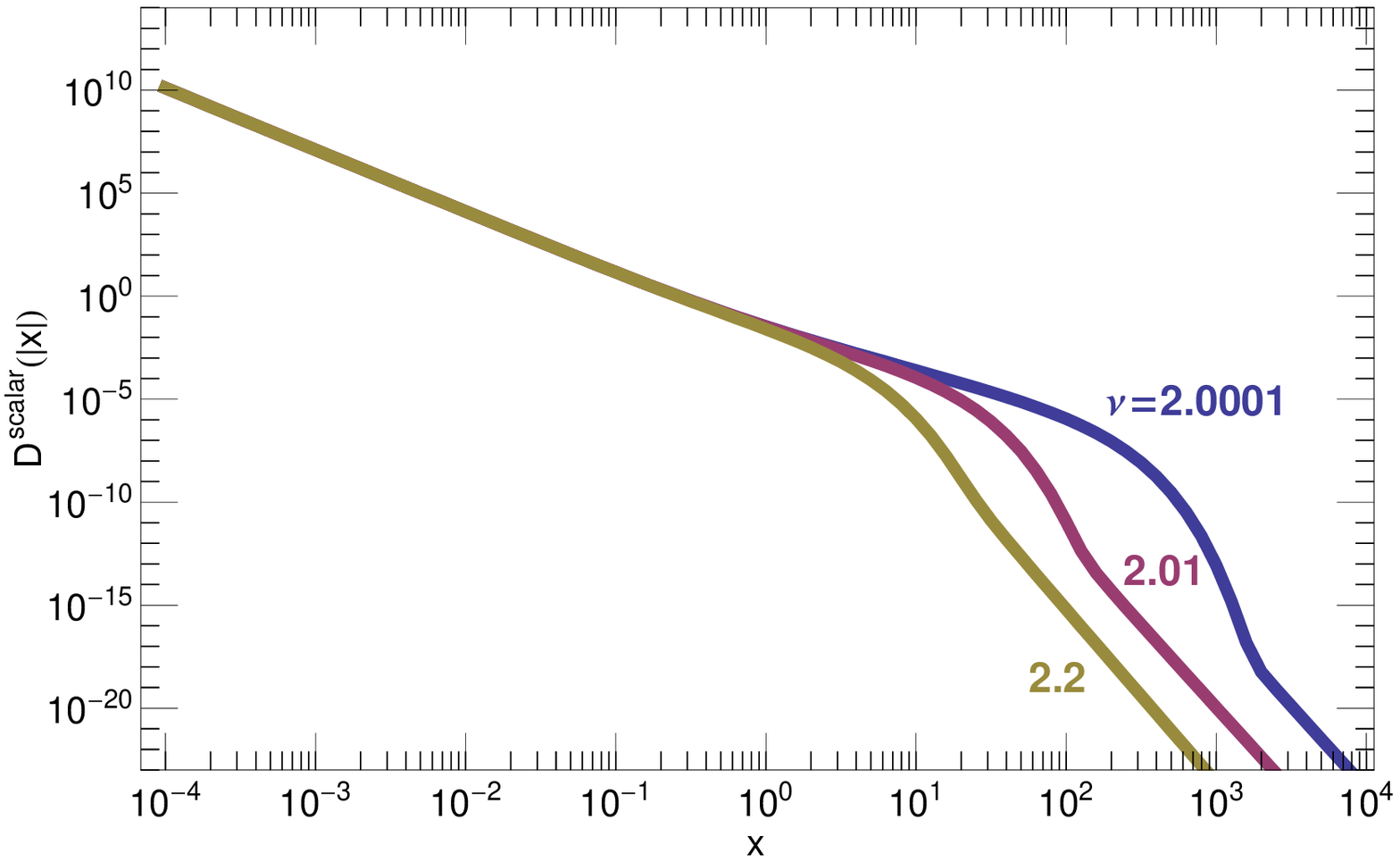}
  \caption{The Euclidean Green's function of the scalar in position space, with the same conventions and notation as Figure 1. The effect of the resonance when $\nu$ is close to 2 is quite pronounced.}
  \label{fig:Greens_position_scalar}
\end{figure}

For comparison, we show in Figure \ref{fig:Greens_position_scalar} the Euclidean-position-space brane-to-brane correlator for a bulk scalar with bulk mass $m_S$. 
Its brane-to-brane propagator in Euclidean momentum-space is given by \cite{Dubovsky:2000am,PerezVictoria:2001pa}
\begin{equation} 
  \label{Euclidean scalar correlator}
  \Delta^{(S)}_E(p) = 
  \frac{1}{2\kappa} \left[\frac{p K_{\nu_S-1}(p/\kappa)}{\kappa
      K_{\nu_S}(p/\kappa)} +  \nu_S -2 \right]^{-1}, 
\end{equation} 
which is almost identical to the RS 2 transverse propagator for a bulk vector field. (In this formula $\nu_S=\sqrt{4+ m^2_S/\kappa^2}$.) 
In position space the correlator is 
\begin{eqnarray} 
D^{\rm scalar}(x) = \int_0^\infty \frac{p^3 dp}{4 \pi^3}
  \int_0^\pi d\theta \sin^2\theta e^{ip x \cos\theta}   
 \Delta_E^{(S)}(p) 
  =   \int_0^\infty \frac{p^3 dp}{4 \pi^2}  
  \frac{J_1(px)}{p x} \Delta_E^{(S)}(p).
 \label{Euclidean position space scalar correlator}
\end{eqnarray} 

\begin{figure}
  \centering
  \includegraphics[angle=0,width=0.90\textwidth]{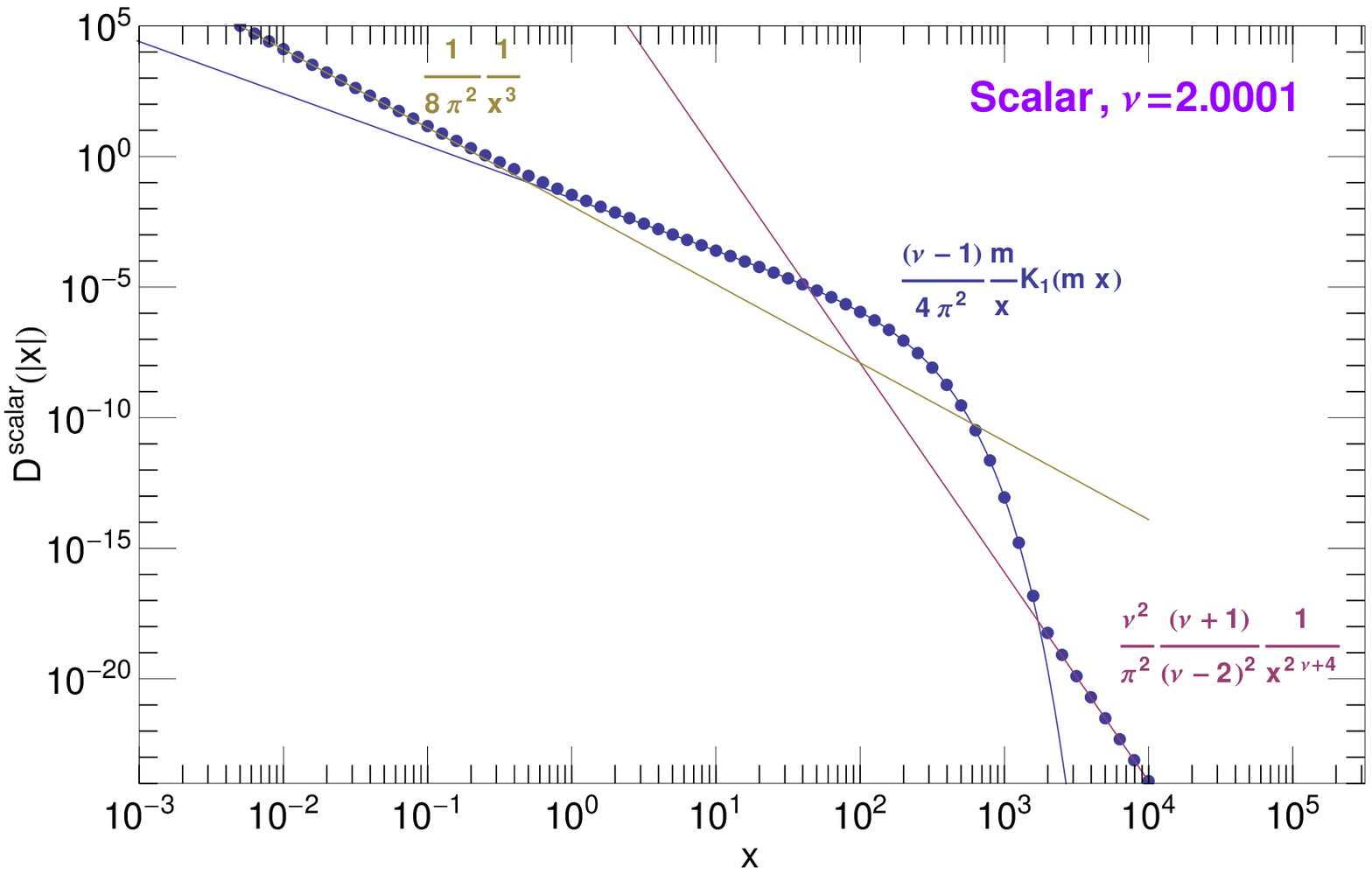}
  \caption{Detail of the Euclidean Green's function of the scalar in position space for $\nu_S=2.0001$, with the same conventions and notation as Figure 1.}
  \label{fig:Greens_position_scalar_detail}
\end{figure}

When the bulk vector or scalar mass is much smaller than the AdS curvature scale then 
the position space correlator has a third regime, intermediate between the two power-law behaviors.  This feature is visible in Figures \ref{fig:Greens_position_vector_resonance} and \ref{fig:Greens_position_vector_resonance_b}  for the vector, and in Figure \ref{fig:Greens_position_scalar_detail} for the scalar. 
For the scalar it is known that in this limit there is a resonance present, bound to the brane 
\cite{Dubovsky:2000am}. 

This small mass limit is discussed further in Section \ref{sec:small mass},  where we show that like the scalar, for the vector there is an intermediate region where the transverse correlator is dominated by a resonance coupled to a CFT. As with the scalar,  here the vector correlator exhibits pure CFT behavior only at very large distances. On the other hand, for vanishing mass the zero mode vector boson 
decouples at low energies \cite{ArkaniHamed:2000ds}, whereas the scalar does not 
\cite{Bajc:1999mh}.

\subsubsection{Large $x$} 

We now wish to understand the large $x$ behavior of the Fourier transform
of these expressions for the vector correlator.
A starting observation is that if a function, or any of its
derivatives, have discontinuities on the real axis, these
singularities dominate the Fourier transform at high frequencies (i.e., at large $x$ in our case).
This
statement is intuitive: ``sharp'' features such as discontinuities,
cusps (discontinuities of the derivative), etc, contain high frequency (i.e., large distance)
components. Quantitatively, ``sharp'' features are points of
nonanalyticity  on the real axis. They can be shown to give a
power law spectrum at high frequencies, while functions analytic on
the real axis give an exponentially decaying spectrum. For an
excellent discussion of this, see \cite{Migdal}, pp. 17-25.

In practice, a ``sharp'' feature (for example a discontinuity in the
third derivative) may be ``concealed'' superimposed on a much larger
``smooth'' (analytic) component. In this case, to understand the
Fourier spectrum at high frequencies (i.e., large distances), the singular part must be
identified and extracted.

Let us see how these observations apply to our case.  Let us
for the moment assume that $2\nu$ is not an integer.  Then,
the integrands in Eqs.~(\ref{eq:position_transverse})
and (\ref{eq:position_longitudinal}) can be formally 
expanded as a power law series in $p/\kappa$, 
 \begin{eqnarray}\label{eq:longitudinal_branetobrane_expand_eucl}
          \Delta^{(L)}_E (p) &\simeq& \frac{1}{2\kappa(\nu^2-1)}\left[
          (1+\nu) +
          \frac{(p/\kappa)^2}{2(\nu-1)} -
          \frac{(p/\kappa)^4}{8(\nu-1)^2(\nu-2)}  + \cdots \right.\nonumber\\
       &&- \left.\frac{2 \nu \Gamma(-\nu)}{\Gamma(\nu)}\left(\frac{p}{2 \kappa} \right)^{2\nu}\left[1+\cdots \right]
     \right],\\
\label{eq:transverse_branetobrane_expand_eucl}
          \Delta^{(T)}_E (p) &\simeq& \frac{1}{2\kappa}\left[
          \frac{1}{\nu-1} -
          \frac{(p/\kappa)^2}{2(\nu-1)^3} +
          \frac{(3\nu-5)(p/\kappa)^4}{8(\nu-1)^5(\nu-2)} + \cdots \right.\nonumber\\
       &&+\left.\frac{2 \nu \Gamma(-\nu)}{(\nu-1)^2\Gamma(\nu)}\left(\frac{p}{2 \kappa} \right)^{2\nu}\left[1+\cdots \right]
     \right].
\end{eqnarray}
For completeness we also provide the low momentum expansion of the scalar Green's function
 (\ref{Euclidean scalar correlator}), 
\begin{eqnarray} 
\Delta^{(S)}_E(p) & =& \frac{1}{2 \kappa} \left[\frac{1}{\nu_S-2} -\frac{(p/\kappa)^2}{2(\nu_S-1)(\nu_S-2)^2} 
+ \frac{3(p/\kappa)^4}{8(\nu_S-1)^2(\nu_S-2)^3} + \cdots \right.  \nonumber \\ 
& & \left. + \frac{2\nu_S \Gamma[-\nu_S]}{(\nu_S-2)^2 \Gamma[\nu_S]} \left(\frac{p}{2 \kappa}\right)^{2 \nu_S} 
\left[1 + \cdots \right] \right]
\label{scalar correlator expand} 
\end{eqnarray}

Then as a
function of $p$, the integrand of $a$, $b$ and $D^{\rm scalar}$ are each a sum of two parts, an
analytic component -- represented by the terms
with integer powers
of $p^2/\kappa^2$ -- and the one with a branch point at zero -- given by the
terms of the form $p^{2\nu+2n}$, $n=0,1,2....$.

Notice that the Fourier transform of the analytic parts is
exponentially suppressed at large $x$. Indeed, each of the terms
in the power expansions is of the form \cite{GradshteynRyzhik}
\begin{equation}
\int_0^\infty dt  J_\beta (a t) t^{\alpha} = 2^\alpha a^{-\alpha-1}
\frac{\Gamma(1/2+\beta/2+\alpha/2)}{\Gamma(1/2+\beta/2-\alpha/2)}
\label{tJint}
\end{equation}
For the analytic terms we have $\alpha=2n+2$, $\beta=1$ for the terms multiplying $J_1(px)$,
and $\alpha=2n+1$, $\beta=2$ for the terms multiplying $J_2(px)$. One can confirm
that for these values the right-side of Eq. (\ref{tJint}) vanishes \footnote{For $n=0$ and $\nu >1$ the integrand multiplying $J_2(px)$ vanishes identically.}. This
means integrating the analytical parts in
Eqs.~(\ref{eq:longitudinal_branetobrane_expand_eucl}), (\ref{eq:transverse_branetobrane_expand_eucl}) and (\ref{scalar correlator expand}) term by term we
obtain zero. Indeed, these are \emph{contact terms}, $\delta(x)$,
$\partial^2\delta(x)$, {\it etc}, vanishing for nonzero $x$. This does
not mean the Fourier transform of the whole analytic function vanishes
for nonzero $x$ -- it does not -- but it does show that at large $x$ the result
falls off faster than any power of $1/x$, {\it i.e.}
exponentially \footnote{An obvious example
is provided by a massive particle in four dimensions. If we expand
the Euclidean propagator
$(p^2+m^2)^{-1} \simeq m^{-2}-m^{-4}p^2+m^{-6}p^4-m^{-8}p^6$ and
Fourier transform it term by term, we get a series of contact
interactions, while by integrating the complete function we get
the well-known answer $1/(2\pi)^2 (m/x) K_1(m x)$. The latter indeed
decays exponentially at large $x$ as $\propto \exp(-m x)$, and $m$
is the distance to the singularity in the complex plane.
\label{fn:mgb} }. 

Next we turn to the non-analytic terms in the expansions. 
The important point here is that the
integral over the \emph{noninteger} powers of $p$ gives a
power law. The lowest such power, $p^{2\nu}$, gives the largest
contribution.
Then the value of $a(x)$, Eq.
(\ref{a}), for large $x$ is the same as the Fourier transform of its leading
non-analytical part.
Explicitly,
using Eqs. (\ref{euclideanT}), (\ref{euclideanL}), (\ref{a}), and (\ref{tJint}),
\begin{eqnarray}
  \label{eq:a_largex}
 a (x)\stackrel{{\rm large}\; x}{\longrightarrow}
\frac{1}{\pi^2}\frac{\nu^2(\nu+2)}{(\nu-1)^2}
\frac{1}{\kappa^{2\nu+1}x^{2\nu+4}}
\label{large-a}
\end{eqnarray}
The same argument can be applied to find the large $x$ behavior
of $b(x)$.
In position space, this becomes
\begin{eqnarray}
  \label{eq:b_largex}
 b(x)\stackrel{{\rm large}\; x}{\longrightarrow}
-\frac{1}{\pi^2}\frac{2\nu^2(\nu+2)}{(\nu-1)^2}
\frac{1}{\kappa^{2\nu+1}x^{2\nu+4}}
\label{large-b}
\end{eqnarray}
Note that $b(x)/a(x) \stackrel{{\rm large}\; x}{\longrightarrow} -2$ as predicted by the AdS-CFT correspondence. We find good agreement in
comparing Eqs.~(\ref{large-a}) and (\ref{large-b})
with the curves in Figure ~\ref{fig:Greens_position} at
large $x$.  The position-space correlator at large distances therefore has all the properties of a CFT vector correlator, providing another validation for the RS2 -CFT correspondence \cite{Gubser:1999vj,ArkaniHamed:2000ds}. 

For the scalar one obtains 
\begin{equation} 
D^{\rm scalar}(x) \stackrel{{\rm large}\; x}{\longrightarrow} \frac{1}{\pi^2} \frac{ \nu^2_S (\nu_S+1)}{(\nu_S-2)^2} \frac{1}{\kappa^{2 \nu_S+1} x^{2 \nu_S +4}}
\label{position space scalar CFT}
\end{equation}
Thus the dimension of the scalar operator in the CFT is $d_S = 2 + \nu_S$, which is correct \cite{Gubser:1998bc,Witten}.
This analytic formula agrees well with the plots in Figures
\ref{fig:Greens_position_scalar} and  \ref{fig:Greens_position_scalar_detail} at large $x$. 

Although 
the 
contact 
interactions are found to be manifestly present at low-momentum, there is an additional subtlety here.  
While the contact terms are seen to dominate the {\em low-energy} scattering amplitude, 
we have seen that interactions
between two sources of the vector field separated by a {\em large distance}
on the brane are dictated by the conformal part of the
interaction. Or in other words, scattering amplitudes at large and fixed impact parameter are dominated by the CFT contribution, not the contact interactions.  The dominance of the contact interactions in scattering amplitudes can be understood by recalling that plane wave scattering, which averages over all impact parameters large and small, receives contributions from all distance scales, even if the external momenta are small. 
The notions of ``low energy'' and ``long
distance'' mean not quite the same thing in this case.

\subsubsection{Short Distance} 

Next we turn to understanding the short distance limits of the correlators. 
To do that, we need 
to consider the limit of large space-like $p$ and use
$lim_{p\rightarrow\infty}p K_{\nu-1}(p/\kappa)/K_{\nu}(p/\kappa)=p$.
Then using again Eq. (\ref{tJint}),
we immediately obtain
\begin{eqnarray}
  \label{eq:position_transverse_flat}
  a (x) &=& \frac{1}{8\pi^2}\int_0^\infty p^3 dp
    \left[\left( \frac{J_1(px)}{(px)}-\frac{J_2(px)}{(px)^2} \right)(p^{-1}+ \cdots) 
     +\frac{1}{m^2_5} \frac{J_2(px)}{(px)^2} (p+ \cdots)
    \right] \nonumber \\ 
    &=  &  \frac{1}{8\pi^2 m^2_5} \frac{3}{x^5}+ \cdots ,\\
  b (x) &=&\frac{1}{8\pi^2 m^2_5}\int_0^\infty p^3 dp
  \left(\frac{ J_1(p x)}{p x}- 4 \frac{J_2(px)}{(px)^2}
    \right) (p + \cdots ) 
     =   -\frac{1}{8\pi^2 m^2_5} \frac{15}{x^5} + \cdots 
\end{eqnarray}
and 
\begin{eqnarray} 
  D^{\rm scalar}(x)  =  
  \int_0^\infty \frac{p^3 dp}{4 \pi^2}  
  \frac{J_1(px)}{p x} (\frac{1}{2p}+ \cdots )
 =  \frac{1}{8 \pi^2} \frac{1}{x^3}  + \cdots 
\end{eqnarray}
The quantitative agreement between these analytical results and the numerical
ones given in  Figures ~\ref{fig:Greens_position}, \ref{fig:Greens_position_scalar}, 
\ref{fig:Greens_position_scalar_detail}, \ref{fig:Greens_position_vector_resonance} and 
\ref{fig:Greens_position_vector_resonance_b} at short distances is excellent. These are the brane-to-brane correlators one expects from a massive vector or scalar boson propagating in flat, five-dimensional space. 

\subsubsection{Technical Remark}

Taken literally, the integrals in the above equations do not converge.
For example, at large $p$ the integrand in
Eq.~(\ref{eq:position_transverse_flat}) behaves as $\sim
p^{3/2} \cos[p x-3 \pi/4]$. In general then,  the
integral in Eq. (\ref{tJint}) converges only for $-Re\; \beta -1 < Re\; \alpha < 1/2, a>0$
\cite{GradshteynRyzhik}. Because of this divergence, 
the integrals are understood to be
regularized with the damping factor $e^{-\epsilon p}$. The regularized integral can be obtained
analytically from  p. 691, Eq. 6.621-1 of \cite{GradshteynRyzhik},
\begin{equation}
\int_0^\infty e^{- \epsilon p} t^\alpha J_\beta (a t)dt =
\frac{\left(a/2 \right)^\beta}{\sqrt{(\epsilon^2+a^2)^{\beta+\alpha+1}}}
\frac{ \Gamma[\beta+\alpha+1]}{\Gamma[\beta+1]} F\left(\frac{\beta+\alpha+1}{2},\frac{\beta-\alpha}{2},\beta+1;\frac{a^2}{\epsilon^2+a^2}\right)
\end{equation}
where $F(a,b,c;z)$ is the hypergeometric function. The limit $\epsilon \rightarrow 0$ of  the right-hand side of the above equation gives Eq. (\ref{tJint}).
 This, of course, is the standard approach in
dealing with Fourier transforms of Green's functions\footnote{For
  example, the same regularization is assumed when a constant in $p$
  space is transformed to yield a contact term in position space:
   for $\alpha=2$,  $\beta=1$, the right-hand-side of (\ref{tJint}) is $\Gamma(1+\alpha/2)/\Gamma(1-\alpha/2)$, which 
  indeed vanishes.}.  The important point is that the same consideration
of convergence applies to our numerical integrals in
Eqs.~(\ref{a}), (\ref{b}), and (\ref{Euclidean position space scalar correlator}).

\subsubsection{Finite Range} 

As we have seen, the low-momentum expansion of the vector and scalar propagators consists of a series of analytic terms and a series of non-analytic terms, of which the leading non-analytic term has a singularity at $p^2=0$. From the general considerations in \cite{Migdal}, one therefore expects that at large distances the Fourier transform to only be power-law suppressed. 

The analytic terms, on the other hand, give rise to delta functions and derivatives of delta functions. This would seem to imply that at short distances the position-space correlator is highly singular. 
But this is incorrect, because this expansion cannot be used to determine whether these contact terms indeed characterize the short distance behavior, simply because the series describes a function that is being expanded about its branch point, which is at $p^2=0$. A simple example is provided by the propagator of a massive particle, which has a finite range of $O(m^{-1})$, that is missed in the Fourier transform of the series expansion about $p^2=0$ (see also footnote \ref{fn:mgb}).
Instead, one needs the Fourier transform of 
the high momentum behavior of the propagators, which as we have seen explicitly have no fundamental contact interactions. 
Indeed this description is confirmed in the region  $0 < x\lesssim1$ shown in  Figures ~\ref{fig:Greens_position} and \ref{fig:Greens_position_scalar}, where the Euclidean propagators make a transition around the scale $x \simeq \kappa^{-1}$ to a less divergent power-law.

\subsubsection{Small bulk vector and scalar masses: $m^2_5, m^2_S  \ll \kappa^2$} 
\label{sec:small mass} 

Finally, we return to the case of $\nu \rightarrow 1$ and $\nu_S \rightarrow 2$, which are not covered by the previous analyses. 
We begin with the vector, returning to the scalar at the end of the Section. 

The appearance of divergences at $\nu=1$ in the asymptotic form of the correlators indicates that our perturbative expansion is breaking down. This is not surprising, since physically this is the limit of vanishing mass for the bulk vector boson, or equivalently, the limit in which the corresponding CFT current is conserved (vanishing anomalous dimension). 
Indeed, inspecting the series expansions Eqs. (\ref{eq:longitudinal_branetobrane_expand_eucl}) and (\ref{eq:transverse_branetobrane_expand_eucl})
reveals $(p/\kappa)^2 \ll \nu^2-1$ and $(p/\kappa)^2 \ll (\nu-1)^2$ are required, respectively. These conditions are obviously impossible to satisfy for any fixed $p$ when $\nu \rightarrow 1$.  
For $\nu$ close to 1 we therefore need to be more careful with the analysis.

To understand the physics, first consider the limit $\nu=1$. Then the bulk gauge symmetry is 
restored and we can ignore the longitudinal propagator since it is gauge-dependent, and focus on the transverse propagator which from Eq. (\ref{euclideanT}) is simply 
\begin{eqnarray} 
  \Delta^{(T)}_E(p) =  
  \frac{1}{2}\left[ \frac{K_{1}(p/\kappa)}{p K_{0}(p/\kappa)} \right]  
  = - \frac{1}{2} \frac{\kappa}{ p^2 \log{(p/ (2\kappa))}} 
  + O(1)+ O(p^2)+ O(1/\log{p^2}),  
 \label{massless transverse}
\end{eqnarray} 
where the second line is valid at low energies $p \ll \kappa$. The Fourier transform of the leading log term gives $x^{-2}$, which is not the correct scaling behavior for the correlator of a current of dimension 3 (which would be $x^{-6}$).
As explained by \cite{WittenSantaBarbara} and \cite{ArkaniHamed:2000ds}, the CFT interpretation of this behavior is instead the following.  In the UV the four-dimensional theory has an external massless gauge boson coupled to a conserved current of a CFT.  In the IR the gauge boson mixes with the vector current of the CFT, producing the non-trivial gauge boson correlator. To see that, note that to next-leading order in perturbation theory the gauge boson correlator is given by the free-theory propagator plus a vacuum bubble insertion of the CFT current-current correlator. The latter insertion causes the coupling to run to zero at $p \rightarrow 0$, which is why the RS 2 model
 has no massless four-dimensional gauge boson in its spectrum. Note that this interpretation gives $1/p^2 \times (p^2 \log p^2) \times 1/p^2 \sim \log p^2 /p^2$ for one such insertion, a form which agrees with Eq. (\ref{massless transverse}) when the $\log p^2$ term can be treated as ``small"; AdS-CFT predicts that summing the bubbles must reproduce
 Eq. (\ref{massless transverse}). 
From Eq. (\ref{massless transverse}) we learn that in this limit the Fourier transform is regular, and that contact interactions are 
indeed present. GIR \cite{GIR} also found contact interactions to be present (and in fact necessary) when the CFT current is conserved. 

Let us now consider $\nu$ close to but not equal to $1$. Recall that this means the bulk vector boson mass $m_5$ is much smaller than the curvature scale. Physically we expect that the description for $m_5 \ll \kappa$ should smoothly map onto the  four dimensional description above having $\nu=1$ $(i.e.,~m_5=0)$. 
This means there should be a light state (actually a resonance), with a mass $m_0$ much less than the curvature scale. When $m^2_5 \rightarrow 0$ this resonance becomes identified with the (stable) massless four dimensional gauge boson state that mixes with the CFT, described above. For 
$m^2_5 >0$ this state is unstable because it couples to the CFT, but for small enough $m_5$ the resonance is expected to be narrow since it must become stable in the limit $m_5=0$.

To find the resonance, we just have to look for a pole in the Minkowski-space transverse propagator $\Delta^{(T)}(p)$. That is,\begin{equation} 
 \frac{p_R}{\kappa} \frac{H^{(1)}_{\nu-1}(p_R/\kappa)}{H^{(1)}_{\nu}(p_R/\kappa)} - (\nu-1) =0 
 \label{resonance-pole}
 \end{equation} 
 with  $p^2_R=m^2_0-i m_0 \Gamma$. 
Moreover, 
the mass and width should satisfy $m_0/ \kappa  \ll 1$, $\Gamma/ m_0 \ll 1$  when $\nu -1 \ll 1$. 
A consistent solution for the resonance mass and width  can be found. It is 
\begin{equation} 
m^2_0 = 2(\nu-1) \kappa^2 \frac{1}{- \log[ m^2_0/(4 \kappa^2 e^{-2 \gamma})]}
\end{equation} 
and is positive-definite in the region where the perturbative expansion can be trusted, {\em i.e.,} $p_R \ll \kappa$. Here $\gamma=0.577216....$ is the Euler-Gamma constant. The 
approximate solution is given by 
\begin{equation} 
m^2_0 \simeq 2 (\nu-1) \kappa^2  \frac{1}{- \log[(\nu-1)/4]}
\label{approximate solution}
\end{equation} 
In this same limit of $m_0 / \kappa \ll 1$  the 
width is  
\begin{equation} 
\frac{\Gamma}{m_0} = \frac{\pi }{ - \log [m^2_0/(4 \kappa^2 e^{-2 \gamma})]}
\end{equation} 
which is automatically narrow in the region $| \log m^2_0/\kappa^2| \gg 1$. Solutions to Eq. (\ref{resonance-pole}) can be found numerically and are found to agree quite well with these approximate analytical results  
in the region where we expect them to.  Finally, since the mass and width of the resonance 
vanish in the limit $\nu \rightarrow 1$, there is a smooth transition from this description to the preceding description of 
$\nu =1$.

\begin{figure}
\includegraphics[angle=0,width=13cm]{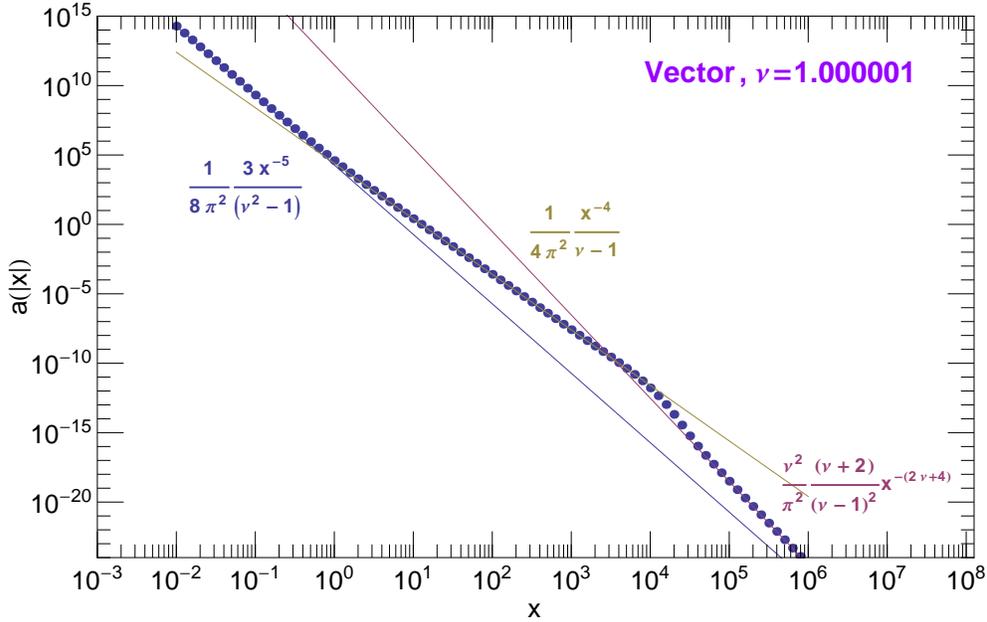}
\caption{The Euclidean Green's function $a(x)$ of the vector boson in position space, with the same conventions and notation as Figure 1. Note  that compared to Figure 1, here the transition to the pure CFT behavior has been pushed out to $x \sim m_0^{-1}$. }
 \label{fig:Greens_position_vector_resonance}
\end{figure}

\begin{figure}
\includegraphics[angle=0,width=13cm]{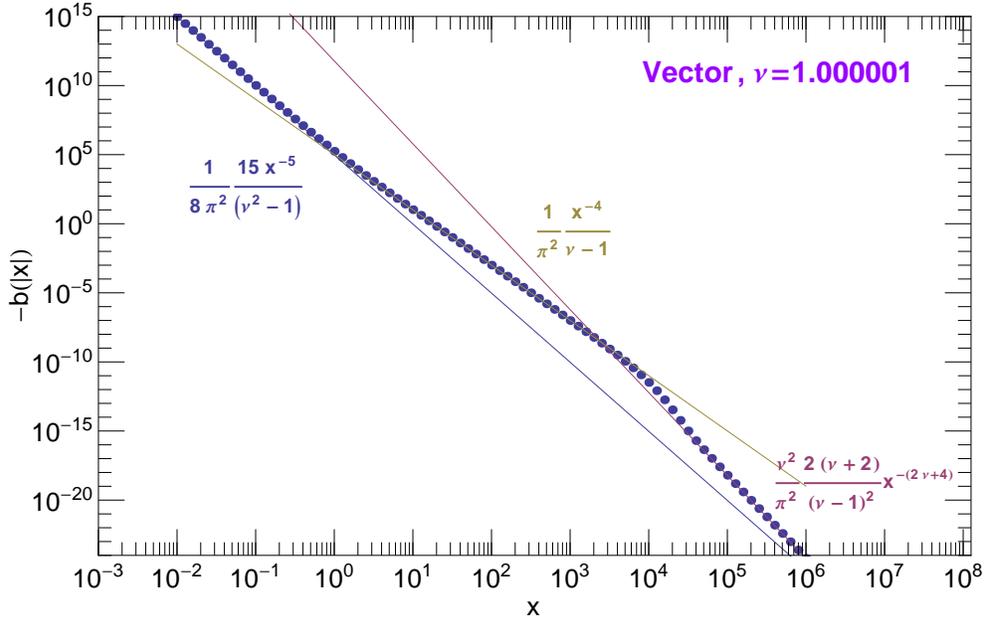}
\caption{The Euclidean Green's function $b(x)$ of the vector boson in position space, with the same conventions and notation as Figure 1. }
 \label{fig:Greens_position_vector_resonance_b}
\end{figure}

In Figures \ref{fig:Greens_position_vector_resonance} and 
 \ref{fig:Greens_position_vector_resonance_b} 
the coefficients $a(x)$ and $b(x)$ of the position space propagator when $\nu = 1.000001$ are shown.
Because of the resonance, the Fourier transform of the transverse propagator exhibits {\em three regimes}.  

\begin{figure}
\includegraphics[angle=0,width=13cm]{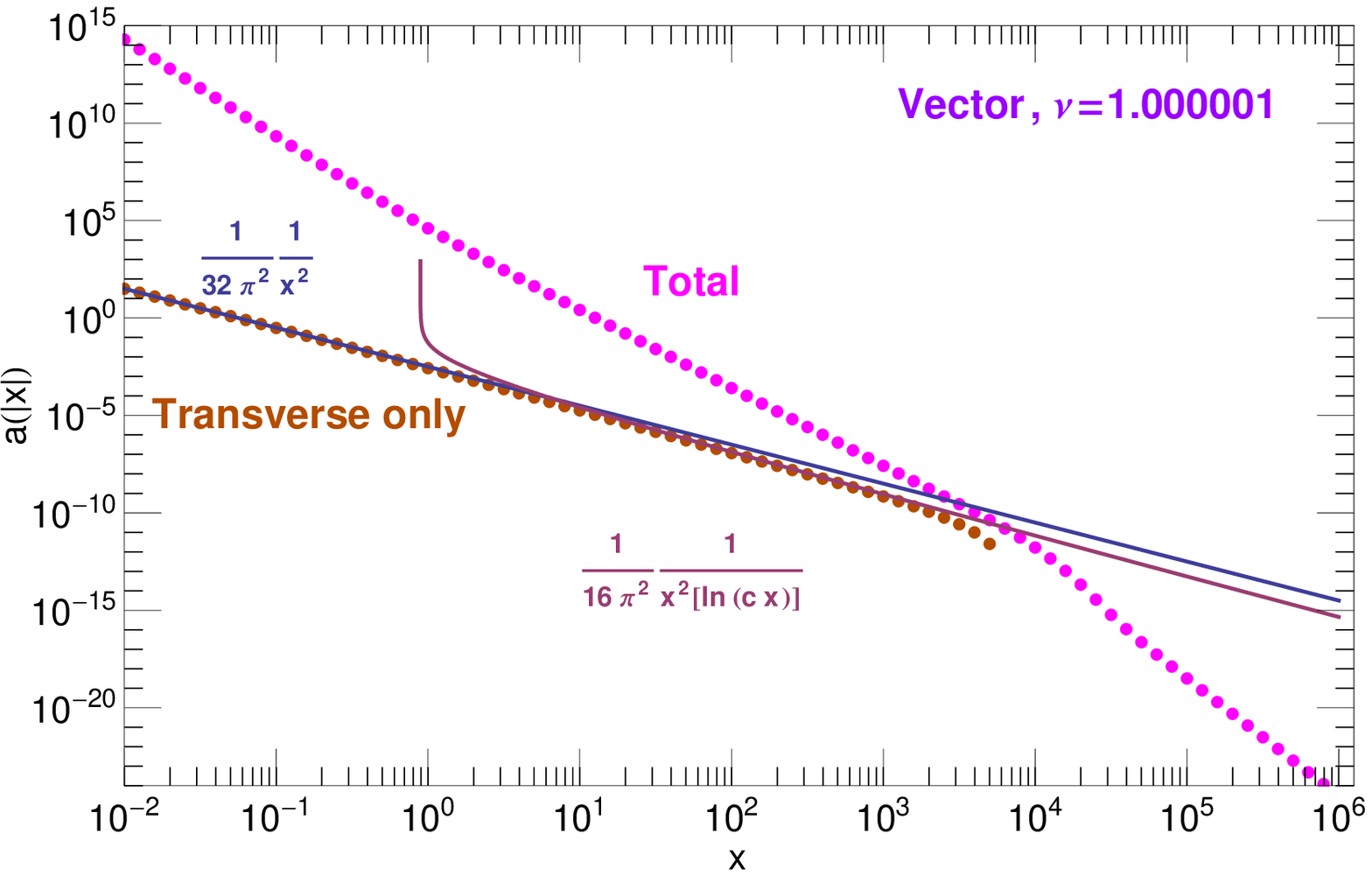}
\caption{The Euclidean Green's function $a(x)$ of the vector boson in position space. Here the contribution of the transverse propagator is also shown, with the same conventions and notation as Figure 1. }
 \label{fig:Greens_position_vector_resonance_transverse_a}
\end{figure}

\begin{figure}
\includegraphics[angle=0,width=13cm]{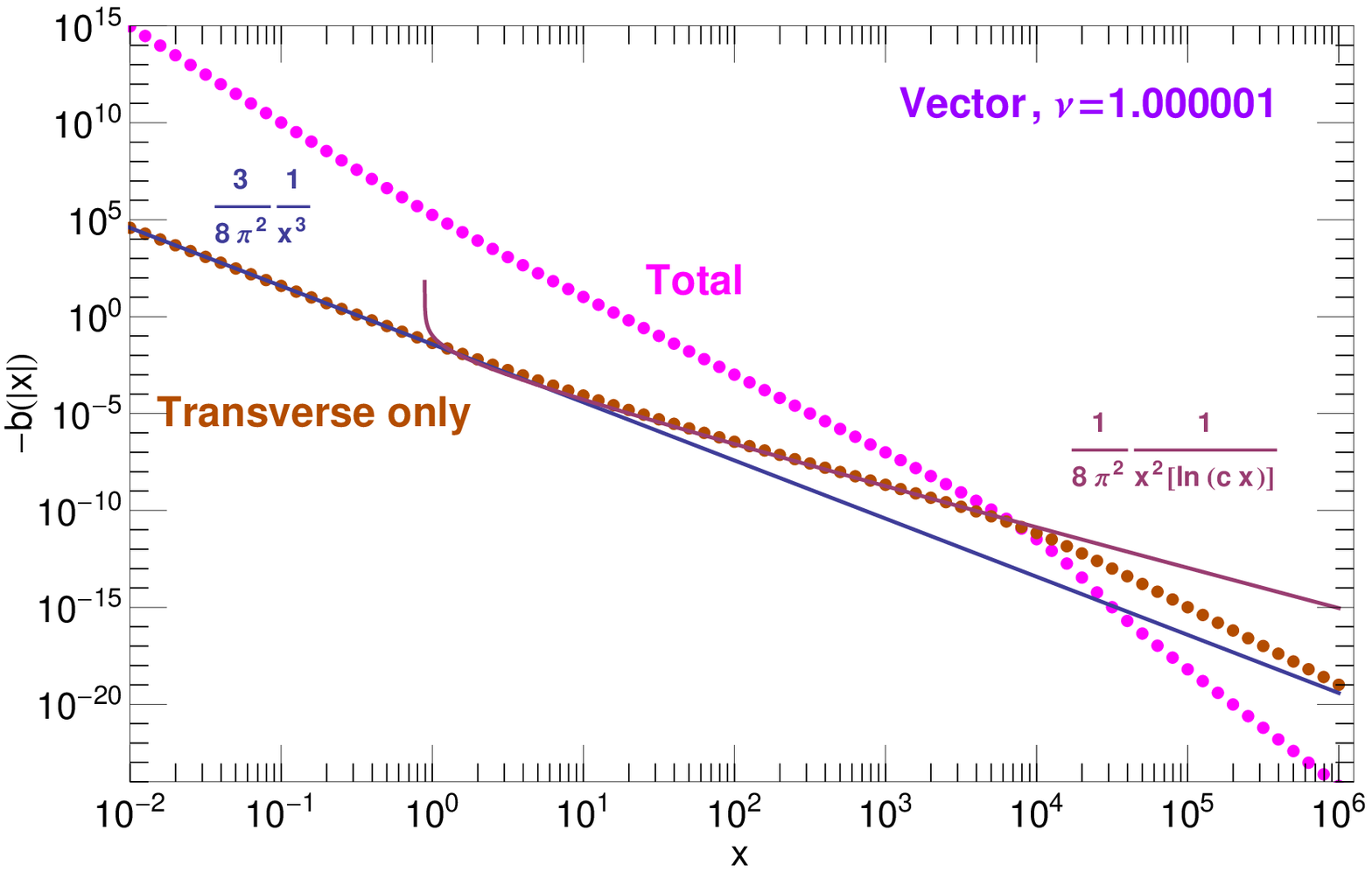}
\caption{The Euclidean Green's function $b(x)$ of the vector boson in position space. Here the contribution of the transverse propagator is also shown, with the same conventions and notation as Figure 1. }
 \label{fig:Greens_position_vector_resonance_transverse_b}
\end{figure}

First, there is the region $x \ll \kappa^{-1} $ described by the flat-space region. 
At distances below the curvature scale the position space correlator is dominated by the longitudinal mode, as explained previously and seen in Figures   
\ref{fig:Greens_position_vector_resonance} and 
 \ref{fig:Greens_position_vector_resonance_b}.  However, one may also be interested in the contribution of the {\em transverse} mode to the position-space correlator, simply because at short distances the mass of the vector boson may be generated by a  Higgs mechanism. If so, then at distances below the inverse Higgs boson mass the longitudinal component becomes gauge-dependent and unphysical. 
For that reason, in  
Figures \ref{fig:Greens_position_vector_resonance_transverse_a} and 
\ref{fig:Greens_position_vector_resonance_transverse_b}  we show the contributions of the {\em transverse} mode to the total position space correlator, again for $\nu=1.000001$. 

To obtain an analytic formula for the contribution of the transverse propagator at short distances, one sets $\nu=1$ and expands 
\begin{eqnarray} 
\Delta^{(T)}_E(p) & = & \frac{1}{2p} \left( 1 +\frac{\kappa}{2 p} + \cdots \right)
\end{eqnarray}
It turns out the next-to-leading order term is needed 
because the leading terms contributing to $a(x)$ at $O(x^{-3})$ cancel. One then has for the contribution of the transverse mode only, 
\begin{eqnarray} 
a^{(T)}(x) & \stackrel{x \rightarrow 0}{=} & \int_0^\infty \frac{p^3 dp}{4 \pi^2}
    \left(\frac{J_1(px)}{(px)} - \frac{J_2(px)}{(px)^2} \right) \left(\frac{1}{2p} \right) \left[1+ \frac{\kappa}{2p} + \cdots \right]   \nonumber \\ 
     & =&  \frac{1}{32 \pi^2} \frac{\kappa}{x^2} + \cdots \\
b^{(T)}(x) &  \stackrel{x \rightarrow 0}{=} &  \int_0^\infty \frac{p^3 dp}{4 \pi^2} \left(\frac{J_1(px)}{px}- 4 \frac{J_2(px)}{(px)^2} \right)
  \left( -\frac{1}{2p} + \cdots  \right) \nonumber \\ 
  & = & \frac{3}{8 \pi^2}\frac{1}{x^3} + \cdots   
\end{eqnarray} 
These expressions agree well with the short-distance behavior of the numerical results presented in Figures \ref{fig:Greens_position_vector_resonance} and 
 \ref{fig:Greens_position_vector_resonance_b}.

Next, there is an intermediate region $\kappa^{-1} \ll x \ll m^{-1}_0$. 
Here the position-space correlator is dominated by its longitudinal component. To see that, note that as 
$\nu \rightarrow 1$ the transverse propagator has the form (\ref{massless transverse}), but the longitudinal propagator is instead 
\begin{equation} 
\Delta^{(L)}_E (p) \stackrel{\nu \rightarrow 1}{\longrightarrow} \frac{1}{2} \frac{1}{\nu-1}  \gg \Delta^{(T)}_E (p)
\end{equation} 
Inserting this solution into  (\ref{a}) and (\ref{b}) one finds 
\begin{eqnarray} 
a(x) & \simeq  & \frac{1}{4 \pi^2} \frac{1}{\nu-1} \frac{1}{\kappa x^{4}} + \cdots \\
b(x) & \simeq & \frac{-1}{\pi^2}  \frac{1}{\nu-1} \frac{1}{\kappa x^{4}} + \cdots 
\end{eqnarray}
which agrees well with the plots in this region of $x$. Since the resonance is in the transverse propagator which is suppressed compared to the longitudinal propagator, 
its effect on the position space correlator  is not noticeable in Figures \ref{fig:Greens_position_vector_resonance} and 
 \ref{fig:Greens_position_vector_resonance_b} . 
 
The transverse contribution to $a(x)$ in this region is shown in 
 Figure \ref{fig:Greens_position_vector_resonance_transverse_a},  and to $b(x)$ in Figure
\ref{fig:Greens_position_vector_resonance_transverse_b}. Good analytic approximations to 
these contributions are obtained 
by beginning with the momentum-space  transverse propagator evaluated for vanishing bulk mass 
($\nu=1$), Eq.(\ref{massless transverse}), 
\begin{equation}
\Delta^{(T)}_E(p)  \simeq  -\frac{1}{2}\frac{\kappa}{p^2 \log[p/(2 \kappa)]} 
\end{equation}
Next, since the Fourier transform of this expression is dominated by $p \sim 1/x$ and $\log p$ is a slowly varying function, we set $p=1/x$ in the logarithm and Fourier transform $p^{-2}$. This gives 
\begin{eqnarray} 
a^{(T)}(x)  & \simeq & \frac{\kappa}{16 \pi^2} \frac{1}{x^2} \frac{1}{ \log[c \kappa x]]} \\
b^{(T)}(x)  & \simeq & \frac{\kappa}{8 \pi^2} \frac{1}{x^2} \frac{1}{\log[c \kappa x]} 
\end{eqnarray} 
These results are also shown in 
Figures \ref{fig:Greens_position_vector_resonance_transverse_a} and 
\ref{fig:Greens_position_vector_resonance_transverse_b}. The numerical constant 
 $c=2 e^{- \gamma}$ is determined by fitting these formulae to the plots. 
These  Figures are 
seen to agree well with the numerical solution for $\kappa^{-1} \ll x \ll m_0^{-1}$.

At $x \sim m_0^{-1}$ one expects a cross-over from the `logarithmic' scaling to 
a region dominated by the resonance. One has $m^{-1}_0
 \sim (\nu-1)^{-1/2} \log ^{1/2} (\nu-1) \kappa^{-1} $ which for 
 $\nu=1.000001$ is roughly 
$ \sim few \times 10^{3} \kappa^{-1} \gg \kappa^{-1}$. The location of this transition is in general agreement with the results seen in Figures  \ref{fig:Greens_position_vector_resonance}--\ref{fig:Greens_position_vector_resonance_transverse_b}.

Finally, for the region $x \gg m^{-1}_0$ the position-space correlator exhibits the CFT behavior, namely that 
$b(x)/a(x)  \stackrel{{\rm large} ~x}{\longrightarrow}  -2$ and $a(x) \sim x^{-6 - 2(\nu-1)}$. As $\nu \rightarrow 1$ the pure CFT behavior is pushed to $x \rightarrow \infty $.

We end this Section by considering the bulk scalar boson when $m^2_S \ll \kappa^2$. Its position space correlator is shown in detail in  Figure  \ref{fig:Greens_position_scalar_detail} for $\nu_S =2.0001$. 
Notice that compared to the simple power-law behavior in Figures \ref{fig:Greens_position_scalar} seen for larger values of $\nu_S$, here there is an additional prominent ``bulge" at intermediate distances. This effect can be simply understood as due to 
 the scalar resonance found in Ref. \cite{Dubovsky:2000am} .
For a bulk scalar having $m^2_S \ll \kappa^2$ there is a resonance present, with a
mass and width given by \cite{Dubovsky:2000am}
\begin{eqnarray} 
m^2_{bs} & = &
 m^2_S/2 \\
 \Gamma_{bs} & = & \frac{\pi}{16} \frac{m^3_{bs}}{\kappa^2}  
\end{eqnarray} 
which  can be obtained by expanding the scalar Green's function (\ref{Euclidean scalar correlator}) at small momentum. Then for $\kappa^{-1} \ll x \ll x_{CFT} $ the position-space Green's function is easily found to be 
\begin{equation} 
D^{\rm scalar}(x) = \frac{1}{4 \pi^2} \frac{m_{bs}}{x} K_1(m_{bs} x) 
\label{position space scalar resonance}
\end{equation} 
where $x_{CFT}$ is defined to be the scale at which the contribution of the CFT begins to dominate. 
Comparing the plot with the analytic formula in this region 
one finds good agreement. 
The cross-over to pure CFT behavior occurs when the resonance contribution to the position-space correlator is comparable to the contribution from the CFT, which upon equating (\ref{position space scalar resonance}) with (\ref{position space scalar CFT}) leads to  $m_5 x_{CFT} \sim \log[\kappa/m_5]$ \cite{Dubovsky:2000am}.
 
The difference between the effect of the scalar and vector resonances on the position space correlator is probably due to the logarithmic running of the coupling of the vector boson to the CFT at energies above its mass. Indeed, the analytic expression for $a(x)$ (Eq. \ref{massless transverse}) corresponding to  $m^2_5=0$ agrees quite well with the numerical results in the region $ \kappa^{-1} < x < m_0^{-1}$.  Also note the transition to the pure CFT behavior is seen to occur at approximately $ x \sim m_0^{-1} \gg \kappa^{-1}$. 
For larger values of $m_5$ below the curvature scale there is still a resonance, but the width is broad and analytic solutions cannot be obtained perturbatively.

\subsection{Generalization to other brane dimensions}
\label{sec:generalD} 

The analysis of perturbative unitarity can be generalized in a straightforward manner to other space-time dimensions. For massive vector  fields propagating on $AdS_{D+1}$ the relation between the order and the vector  boson bulk mass becomes 
$\nu=\sqrt{m^2_{D+1}+ (D-2)^2/4}$ \cite{ADSCFT_Vector} and the corresponding operator dimension is $d_V=D/2 + \nu$ \cite{ADSCFT_Vector}. 
Next, we find that the unitarity condition for forward scattering of brane localized states through an intermediate bulk gauge boson  
remains $m^2_{D+1} \geq 0$, implying that $\nu \geq (D-2)/2$. Taken together these relations and the bulk unitarity condition $\hbox{Im}{\cal A} \geq0$ imply the correct unitarity bound \cite{Minwalla:1997ka} \begin{equation} 
d_V \geq D-1 
\end{equation} on primary vector operators in $D$ space-time dimensions.

Next we turn to generalizing the properties of the spectral representation. For the transverse mode one has in $D$ space-time dimensions 
\begin{equation}
 \langle {\cal O}_{\mu}(p) {\cal O}_{\nu}(-p) \rangle
 \sim  \left[
       - \eta_{\mu\nu} +\frac{p_\mu p_\nu}{p^2}
    \right] \int^{\infty}_0 d M^2 \frac{(M^2)^{d_V-D/2}}{p^2 -M^2 + i \epsilon}   
  \end{equation} 
The only change from $D=4$ is a dependence on $D$ in the power; this is needed in order that in position-space the correlator have scaling dimension $d_V$. Observe that the integral converges in the UV only if 
\begin{equation} 
d_V < D/2
\end{equation} 
If $d_V$ is larger than this value, then the integral must be regularized and this will lead to contact interactions. 

To see that the contact interactions are important for values of $d_V$ precisely above this upper limit, note that in the scattering amplitude the leading non-analytic part will scale as 
\begin{equation} 
{\cal A} \sim p^{2 \nu}
\end{equation} 
where 
\begin{equation} 
\nu = d_V-D/2
\end{equation}
The contact interactions begin in general at $p^0$ and therefore dominate if $\nu >0$, or in other words, 
$d_V > D/2$. 

Combined with the unitarity bound, we see that the spectral integral converges and the vector contact interactions are subdominant only for 
\begin{equation} 
D-1 \leq d_V < D/2 
\end{equation} 
which cannot be satisfied for any $D \geq 2$. For (gauge-invariant, primary) vector operators then, contact interactions appear to be always relevant to low-energy scattering amplitudes. 

It is instructive to repeat this exercise for scalar operators. The unitarity bound is \cite{Minwalla:1997ka}
\begin{equation} 
d_S \geq (D-2)/2
\end{equation} 
which is weaker than for vector operators. 
The conditions that the spectral integral is convergent in the UV and that contact interactions are subdominant remains unchanged, 
\begin{equation} 
d_S < D/2
\end{equation} 
Thus for scalars, the window where unitarity is obeyed, the spectral integral is 
convergent in the UV, and contact interactions are subdominant (or nonexistent) is 
\begin{equation} 
(D-2)/2 \leq d_S \leq D/2 
\end{equation} 
Note that for $D=2$ the window is $0 \leq d_S < 1$, which is precisely the range of 
scalar operator dimensions considered by Georgi and Kats in their $D=2$ unparticle example \cite{Georgi:2008pq}. It is therefore not surprising that they do not find any contact interactions. 
But given the above general discussion on vector operators, we do expect contact interactions in this model if 
the ``SM" particles are coupled to vector operators of the Sommerfeld model, rather than to 
scalar operators. It would be interesting to explore this further. 

\section{Lykken-Randall model} 
\label{sec:tensionlessbrane}

In this Section we consider the Lykken-Randall 
model \cite{lykken-randall}, which describes the RS 2 model with in addition a 
``tensionless brane'' or ``probe brane'' (LR brane) located  in the infra-red at $z = z_{SM} > \kappa^{-1}$.  In fact, the location of the UV brane will not be essential to the following discussion and it can be decoupled. We will assume that all the SM degrees of freedom are localized on the LR brane. As we shall see, this scenario provides another realization of ``unparticle-stuff''. We will see that the following properties of unparticles -  the phase of the unparticle propagator, the tensor structure of the CFT propagator, the unitarity bounds on the operator dimensions, and dimensional transmutation - all emerge naturally. Contact interactions will also be found. 

There are two factors that motivate consideration of this model. First,  
from the CFT description the SM particles are composite fields of the breaking of $CFT_1 \rightarrow CFT_2 \times SM$ at the scale $\Lambda_{{\cal U}}= 1/z_{SM}$ \cite{ArkaniHamed:2000ds}.  Note that this  ``$CFT_2$'' plays the role of the hidden sector CFT in the unparticles scenario, whereas here the ``$CFT_1$'' is in the unparticle scenario the UV completion of the SM, the hidden sector CFT and their interactions. Above the scale $\Lambda_{{\cal U}}$ the transition is sharp, since for $z_{UV}<  z< z_{SM}$ the geometry is engineered to be AdS. 
Likewise, the $CFT_2$ is described 
in the bulk by those modes living in the region $z > z_{SM}$. Because both the SM fields and the states in the low-energy CFT  ``$(CFT_2)$" arise from the same high-energy CFT, effective field theory suggests that interactions  between these two sectors in the form of higher dimension operators will be generated at the ``transmutation scale" $\Lambda_{{\cal U}}$. Thus from the CFT point of view, interactions between the  SM fields and low-energy CFT operators  are {\em generically} expected. From the brane point of view, these interactions are modeled by introducing explicit couplings between SM operators localized on the LR brane and fields living in the bulk. 

That interactions exist between probe brane localized observers and the hidden sector CFT has already been previously noted in the literature. 
Such observers see two kinds of 
gravitational corrections to Newton's law \cite{lykken-randall}: a universal correction caused by the bulk gravitational modes \cite{RSII}; and a probe brane-specific correction \cite{lykken-randall}. 
The latter correction, not present in the original RS 2 model,  was interpreted in \cite{ArkaniHamed:2000ds} as due to the presence of $z^{-1}_{SM}$ suppressed contact interactions between the stress-tensors of the Standard Model and the $CFT_2$.

Finally, the ``transmutation scale" in the RS 2 model occurs at the AdS curvature scale $\kappa$. But since the curvature scale together with the five-dimensional Planck mass determines the four dimensional Planck mass,  it is desirable to construct a model in which the transmutation scale is unrelated to these other ones. The LR model does just that.

\subsection{Transverse Mode} 
\label{LR transverse mode} 

In the region $\kappa^{-1} < z < z_{SM}$ the solution for the transverse propagator is 
\begin{equation} 
\Delta ^{(T)} _<(p,z) =  a^{(1)}_T(p) z \left( J_{\nu}(pz) + b(p) Y_{\nu}(pz) \right) 
\end{equation} 
where as before $p=\sqrt{p^2}$ and 
$\nu = \sqrt{1+ m^2_5 /\kappa^2}$. Throughout we will assume that $m^2_5 /\kappa^2 \geq -1$ in order that $\nu$ is purely real. (Thus the Bessel functions $J_{\nu}(x)$ and $Y_{\nu}(x)$ are real.) Later it will be shown that in order for SM-SM scattering amplitudes on the LR brane to be unitary  the stronger condition $m^2_5 \geq 0$ is required. 

The factor $b(p)$ is fixed by 
the Neumann boundary condition at the UV brane ($z =\kappa^{-1}$) to be 
\begin{equation} 
b(p)= -\frac{(1-\nu) J_{\nu}(p /\kappa)+ p /\kappa J_{\nu-1}(p/\kappa)}{(1-\nu) 
Y_{\nu}(p /\kappa)+ p /\kappa Y_{\nu-1}(p/\kappa)} 
\label{b(p)}
\end{equation} 
and by inspection is real.
This factor is temporarily neglected in what follows in order to simplify the algebra. It will be reinstated in the final result below, Eq. (\ref{LRtransverse}). Physically,  
we can neglect this contribution to the Green's function for $p /\kappa \ll 1$ since  in this limit $ b(p) \simeq (p /\kappa)^{2 \nu} \ll 1$. Moreover, since we will evaluate the 
Green's function at 
$z=z_{SM} \gg \kappa^{-1}$, such contributions are always subdominant. (If $p z_{SM} \simeq O(1)$ the contribution of the $b Y_\nu$ term is suppressed by $(p/\kappa)^{2 \nu}$
and if $p z_{SM} \ll 1$ it is suppressed by $( \kappa z_{SM})^{-2 \nu}$.)
Therefore to a good approximation 
\begin{equation} 
\Delta^{(T)} _<(p,z) =  a^{(1)}_T(p) z J_{\nu}(pz) 
\end{equation}
The overall normalization factor will be fixed by the boundary condition at the location of the LR brane. 

We next solve the bulk equation of motion in the 
region $z > z_{SM}$. The general solution for the transverse mode satisfying the outgoing wave boundary condition is simply 
\begin {equation} 
\Delta ^{(T)} _>(p,z) = a^{(2)}_T(p)  z H^{(1)}_{\nu}(pz)  
\end{equation} 
Continuity of $\Delta^{(T)}$ at $z=z_{SM}$ implies
\begin{equation} 
a^{(1)}_T(p) = a^{(2)}_T(p) \frac{H^{(1)}_{\nu}(p z_{SM})}{J_\nu (p z_{SM})} 
\end{equation} 
Finally, the modified Neumann boundary condition (\ref{ATbc}) at $z= z_{SM}$ 
\begin{equation} 
\left[ \partial_z \Delta^{(T)}_> -  \partial_z \Delta^{(T)}_<  \right]  |_{z=z_{SM}}= a_{SM}     
\end{equation} 
fixes $a^{(2)}_T(p)$ to be  
\begin{equation} 
a^{(2)}_T(p) = -i \frac{\pi }{2} a_{SM} J_{\nu}( p z_{SM}) 
\end{equation} 
(Recall that the dependence on the UV brane has been dropped here.)
The transverse propagator for $p /\kappa  \ll 1$ and with the source localized at $z=z_{SM}$ is therefore (using the notation of (\ref{eq:corr_decompose}))
\begin{equation} 
\Delta^{(T)}_>(p,z) = - i \frac{\pi }{2} a_{SM}  J_{\nu}( p z_{SM}) z H^{(1)}_{\nu}(p z) 
\label{LRtransverse:neglectb(p)}
\end{equation} 


The effect of the UV boundary condition can  be kept. Keeping $b(p)$ given by (\ref{b(p)})  one finds 
more generally that 
\begin{eqnarray} 
\Delta^{(T)}_>(p,z) &=&  - i \frac{\pi  z}{2} a_{SM} \frac{H^{(1)}_{\nu}(p z)}{1+ i b(p)} \left( J_{\nu}(p z_{SM}) + b(p) Y_{\nu}(p z_{SM}) \right)  \label{LRtransverse-compact} \\
&=& \frac{\pi z}{2} a_{SM} \frac{H^{(1)}_{\nu}(p z)}{H^{(1)}_{\nu}(p /\kappa )} 
\left[p /\kappa H^{(1)}_{\nu-1}(p /\kappa)/ H^{(1)}_{\nu}(p /\kappa )-(\nu-1) \right]^{-1}  \nonumber \\
& & \times \left(p /\kappa \left[ J_{\nu}(p z_{SM}) Y_{\nu-1}(p /\kappa) - J_{\nu-1}(p /\kappa) Y_{\nu}(p z_{SM}) \right]  \right. \nonumber \\ 
& & \left. - (\nu-1) \left[ J_{\nu}(p z_{SM}) Y_{\nu}(p /\kappa) - J_{\nu}(p /\kappa) Y_{\nu}(p z_{SM}) \right] \right) 
\label{LRtransverse}
\end{eqnarray} 
To summarize, this is the brane-to-bulk  transverse propagator evaluated in the bulk at $z>z_{SM}$ for a source localized on the LR brane at $z=z_{SM}$. In the limit $p /\kappa\ll 1$ one recovers (\ref{LRtransverse:neglectb(p)}). The brane-to-brane propagator is obtained by setting $z=z_{SM}$.

As a check on this computation, we can consider sending $z_{SM} \rightarrow \kappa^{-1}$. In this limit we should recover the RS2 transverse propagator (\ref{eq:greens_transverse_2}) with $c_T(p)$ given by 
(\ref{eq:greens_transverse_cp}).  Using the identity 
\begin{equation} 
J_{\nu}(x) Y_{\nu-1}(x) - J_{\nu-1}(x) Y_{\nu}(x) = \frac{2}{\pi}\frac{1}{x} 
\end{equation} 
and (\ref{LRtransverse}) gives 
\begin{equation} 
\Delta^T_>(p,z>z_{SM} \rightarrow \kappa^{-1}) = z   \frac{H^{(1)}_{\nu}(p z)}{H^{(1)}_{\nu}(p /\kappa)} 
\left[p /\kappa H^{(1)}_{\nu-1}(p /\kappa)/H^{(1)}_{\nu}(p /\kappa)-(\nu-1) \right]^{-1}
\end{equation} 
which agrees with (\ref{eq:greens_transverse_2}) and 
(\ref{eq:greens_transverse_cp}) (up to an irrelevant normalization of 1/2). 

Finally, taking the limit $z \rightarrow  z_{SM}$ of (\ref{LRtransverse:neglectb(p)}) gives the transverse brane-to-brane propagator in the approximation of neglecting the UV dependence,  
\begin{equation} 
\Delta^{(T)}(p) = - i \frac{\pi z_{SM}}{2}  a_{SM} J_{\nu}( p z_{SM})  H^{(1)}_{\nu}(p z_{SM}) 
\label{transverseb2bprop}
\end{equation} 

In the low energy limit $p z_{SM} \ll 1$ one obtains from (\ref{transverseb2bprop}) 
\begin{eqnarray} 
\Delta ^{(T)}(p) & =  &  z_{SM} a_{SM} \left[ - \frac{1}{2 \nu} - \frac{1}{4} \frac{1}{\nu (\nu^2-1)}  (p z_{SM})^2 + \cdots \right. \nonumber \\ 
& & \left.  - i \frac{\pi}{2 ( \Gamma[\nu+1])^2} (1+ i \cot \pi \nu) \left(\frac{p z_{SM}}{2}\right)^{2 \nu} 
+ \cdots \right]
\label{lowenergyT}
\end{eqnarray} 

\subsection{Longitudinal Mode} 
\label{sec:tensionless brane:longmode}

The equation for $\Delta_5$ in the bulk is 
\begin{equation} 
-p^2 \Delta_5 - \partial _z (a^{-3} \partial_z (a^3 \Delta_5)) + m^2_5 a^2 \Delta_5=0
\end{equation} 

In the region $\kappa^{-1} < z < z_{SM}$ the solution
having a Dirichlet boundary condition (\ref{LRUV5}) at $z=\kappa^{-1}$ is 
\begin{equation} 
\Delta^<_5(p,z) = a^{(1)}_L(p) z^2 \left( J_{\nu}(pz) - \frac{J_{\nu}(p/\kappa)}{Y_{\nu}(p/\kappa)} Y_{\nu}(pz)\right)
\end{equation}
As with the transverse mode, for $p/ \kappa \ll 1$ and $ z_{SM} \gg \kappa^{-1}$ 
we can to a good approximation drop the second term, such that
\begin{equation}
\Delta^<_5(p,z) = a^{(1)}_L(p) z^2  J_{\nu}(pz) 
\end{equation} 
Later we will restore the dependence on the UV boundary condition (see Eq. (\ref{LRlongitudinal}) below). 

In the region $z > z_{SM}$ the solution satisfying the outgoing wave condition is 
\begin{equation} 
\Delta^>_5(p,z) = a^{(2)}_L(p) z^2  H^{(1)}_{\nu}(pz) 
\end{equation}

Substituting the solutions for $\Delta_5$ in the two regions into the boundary conditions (\ref{a5bc}) and (\ref{da5bc}) 
fixes $a^{(1)}_L(p)$ and $a^{(2)}_L(p)$. One finds 
\begin{equation} 
a^{(2)}_L(p) = -i \frac{\pi }{2} \frac{1}{ z^2_{SM} a_{SM}} \frac{1}{m^2_5} J_{\nu}(p z_{SM}) \left[1+ \nu
- p z_{SM} \frac{J_{\nu-1}(p z_{SM})}{J_{\nu}(p z_{SM})} \right] 
\end{equation} 
and therefore for $z > z_{SM}$, 
\begin{equation} 
 \Delta^>_5(p,z) =   -i \frac{\pi}{2} \frac{z^2}{z^2_{SM}} \frac{1}{a_{SM} m^2_5}  H^{(1)}_{\nu} (pz) J_{\nu}(p z_{SM}) \left[1+ \nu
- p z_{SM} \frac{J_{\nu-1}(p z_{SM})}{J_{\nu}(p z_{SM})} \right] 
\end{equation} 

To obtain $\Delta^{(L)}$ we use (\ref{L5 relation}) in the bulk region $z > z_{SM}$
to obtain 
\begin{eqnarray} 
\Delta^{(L)}(p,z>z_{SM}) &=& - i\frac{\pi}{2}\frac{1}{a_{SM} z_{SM} m^2_5} J_{\nu}(p z_{SM}) 
  \left[1+ \nu
- p z_{SM} \frac{J_{\nu-1}(p z_{SM})}{J_{\nu}(p z_{SM})} \right] 
\nonumber \\ 
& & \times \frac{z}{z_{SM}} \left[pz H^{(1)}_{\nu-1}(pz) - (1+\nu) H^{(1)}_{\nu} (pz) \right] 
\label{LRLong-neglectUV}
\end{eqnarray}
If the dependence on the UV brane is restored, one finds the following more complicated expression,
\begin{eqnarray} 
\Delta^{(L)}(p,z>z_{SM})  
& =&  \frac{1}{m^2_5}\frac{z}{z_{SM}} \frac{1}{a_{SM} z_{SM}} \frac{H^{(1)}_{\nu}(pz)}{H^{(1)}_{\nu}(p/\kappa)} \left( pz H^{(1)}_{\nu-1}(pz)/H^{(1)}_{\nu}(pz)-(\nu+1) \right) \nonumber \\ 
& & \times \left(\left[J_{\nu}(p z_{SM}) Y_{\nu}(p /\kappa) - J_{\nu}(p /\kappa) Y_{\nu}(p z_{SM}) \right](1+\nu)  \right. \nonumber \\ 
& & \left. + p z_{SM} \left[ J_{\nu}(p /\kappa) Y_{\nu-1}(p z_{SM})- Y_{\nu}(p /\kappa) J_{\nu-1}(p z_{SM}) \right] \right)
\label{LRlongitudinal}
\end{eqnarray} 
which is equivalent to (\ref{LRLong-neglectUV}) in the limit $p /\kappa\ll 1$. The brane-to-brane propagator is obtained by setting $z=z_{SM}$. 


Next we check this result against the longitudinal propagator in the RS 2 model. The limit $z_{SM} 
\rightarrow \kappa^{-1}$ is straightforward and gives 
\begin{eqnarray} 
\Delta^{(L)}(p,z>\kappa^{-1}) & =&  \frac{\kappa}{m^2_5} z\frac{  H^{(1)}_{\nu}(pz)}{H^{(1)}_{\nu}(p/\kappa)} \left( pz H^{(1)}_{\nu-1}(pz)/H^{(1)}_{\nu}(pz)-(\nu+1) \right) 
\end{eqnarray} 
which agrees with 
(\ref{eq:DeltaL_answer}) 
up to an irrelevant factor of 1/2 (which is the same discrepancy  
the transverse propagator (\ref{LRtransverse}) has with the 
previous computations (\ref{eq:greens_transverse_2}) and 
(\ref{eq:greens_transverse_cp}), so their ratio agrees). 

Finally, taking the limits $z \rightarrow z_{SM}$ and $p /\kappa \ll 1$ 
gives the longitudinal brane-to-brane 
propagator in limit that the UV dependence is dropped, 
\begin{eqnarray} 
\Delta^{(L)}(p) &=& i\frac{\pi}{2 m^2_5} \frac{1}{a_{SM} z_{SM}} J_{\nu}(p z_{SM}) 
  \left[1+ \nu
- p z_{SM} \frac{J_{\nu-1}(p z_{SM})}{J_{\nu}(p z_{SM})} \right] 
\nonumber \\ 
& & \times  \left[(1+\nu) H^{(1)}_{\nu} (p z_{SM}) 
-pz_{SM} H^{(1)}_{\nu-1}(p z_{SM}) \right] 
\label{longitudinalb2bprop}
\end{eqnarray}

In the low-energy limit $p z_{SM} \ll 1$ one obtains from (\ref{longitudinalb2bprop})
\begin{eqnarray} 
\Delta^{(L)}(p) & = &  \frac{1}{m^2_5} \frac{1}{a_{SM} z_{SM}}  \left[ -\frac{\nu^2-1}{2 \nu} + \frac{1}{4} \frac{1}{\nu} (p z_{SM})^2 + \cdots \right. \\ 
& & \left. + i \frac{\pi}{2 (\Gamma[\nu+1])^2} (1 + i \cot \pi \nu)
\left(\nu-1\right)^2 \left(\frac{p z_{SM}}{2} \right)^{2 \nu} + \cdots \right]
\label{lowenergyL}
\end{eqnarray} 
where only the first few terms have been shown. 

\subsection{Non-analytic or CFT terms} 

At first sight it appears difficult for 
the non-analytic terms in the transverse and longitudinal propagators to combine 
at low-energies into the form required by conformal invariance; they don't even appear to be of the same magnitude. This isn't the case though. 
The transverse propagator (\ref{transverseb2bprop}) 
scales as $a_{SM} z_{SM} = \kappa^{-1}$, whereas the 
longitudinal propagator (\ref{longitudinalb2bprop}) scales as $m^{-2}_5/( a_{SM} z_{SM}) = (\nu^2-1)^{-1} \kappa^{-1}$. 
So they are indeed parametrically the same size. 

Next we combine the low-energy expansions,  (\ref{lowenergyT}) and (\ref{lowenergyL}). 
After some short algebra one has 
\begin{eqnarray} 
\Delta_{\mu \nu}(p) & =& \left( - \eta _{\mu \nu} + \frac{p _\mu p_\nu}{p^2} \right) \Delta^{(T)}(p) 
- \frac{p_\mu p_\nu}{p^2} \Delta^{(L)}(p) \nonumber \\ 
& = &\cdots +  \frac{\pi \kappa^{-1} }{2 \Gamma[\nu+1]^2} (i- \cot \pi \nu) 
 \left(\frac{p z_{SM}}{2} \right)^{2 \nu} \left( \eta_{\mu \nu} - \frac{2 \nu}{\nu+1}  \frac{p _\mu p_\nu}{p^2} \right) + \cdots \nonumber \\
 & =&  \cdots +\frac{\pi \kappa^{-1} }{2 \Gamma[\nu+1]^2} \frac{\exp[-i \pi \nu]}{\sin [\pi \nu]} 
 \left(\frac{p z_{SM}}{2} \right)^{2 \nu} \left(- \eta_{\mu \nu} + \frac{2 \nu}{\nu+1}  \frac{p _\mu p_\nu}{p^2} 
\right) +\cdots \end{eqnarray} 
 where the local and subleading non-analytic terms are not shown. Note that the non-analytic terms have the correct tensor structure and mass dimension to describe a CFT current-current correlator where the  CFT current has scaling dimension $d_V= 2 + \nu$. Also note that it has the correct unparticle phase.

Next notice the real part of the non-analytic terms has the same divergence at integer $d_V$ as 
the unparticle propagator (\ref{eq:GIRcorr}). However, as with the RS2 propagator, here those divergences are cancelled by the contact terms. This must be the case since the LR propagators do not exhibit any pathology at integer $d_V$, simply because the Bessel functions are entire functions of their order. 

Finally, the scale suppressing this interaction is
$1/z_{SM}$, as expected. The transmutation scale is therefore $\Lambda_{{\cal U}} \simeq 1/z_{SM}$, which we note can be hierarchically smaller than the AdS curvature scale $\kappa$. 
 
 \subsection{Contact Interactions} 
 
Inspecting the expansions for both the transverse and longitudinal propagators indicates that local terms are present, with the scale of these interactions set by $\Lambda_{{\cal U}}= 1/z_{SM}$. In the effective field theory at energies below $1/z_{SM}$ these local terms are given by operators of dimensions 6, 8 and higher  involving only SM currents, so that 
\begin{equation} 
{\cal L}_{eff} \sim  {\cal L}_{SM} + {\cal L}_{CFT_2}+ e j_{\mu,SM} {\cal O}^{\mu}_{CFT_2} + \frac{1}{\Lambda^2_{{\cal U}}}
e^2 j_{SM} \left( c_0+ c_1 p^2/\Lambda^2_{{\cal U}} + \cdots \right) j_{SM}
\end{equation} 
The existence and scale of these interactions is consistent with the picture that the SM fields are composites of a CFT that breaks at the scale $1/z_{SM}$. Indeed, on the basis of effective field theory we expect higher dimension operators to appear at this scale. 

\subsection{Unitarity}

Here we shall show that 
\begin{equation} 
m^2_5  \geq 0
\label{LRunitarity}
\end{equation} 
is a sufficient condition for brane-to-brane scattering amplitudes on the LR brane to preserve unitarity. 

As before, there is both an $s-$channel and a $t-$channel contribution. It turns out the $t-$channel contribution is purely real, so it does not contribute to the imaginary part. Physically this is reasonable, since there is no particle production in this channel. To check this statement mathematically requires analytically continuing the two brane-to-brane propagators (\ref{LRtransverse}) and 
(\ref{LRlongitudinal}) (setting $z=z_{SM}$) to Euclidean space. Although at first glance the Euclidean propagators may not appear to be real, we have confirmed that indeed they are. That leaves the $s-$channel contribution. 

First lets look at the transverse brane-to-brane propagator. 
It turns out that for this purpose a more convenient expression than (\ref{LRtransverse}) for the brane-to-brane propagator is instead (\ref{LRtransverse-compact}) 
\begin{equation} 
\Delta^{(T)}(p) = - i \frac{\pi}{2}( a_{SM} z_{SM}) \frac{H^{(1)}_{\nu}(p z_{SM})}{1+ i b(p)} \left( J_{\nu}(p z_{SM}) + b(p) Y_{\nu}(p z_{SM}) \right) 
\end{equation} 
where $b(p)$ is purely real. 

Recall that the unitarity condition 
$\hbox{Im}{\cal T} \geq 0$ requires $\hbox{Im} \Delta^T(p) \leq 0$, where we follow our convention in  (\ref{eq:corr_answer_combined}) for  the overall signs of the transverse and longitudinal propagators.
Using $H^{(1)}_{\nu}(x) = J_{\nu}(x) + i Y_{\nu}(x)$, 
\begin{eqnarray} 
\Delta^{(T)}(p) &=& -i  \frac{\pi}{2} \frac{a_{SM} z_{SM}}{1+b(p)^2}  \left[J_{\nu}( p z_{SM}) + b(p) Y_{\nu}(p z_{SM}) \right]^2  \nonumber \\
& & \! \! \! \! \! - \frac{\pi}{2} \frac{a_{SM} z_{SM}}{1+b(p)^2}  \left(J_{\nu}( p z_{SM})+ b(p) Y_{\nu}(p z_{SM}) \right) \left[ b(p) J_{\nu}(p z_{SM}) - Y_{\nu}(p z_{SM}) \right]  
\end{eqnarray} 
which indeed satisfies $\hbox{Im} \Delta^{(T)}(p) \leq 0$.

Next, the longitudinal propagator must satisfy $\hbox{Im} \Delta^L(p) \geq 0$. From (\ref{LRlongitudinal}) the longitudinal propagator can be written, after some 
rearrangement of terms,  as 
\begin{eqnarray} 
\Delta^{(L)}(p,z_{SM}) &=&
 \frac{1}{m^2_5} \frac{1}{a_{SM} z_{SM}} \frac{J_{\nu}(p/\kappa)-i Y_{\nu}(p /\kappa)}{\sqrt{\left[J_{\nu}(p/\kappa)\right]^2 + \left[Y_{\nu}(p /\kappa)\right]^2}} \left( pz_{SM} H^{(1)}_{\nu-1}(pz_{SM})-(\nu+1) H^{(1)}_{\nu}(pz_{SM})\right) \nonumber \\ 
& & \times \left(\left[J_{\nu}(p z_{SM}) Y_{\nu}(p /\kappa) - J_{\nu}(p /\kappa) Y_{\nu}(p z_{SM}) \right](1+\nu)  \right. \nonumber \\ 
& & \left. + p z_{SM} \left[ J_{\nu}(p /\kappa) Y_{\nu-1}(p z_{SM})- Y_{\nu}(p /\kappa) J_{\nu-1}(p z_{SM}) \right] \right)
\end{eqnarray} 
so that 
\begin{eqnarray} 
\hbox{Im} \Delta^{(L)}(p,z_{SM}) &=& 
 \frac{1}{m^2_5} \frac{1}{a_{SM} z_{SM}} \frac{1}{\sqrt{\left[J_{\nu}(p/\kappa)\right]^2 + \left[Y_{\nu}(p /\kappa)\right]^2}} 
  \nonumber \\ 
  & & \times \left( J_{\nu}(p/\kappa) [ pz_{SM} Y_{\nu-1}(pz_{SM})-(\nu+1) Y_{\nu}(pz_{SM})] 
  \right. \nonumber \\ 
  & & \left. - Y_{\nu}(p/\kappa)  [ pz_{SM} J_{\nu-1}(pz_{SM})-(\nu+1) J_{\nu}(pz_{SM})]
 \right) \nonumber \\ 
 & & \times  \left(\left[J_{\nu}(p z_{SM}) Y_{\nu}(p /\kappa) - J_{\nu}(p /\kappa) Y_{\nu}(p z_{SM}) \right](1+\nu)  \right. \nonumber \\ 
& & \left. + p z_{SM} \left[ J_{\nu}(p /\kappa) Y_{\nu-1}(p z_{SM})- Y_{\nu}(p /\kappa) J_{\nu-1}(p z_{SM}) \right] \right) \nonumber \\
& = & \frac{1}{m^2_5} \frac{1}{a_{SM} z_{SM}} \frac{1}{\sqrt{\left[J_{\nu}(p/\kappa)\right]^2 + \left[Y_{\nu}(p /\kappa)\right]^2}} 
  \nonumber \\ 
  & & \times  \left(\left[J_{\nu}(p z_{SM}) Y_{\nu}(p /\kappa) - J_{\nu}(p /\kappa) Y_{\nu}(p z_{SM}) \right](1+\nu)  \right. \nonumber \\ 
& & \left. + p z_{SM} \left[ J_{\nu}(p /\kappa) Y_{\nu-1}(p z_{SM})- Y_{\nu}(p /\kappa) J_{\nu-1}(p z_{SM}) \right] \right)^2
\end{eqnarray} 
which satisfies the unitarity condition $\hbox{Im} \Delta^{(L)}(p) \geq 0$ provided $m^2_5 \geq 0$, 
which is what we wanted to show.

\subsection{High Energy Limit} 

To simplify the presentation we work in Euclidean space $p^2 >0 $ with signature $(++++)$ and drop the dependence of the UV boundary (i.e., $p /\kappa  \ll1$). Then dropping irrelevant overall factors, 
\begin{equation} 
\Delta^{(T)}_E(p,z= z_{SM}) \propto   K_{\nu}( p z_{SM}) I_{\nu}(p z_{SM}) 
\label{LRtransverseE:neglectb(p)}
\end{equation} 
In the high-energy limit $ p z_{SM} \gg 1$ this reduces to 
\begin{equation} 
\Delta^{(T)}_E(p,z_{SM})  \propto  \frac{1}{\sqrt{p^2}} +\cdots 
\end{equation} 
For the longitudinal mode one has in this limit 
\begin{eqnarray} 
\Delta^{(L)}_E(p,z= z_{SM}) & \propto &  \frac{1}{m^2_5} K_{\nu}(p z_{SM}) I_{\nu}(p z_{SM})   \nonumber \\
& & \times 
\left[ z \frac{d}{d z} \log K_{\nu}(p z) \right] \left[  z \frac{d}{d z} \log I_{\nu}(p z) \right] \large|_{z=z_{SM}} +\cdots \nonumber \\
 & \propto  & \frac{\sqrt{p^2}}{m^2_5} + \cdots 
\end{eqnarray} 
Neither of these results are analytic in $p$, demonstrating
that, as in RS 2,  the ``contact" interactions seen at $p \ll z^{-1}_{SM}$ are not contact at all, but are resolved at energies above this scale (i.e., at distance scales of  the order of $x^\mu \sim z_{SM}$). 


\section{Conclusions}
\label{conclusions}

\emph{Summary of results}.

\begin{itemize} 
\item We have derived the (tree-level) propagator for a massive vector
  boson in the RS 2 background, evaluated for observers
  living on the UV brane or, more generally, a probe brane
  (``Lykken-Randall'' model). The results, given in
  Sect.~\ref{sec:greensfunction} and in Eqs. (\ref{LRtransverse}) and
  (\ref{LRlongitudinal}) of Sect.~\ref{sec:tensionlessbrane}, include
  both the longitudinal and transverse components. As far as we know,
  these expressions have not previously appeared in the literature. 

\item We have presented a comprehensive analysis of this propagator,
  in particular showing that the required properties of unparticles
  listed in the Introduction are all fulfilled.

\item The propagator does not have a CFT form. Rather, at low energies
  it is dominated by short-distance interactions and contains a
  subdominant nonanalytic (CFT) piece. At high energies, the
  propagator has the form expected in flat five-dimensional space.

\item The nonanalytic piece, in addition to the obvious power-law
  dependence on the momentum, has also all the other properties
  expected of a CFT-mediated interaction. The phase agrees with that
  of \cite{GeorgiII} and the CFT tensor structure
  \cite{Schreier1971,ADSCFT_Vector,GIR} is reproduced upon combining
  the transverse and longitudinal components. 

\item  The imaginary part of the
  propagator is related to the rate of escape of the vector particles
  into the bulk \cite{Dubovsky:2000am}. Requiring that this rate be
  nonnegative, particularly for the longitudinal polarization, gives
  $m_{5V}^2\ge0$ for the bulk mass of the vector.  This condition, to
  the best of our knowledge, has not been previously discussed. It
  should be contrasted with the well-known result for a scalar in the
  AdS background, in which case negative values of $m_{5S}^2$ are
  allowed \cite{Breitenlohner:1982bm}. The bound $m_{5V}^2\ge0$
  generalizes unchanged to $D$ spacetime dimensions on the brane.  For
  the nonanalytic piece, it implies the lower bound on the conformal
  dimension $d_V\ge D-1$, reproducing the $D$ dimensional
  generalization \cite{Minwalla:1997ka} of Mack's unitarity bound
  \cite{Mack}.

\item For the RS 2 model, we have also presented a detailed analysis
  of the vector and scalar correlators in position space.  As far as
  we know, such an analysis is also new.  The ``contact'' terms of the
  low-energy expansion are seen to be resolved at short distances. The
  propagator exhibits two limiting regimes: a flat five-dimensional
  regime at short distances and a \emph{pure} CFT regime at long
  distances.
%
  The transition between these regimes
  deserves some discussion. For large values of the bulk mass, it
  occurs rather abruptly, at distances $\sim\kappa^{-1}$, both for the
  vector and scalar cases. In contrast, when $m_5\ll\kappa$, the
  transition regime becomes extended and pure CFT sets in at
  larger distances. In fact, as the bulk mass is taken to zero, the pure CFT
  regime is pushed off to infinity.
  The vector and scalar propagators behave quite differently in the
  transition regime. The scalar interaction is dominated by a (near
  zero-mode) state bound to the brane, and thus is essentially
  four-dimensional. The vector transition is instead
  characterized by a (weakly) bound mode mixed with the bulk KK
  states.

\end{itemize}

\emph{Discussion}. Finally, it is worth discussing two additional
aspects of our analysis: its connection to the AdS/CFT correspondence
and possible extensions involving nontrivial multi-point functions.

First, on the connection to the AdS/CFT correspondence.  The latter,
as proposed in \cite{Maldacena}, connects supergravity (string theory)
on the AdS$_5\times S_5$ background to a highly supersymmetric (${\cal
  N}=4$) $SU(N_c)$ super-Yang-Mills theory at large $N_c$
\footnote{Dual theories with fewer supersymmetries (with
  AdS$_5\times X_5$ on the gravity side) have also been discussed. See
  \cite{KlebanovReview} for an overview.}. It should be clear to the
reader that what we consider here is not literally the same: the bulk
fields in our case do not come in supermultiplets and the KK states
associated with the compact $S_5$ ($X_5$) coordinates do not show up
at the scale of the AdS curvature. In fact, our constructions are
intended as field theory models in a putative curved background.
At the same time, a qualitative
connection between the models we study and Yang-Mills theories with
large $N_c$ and large 't Hooft coupling is expected. In this sense,
our analysis extends the results found by Ref.\cite{GIR} at weak
coupling to this regime of parameters.


Even without specifying the exact dual CFT, as mentioned in
Sect.~\ref{sect:leterature}, one knows to expect a field theory on
AdS$_{d+1}$ to be linked to some conformal field theory on the
boundary \cite{Witten,ADSCFT_Vector}. It is then perhaps not
surprising that a CFT shows up in the RS 2 (LR) models. As we stressed
already, however, the important point is that, for a bulk vector in the
RS 2 background, this CFT is \emph{subdominant} to short-distance
interactions. Tracing back to the
derivations of \cite{Witten,ADSCFT_Vector,Aharony:1999ti}, one notices
a important difference in our procedure compared to what was done
there. In \cite{Witten,ADSCFT_Vector,Aharony:1999ti}, as the brane is
taken to the boundary of the AdS space, the dominant short-distance
interactions become point-like and, if one wants to normalize to the
CFT piece, infinite in strength.  They are then subtracted out by
local counter-terms. In contrast, in our analysis, the brane is at a
fixed position and the short-distance terms are physical. They capture
the fact that the field is largely (but not completely) bound to the brane.
They \emph{must} be kept, and play a crucial role in
phenomenology \cite{GIR}.

The RS 2/AdS/CFT connection has been extensively discussed before,
although, it is interesting to note, usually in what we called above
the transitional regime. Specifically, as the dual 4d description one
considers a photon mixing with the CFT via a sequence of the CFT
bubble insertions on the photon line
\cite{WittenSantaBarbara,Gubser:1999vj,GiddingsKatzRandall,ArkaniHamed:2000ds}.
This \emph{quantum} effect in the CFT picture is characteristically
captured by the \emph{classical} computations in the AdS background
and is clearly seen in our analysis, as outlined above.

The connection can be also easily seen in the large-distance behavior
of the RS 2 propagator, where, as we saw, the CFT interaction shows up
directly. This regime is generically present, with the exception of
the strictly massless case ({\it i.e.}, for general CFT dimension
$d_V\ne3$). In fact, for $m_5 \sim \kappa$, pure CFT is what one sees
more or less immediately at distances $\kappa^{-1}$, as the theory
transitions out of the flat 5d regime. In the four-dimensional
description, this means the ``photon'' is massive and, at large
distances, can be integrated out leaving pure CFT.

That the CFT behavior dominates at large distance raises a puzzle; for
one might conclude that the CFT dominates low-energy scattering
amplitudes. Yet, we saw that for vector operators the non-analytic
contribution is always subdominant in momentum space. So how can these
two statements both be correct?  The point is to distinguish between
plane-wave scattering and scattering at fixed impact parameter.
Plane-wave scattering amplitudes averages over all impact parameters,
so all distance scales contribute. By contrast, scattering amplitudes
at fixed and large enough impact parameter are dominated by the
interactions at that distance scale, and therefore by the CFT.

Finally, it is important to consider the theory beyond its propagator.
It is well-known that, in the case of a scalar CFT operators,
two-point and three-point functions are fixed by conformal symmetry,
up to constants. Any realization of a CFT must, therefore, lead to the
same form and it is reassuring that the models we consider obey the
required unparticle properties\footnote{The presence and dominance of
  the contact terms also follows from general principles, those of
  effective field theory and dimensional arguments
  \protect{\cite{Friedland:2009iy}}.}.  At the same time, four-point
functions and on (and, in the case of bulk QED, already the
three-point function \cite{OsbornPetkou,ADSCFT_Vector}) are not
uniquely fixed by the symmetry. By studying these, we can learn what
kinds of CFTs are obtained with the RS 2/LR realizations ({\it cf.}
\cite{ADSCFT_Vector}). This is also important from the
phenomenological point of view, to describe the ``decay'', or
``showering'' of unparticles back to the Standard Model states.

The higher-point functions in AdS are obtained by adding field
interactions in the bulk, yielding the Witten diagrams \cite{Witten}.
Various aspects of these calculations have been since extensively
studied, \cite{Volovich1998,seiberg,ADSCFT_Vector}. In the case of the
scalar field in the RS 2 model, a three-point function is analyzed in
\cite{PerezVictoria:2001pa} and both the contact terms and the
noncontact CFT interaction are discussed. In the setup of the
Super-Yang-Mills theories at large, but finite $N_c$, three-point
functions appear, scaling as $1/N_c$ and independent of the 't Hooft
coupling \cite{seiberg}. See \cite{MaldacenaTASI,dhokerfreedmanlectures} for a
clear review and further details.

From the point of view of experimental signatures, an extremely
important observation is that the shape of the showers is expected to
be qualitatively different \cite{Strassler:2008bv} in the AdS-based
models and in the weakly coupled QCD-like CFTs. For CFTs that have an
AdS dual description, the shower is more spherical, and less like a
QCD-jet \cite{Strassler:2008bv,Hofman:2008ar,Csaki:2008dt}.  Refs.
\cite{Polchinski:2002jw,Csaki:2008dt} have also investigated certain
features of gauge theories as the 't Hooft coupling and numbers of
colors $N_c$ are varied.

This illustrates the following basic point: different regimes of
unparticles are possible. The RS 2 and LR models considered here
capture one such regime. We, therefore, would like to stress that the
models we considered represent \emph{realizations} of unparticle
physics. The RS2/unparticle relation is not to be viewed as a one-to-one
correspondence or duality.

Characterizing signatures of conformal fields theories will continue
to be a fascinating subject and we hope to contribute more to it in
future work.

\acknowledgments{The authors would like to thank T. Bhattacharya,
  Y.Kats, Y. Nomura, L. Randall, M.~Perez-Victoria, Yu. Shirman, S.
  Thomas and L. Vecchi for valuable comments. This work was
  supported by the U.S.  Department of Energy at Los Alamos National
  Laboratory under contract No. DE-AC52-06NA25396.}

\end{document}